\documentclass[11pt,letterpaper]{article}

\usepackage[utf8x]{inputenc}%
\usepackage{tcolorbox,fancybox}%
\usepackage{physics}
\usepackage{cite}
\usepackage{moresize}
\usepackage{ntheorem}
\usepackage{mathtools}

\usepackage[english]{babel}
\usepackage{amsmath,amssymb,amsfonts}
\usepackage{graphicx}
\usepackage{booktabs}
\usepackage{color,xcolor}
\usepackage[numbers,sort&compress]{natbib}

\usepackage{tikz}
\usepackage[compat=1.1.0]{tikz-feynman}
\usetikzlibrary{positioning}
\usetikzlibrary{decorations.text}
\usetikzlibrary{decorations.pathmorphing}

\usetikzlibrary{calc}
\usetikzlibrary{shapes.misc}
\tikzset{
        >=latex,
    photon/.style={decorate, decoration={snake}, draw=black, thick},
    fermionnoarrow/.style={draw=black, postaction={decorate}, thick},
    scalar/.style={draw=black, postaction={decorate}, decoration={markings,mark=at position .55 with {\arrow{>}}}, thick, dashed},
    scalarnoarrow/.style={draw=black, postaction={decorate},  thick, dashed},
    fermion/.style={draw=black, postaction={decorate},decoration={markings,mark=at position .55 with {\arrow{>}}}, thick},
    gluon/.style={decorate, draw=black, decoration={coil,amplitude=4pt, segment length=5pt}, thick},
    vertex/.style={draw,shape=circle,fill=black,minimum size=3pt,inner sep=0pt},
    fillvertex/.style={draw,shape=circle,fill=black,minimum size=5pt,inner sep=0pt},
    openvertex/.style={draw,shape=circle,minimum size=5pt,inner sep=0pt},
    blob/.style={draw=red,shape=circle,fill=red,minimum size=6pt,inner sep=0pt},
    redvertex/.style={draw=red,shape=circle,fill=red,minimum size=3pt,inner sep=0pt},
    cross/.style={cross out, draw=black,thick, minimum size=5pt, inner sep=0pt, outer sep=0pt}
}

\usepackage{physics}
\usepackage{orcidlink}
\usepackage{slashed}
\usepackage{mathrsfs}
\usepackage{enumerate}
\usepackage{mdframed}
\usepackage{url}
\usepackage[makeroom]{cancel}
\usepackage{geometry}

\usepackage{pifont}

\newtheorem*{thm-non}{Theorem}
\newtheorem*{conj}[thm-non]{Conjecture}

\newtheorem*{define}[thm-non]{Definition}

\usepackage{hyperref} 
\hypersetup{
    colorlinks=true,       
    linkcolor=red,          
    citecolor=blue,        
    filecolor=magenta,      
    urlcolor=purple           
}
\usepackage[all]{hypcap} 

\def\beqn{\begin{eqnarray}}
\def\eeqn{\end{eqnarray}}
\def\beqs{\begin{subequations}}
\def\eeqs{\end{subequations}}
\def\beq{\begin{equation}}
\def\eeq{\end{equation}}
\def\ba{\begin{array}}
\def\ea{\end{array}}

\def\non{\nonumber\\}

\def\hf{\frac{1}{2}}
\def\[{\left[}
\def\]{\right]}
\def\({\left(}
\def\){\right)}
\newcommand\para{\paragraph{}}

\def\gSU{\rm SU}

\def\gSO{\rm SO}

\newcommand{\rep}[1]{\mathbf{#1}}
\newcommand{\repb}[1]{\mathbf{\overline{#1}}}

\def\Bc{\mathcal{B}}

\def\Dc{\mathcal{D}}
\def\Ec{\mathcal{E}}
\def\Fc{\mathcal{F}}
\def\Gc{\mathcal{G}}

\def\Lc{\mathcal{L}}
\def\Mc{\mathcal{M}}
\def\Nc{\mathcal{N}}
\def\Oc{\mathcal{O}}
\def\Pc{\mathcal{P}}
\def\Qc{\mathcal{Q}}
\def\Rc{\mathcal{R}}

\def\Tc{\mathcal{T}}
\def\Uc{\mathcal{U}}

\def\Xc{\mathcal{X}}
\def\Yc{\mathcal{Y}}

\def\DG{\mathfrak{D}}  \def\dG{\mathfrak{d}}
\def\EG{\mathfrak{E}}  \def\eG{\mathfrak{e}}

  \def\nG{\mathfrak{n}}

  \def\sG{\mathfrak{s}}
  
\def\UG{\mathfrak{U}}  \def\uG{\mathfrak{u}}

\title{
{\bf Further study of the maximally symmetry breaking patterns in an ${\rm SU}(8)$ theory} \\
\author{\large Ning Chen$^1$\,\orcidlink{0000-0002-0032-9012}, Zhiyuan Chen$^2$\,\orcidlink{0009-0004-7518-5828}, Zhanpeng Hou$^3$\,\orcidlink{0000-0002-6035-368X}, Zhaolong Teng$^4$\, \orcidlink{0000-0002-7141-2331}, Bin Wang$^5$\,\orcidlink{0009-0001-8556-7115} }
\date{\small \it
$^1\, ^2 \, ^3 \, ^4 \, ^5$School of Physics, Nankai University, Tianjin, 300071, China \\
}
}

\begin{document}

\maketitle
\setlength{\parskip}{0.2ex}

\begin{abstract}
\bigskip
An ${\rm SU}(8)$ theory was previously found to be the minimal simple gauge group where all three-generational Standard Model (SM) fermions can be nontrivially embedded. 
It is maximally broken into a subgroup of ${\rm SU}(8)\to \Gc_{441}\equiv {\rm SU}(4)_s \otimes {\rm SU}(4)_W \otimes {\rm U}(1)_{X_0}$ at the grand unified theory scale by the ${\rm SU}(8)$ adjoint Higgs field of $\rep{63_H}$.
Gauge symmetries in the strong and the weak sectors are extended by one and two ranks, respectively.
The sequential strong-weak-weak (SWW) symmetry breaking stages were found to generate the observed hierarchical SM quark/lepton masses as well as the Cabibbo-Kobayashi-Maskawa mixing pattern with the precise flavor identifications~\cite{Chen:2023qxi,Chen:2024cht}.
We further study the possible weak-strong-weak and weak-weak-strong symmetry breaking patterns, and compare with the results that we have obtained by following the SWW sequence.
The two-loop renormalization group equations following both patterns are analyzed, where we cannot achieve the gauge coupling unification in the field theory framework.
Through these analyses, we suggest the gauge coupling unification to be interpreted in the context of the affine Lie algebra.
\end{abstract}

\vspace{7.5cm}
{\emph{Emails:}\\  
$^{\,1}$\url{chenning_symmetry@nankai.edu.cn}\\
$^{\,2}$\url{chenzhiyuan@mail.nankai.edu.cn}\\
$^{\,3}$\url{houzhanpeng@mail.nankai.edu.cn}\\
$^{\,4}$\url{tengcl@mail.nankai.edu.cn}\\
$^{\,5}$\url{wb@mail.nankai.edu.cn}\\
 }

\thispagestyle{empty}  
\newpage  
 
\setcounter{page}{1}  

\vspace{1.0cm}
\eject
\tableofcontents

\section{Introduction}
\label{section:intro}
%
%

\para
The grand unified theory (GUT) was proposed to unify all three fundamental symmetries described by the Standard Model (SM).
Two original versions that have been widely studied over decades are the $\gSU(5)$ Georgi-Glashow theory~\cite{Georgi:1974sy} and the $\gSO(10)$ Fritzsch-Minkowski theory~\cite{Fritzsch:1974nn}, as well as their supersymmetric extensions~\cite{Dimopoulos:1981zb}.
However, the minimal unified frameworks cannot address much on the existing flavor puzzles of all three-generational SM fermions, unless further ingredients are included~\cite{Georgi:1979ga,Ellis:1979fg,Chen:2021zty,King:2021fhl,Ding:2021zbg,Ding:2021eva,deMedeirosVarzielas:2023ujt,deAnda:2023spb}.

\para
From the experimental perspective, the indispensable guidance to unveil the flavor puzzle (particularly for quarks and leptons) to date is the discovery~\cite{ATLAS:2012yve,CMS:2012qbp} and the measurements~\cite{CMS:2022dwd,ATLAS:2022vkf} of one single SM Higgs boson.
The LHC measurements have confirmed that the Yukawa couplings at the electroweak (EW) scale between the SM Higgs boson and the $(t\,,b\,,\tau\,,\mu)$ are consistent with the SM predictions.
Based on the experimental facts, one natural conjecture is that the hierarchical Yukawa couplings of the SM Higgs boson originate from the flavor nonuniversalities to all three generations beyond the SM.
Obviously, the minimal $\gSU(5)$ or $\gSO(10)$ theories only allow the simply repetitive flavor structure.
In an earlier study by Georgi~\cite{Georgi:1979md}, he proposed to extend the gauge group beyond the minimal ${\rm SU}(5)$ so that three-generational SM fermions can be nontrivially embedded.
Consequently, three-generational SM fermions must transform differently in the UV-complete theory.
In the original third law of the $\gSU(N)$ unified theory, Georgi determined to have nonrepetitive ${\rm SU}(N)$ irreducible representations (IRs), which lead to a minimal three-generational ${\rm SU}(11)$ framework with the chiral fermions of
\beqn\label{eq:Georgi_SU11}
\{ f_L \}_{ {\rm SU}(11)}^{n_g=3}&=&  \[ 11\,,4 \]_{\rep{F}}  \oplus  \[ 11\,, 8\]_{\rep{F}}  \oplus  \[ 11\,,9 \]_{\rep{F}}  \oplus \[ 11\,,10 \]_{\rep{F}}  \,.
\eeqn 
In recent studies, we introduced a concept of chiral irreducible anomaly-free fermion sets (IRAFFSs), which reads~\cite{Chen:2023qxi}:
\begin{define}\label{def:IRAFFS}

A chiral IRAFFS is a set of left-handed antisymmetric fermions of $\sum_\Rc m_\Rc \, \Fc_L(\Rc)$, with $m_\Rc$ being the multiplicities of a particular fermion representation of $\Rc$.
Obviously, the anomaly-free condition reads $\sum_\Rc m_\Rc \, {\rm Anom}(  \Fc_L(\Rc) ) =0$.
We also require the following conditions to be satisfied for a chiral IRAFFS:
\begin{itemize}

\item the greatest common divisor (GCD) of the $\{ m_\Rc \}$ should satisfy that ${\rm GCD} \{  m_\Rc \} =1$;

\item the fermions in a chiral IRAFFS can no longer be removed, which would otherwise bring nonvanishing gauge anomalies;

\item there should not be any singlet, self-conjugate, adjoint fermions, or vectorial fermion pairs in a chiral IRAFFS.

\end{itemize}

\end{define}
Correspondingly, Georgi's 1979 third law~\cite{Georgi:1979md} can be reformulated as follows~\cite{Chen:2023qxi}.
\begin{conj}[Chen]

Only distinctive chiral IRAFFSs are allowed in the GUT.

\end{conj}
This leads to an ${\rm SU}(8)$ theory with the minimal chiral fermions of
\beqn\label{eq:SU8_3gen_fermions}
\{ f_L \}_{ {\rm SU}(8)}^{n_g=3}&=& \Big[ \repb{8_F}^\omega \oplus \rep{28_F} \Big] \bigoplus \Big[ \repb{8_F}^{ \dot \omega } \oplus \rep{56_F} \Big] \,,~ {\rm dim}_{ \mathbf{F}}= 156\,, \non
&& \Omega \equiv ( \omega \,, \dot \omega ) \,, ~ \omega = ( 3\,, {\rm IV}\,, {\rm V}\,, {\rm VI}) \,, ~  \dot \omega = (\dot 1\,, \dot 2\,, \dot {\rm VII}\,, \dot {\rm VIII}\,, \dot {\rm IX} ) \,,
\eeqn
with undotted/dotted indices for the $\repb{8_F}$'s in the rank-two chiral IRAFFS and the rank-three chiral IRAFFS, respectively.
The Roman numbers and the Arabic numbers are used for the heavy partner fermions and the SM fermions.
The same chiral fermions in Eq.~\eqref{eq:SU8_3gen_fermions} were previously proposed in Ref.~\cite{Barr:2008pn}, while a partition based on the chiral IRAFFSs was not acknowledged.

\para
The ${\rm SU}(8)$ gauge group has rank seven, which predicts three intermediate symmetry breaking stages between the GUT scale and the EW scale.
The adjoint Higgs field of $\rep{63_H}$ achieves the maximally breaking pattern of ${\rm SU}(8)\to \Gc_{441}$ through its vacuum expectation value (VEV) of
\beqn\label{eq:63H_VEV}
\langle  \rep{63_H}\rangle &=&\frac{1}{ 4 } {\rm diag}(- \mathbb{I}_{4\times 4} \,, +\mathbb{I}_{4\times 4} ) v_U \,,
\eeqn
according to Li~\cite{Li:1973mq}.
The ${\rm U}(1)_{X_0}$ charges of the ${\rm SU}(8)$ fundamental representation are defined as follows:
\beqn\label{eq:X0charge}
\hat X_0( \rep{8} ) &\equiv& {\rm diag} ( \underbrace{ - \frac{1}{4}  \mathbb{I}_{4\times 4}  }_{ \rep{4_s} }\,, \underbrace{ + \frac{1}{4}  \mathbb{I}_{4\times 4}  }_{ \rep{4_W} } )\,.
\eeqn
In this pattern, the gauge symmetries in the strong and the weak sectors are extended by one and two ranks beyond the SM gauge groups, respectively.
The sequential symmetry breaking patterns may follow the strong-weak-weak (SWW)~\cite{Chen:2023qxi,Chen:2024cht,Chen:2024deo}, the weak-strong-weak (WSW), and the weak-weak-strong (WWS) sequences. 
These cannot be determined purely by group theory.
Other different symmetry breaking patterns of the ${\rm SU}(8)$ group were previously described in Refs.~\cite{Chakrabarti:1980bn,Ma:1981pr,Adler:2014eha,Adler:2014pga,Adler:2015dba,Adler:2016tqy}.

\para
Through the previous analyses of the SWW symmetry breaking pattern~\cite{Chen:2023qxi,Chen:2024cht}, we found (i) one unique SM Higgs boson responsible for the electroweak symmetry breaking (EWSB) by looking for the origin of the global $\widetilde{ {\rm U}}(1)_{B-L}$ symmetry, (ii) the hierarchical Yukawa couplings to all three-generational SM quarks/leptons due to the intrinsic symmetry properties, and (iii) the reasonable Cabibbo-Kobayashi-Maskawa (CKM) mixing pattern~\cite{Cabibbo:1963yz,Kobayashi:1973fv} in the quark sector.
Three intermediate symmetry breaking scales beyond the EW scale enter into the SM quark/lepton mass matrices, and hence are set by requiring a reasonable matching with the experimental measurements.
Meanwhile, the renormalization group equations (RGEs) of three gauge couplings in the SWW symmetry breaking pattern together with three intermediate scales could not achieve the unification in the field theory context~\cite{Chen:2024deo}.
In this work, we further analyze two other possible WSW and WWS symmetry breaking patterns of the ${\rm SU}(8)$ theory, where the extended weak symmetries breaking pattern of $\Gc_{441} \to \Gc_{431}$~\footnote{An effective $\Gc_{431}$ was previously studied in Ref.~\cite{Gao:1980up}.} happened at the first stage.
Within each symmetry breaking pattern, we shall look for whether the reported hierarchical SM quark/lepton masses as well as the CKM mixing patterns following the SWW sequence in Ref.~\cite{Chen:2024cht} can be reproduced.
Furthermore, we shall derive the RGEs in both the WSW and the WWS symmetry breaking patterns to look for the gauge coupling unification.

\para
The rest of the paper is organized as follows.
In Sec.~\ref{section:pattern}, we review the ${\rm SU}(8)$ framework, where three-generational SM fermions are nontrivially embedded, and hence they transform differently in the UV theory. 
We focus on the WSW and the WWS symmetry breaking patterns, define various gauge ${\rm U}(1)$ and nonanomalous global $\widetilde{\rm U}(1)_T$ charges at different stages, and decompose all ${\rm SU}(8)$ chiral fermions and Higgs fields accordingly.
A set of $d=5$ operators that generate all light SM quark/lepton mass terms other than the top quark mass are collected according to Ref.~\cite{Chen:2024cht}.
In Secs.~\ref{section:WSW_pattern} and \ref{section:WWS_pattern}, we collect the vectorlike fermion masses and the benchmark points in both the WSW and the WWS symmetry breaking patterns. 
In Sec.~\ref{section:RGE}, we obtain the RGEs of the minimal ${\rm SU}(8)$ theory by following both WSW and WWS sequences based on the suggested benchmark points.
The RGE behaviors in none of the symmetry breaking patterns are likely to achieve the conventional gauge coupling unification, and their behaviors also match with what we have found in the SWW sequence~\cite{Chen:2024deo} by and large.
We conclude and make future perspective in Sec.~\ref{section:conclusion}, where we suggest that the gauge coupling unification in the minimal ${\rm SU}(8)$ theory to be interpreted in terms of the affine Lie algebra.
In Appendix~\ref{section:SWW_results}, we collect the SM quark/lepton mass matrices from the SWW symmetry breaking pattern that we have obtained in Refs.~\cite{Chen:2023qxi,Chen:2024cht}.
In Appendixes~\ref{section:WSW_process} and \ref{section:WWS_process}, we present details in both WSW and WWS symmetry breaking patterns, where we describe the procedures to integrate out the massive vectorlike fermions, and also derive the SM quark/lepton mass terms based on a set of $d=5$ fermion bilinear operators and the irreducible Higgs mixing operators.
In order to reproduce the SM quark/lepton mass hierarchies and the CKM mixing pattern in Ref.~\cite{Chen:2024cht}, the SM flavor identifications are modified for the first and second generations accordingly..

\section{The ${\rm SU}(8)$ theory and possible symmetry breaking patterns}
\label{section:pattern}

\subsection{Overview}

\para
The ${\rm SU}(8)$ theory was formulated by requiring several distinctive chiral IRAFFSs that can lead to three-generational SM fermions at the electroweak scale~\cite{Chen:2023qxi,Chen:2024cht}.
The nonanomalous global Dimopolous-Raby-Susskind (DRS) symmetries~\cite{Dimopoulos:1980hn} from fermions in Eq.~\eqref{eq:SU8_3gen_fermions} are
\beqn\label{eq:DRS_SU8}
\widetilde{ \Gc}_{\rm DRS} \[{\rm SU}(8)\,, n_g=3 \]&=& \Big[ \widetilde{ {\rm SU}}(4)_\omega  \otimes \widetilde{ {\rm U}}(1)_{T_2} \Big]  \bigotimes \Big[ \widetilde{ {\rm SU}}(5)_{\dot \omega } \otimes \widetilde{ {\rm U}}(1)_{T_3}   \Big]  \,,
\eeqn
and we also denote the anomalous global Peccei-Quinn (PQ) symmetries~\cite{Peccei:1977hh} as
\beqn\label{eq:PQ_SU8}
\widetilde{ \Gc}_{\rm PQ} \[{\rm SU}(8)\,, n_g=3 \]&=& \widetilde{ {\rm U}}(1)_{{\rm PQ}_2} \bigotimes \widetilde{ {\rm U}}(1)_{ {\rm PQ}_3}  \,.
\eeqn
Since the global $B-L$ symmetry should be identical for all three generations, we further require a common nonanomalous $\widetilde{ {\rm U}}(1)_{T}\equiv \widetilde{ {\rm U}}(1)_{T_2} = \widetilde{ {\rm U}}(1)_{T_3}$ between two chiral IRAFFSs.
The anomalous global $\widetilde{ {\rm U}}(1)_{\rm PQ}$ charges are assigned such that
\beqn\label{eq:PQcharges_SU8}
&&  p : q_2  \neq -3 : +2  \,, \quad p  : q_3 \neq -3  : +1  \,.
\eeqn
%
%

\begin{table}[htp]
\begin{center}
\begin{tabular}{c|cccc}
\hline\hline
 Fermions &  $\repb{8_F}^\Omega$ &  $\rep{28_F}$  &  $\rep{56_F}$  &     \\[1mm]
\hline
$\widetilde{ {\rm U}}(1)_T$ &  $-3t$  &  $+2t$  & $+t$ &      \\[1mm]
$\widetilde{ {\rm U}}(1)_{\rm PQ}$ &  $p$  &  $q_2$  & $q_3$ &      \\[1mm]
\hline
Higgs  &  $\repb{8_H}_{\,, \omega }$  & $\repb{28_H}_{\,, \dot \omega }$   & $\rep{70_H}$ &  $\rep{63_H}$    \\[1mm]
\hline
$\widetilde{ {\rm U}}(1)_T$ &  $+t$  &  $+2t$   &  $-4t$  &  $0$    \\[1mm]
$\widetilde{ {\rm U}}(1)_{\rm PQ}$ &  $-(p+q_2)$  &  $-(p+q_3 )$  & $-2q_2$ &  $0$ \\[1mm]
\hline\hline
\end{tabular}
\end{center}
\caption{The nonanomalous $\widetilde{ {\rm U}}(1)_T$ charges and the anomalous global $\widetilde{ {\rm U}}(1)_{\rm PQ}$ charges for the $\gSU(8)$ fermions and Higgs fields.}
\label{tab:U1TU1PQ}
\end{table}%

\para
The most general gauge-invariant Yukawa couplings at least include the following renormalizable and nonrenormalizable terms:~\footnote{The term of $\rep{56_F} \rep{56_F} \rep{28_H}  + {\rm H.c.}$ vanishes due to the antisymmetric property~\cite{Barr:2008pn}. Instead, only a $d=5$ nonrenormalizable term of $\frac{1}{ M_{\rm pl} } \rep{56_F}  \rep{56_F}  \repb{28_{H}}_{\,,\dot \omega }^\dag  \rep{63_{H}} $ is possible to generate masses for vectorlike fermions in the $\rep{56_F}$. Since it transforms as an $\widetilde{ {\rm SU}}(5)_{\dot \omega }$ vector and carries nonvanishing $\widetilde {\rm U}(1)_{\rm PQ}$ charge of $p+3q_3 \neq 0$ from Eq.~\eqref{eq:PQcharges_SU8}, it is only possible due to the gravitational effect.}
\beqn\label{eq:Yukawa_SU8}
-\Lc_Y&=& Y_\Bc  \repb{8_F}^\omega  \rep{28_F}  \repb{8_{H}}_{\,,\omega }  +  Y_\Tc \rep{28_F} \rep{28_F} \rep{70_H} \non
&+&  Y_\Dc \repb{8_F}^{\dot \omega  } \rep{56_F}  \repb{28_{H}}_{\,,\dot \omega }   + \frac{ c_4 }{ M_{\rm pl} } \rep{56_F}  \rep{56_F}  \repb{28_{H}}_{\,,\dot \omega }^\dag  \rep{63_{H}} + H.c.\,,
\eeqn
with the reduced Planck scale of $M_{\rm pl}= ( 8 \pi G_N)^{-1/2}= 2.4 \times 10^{18}\, {\rm GeV}$.
All renormalizable Yukawa couplings and nonrenormalizable Wilson coefficients are expected to be $(Y_\Bc \,, Y_\Tc\,, Y_\Dc\,, c_4 )\sim\Oc(1)$.
Altogether, we collect the ${\rm SU}(8)$ Higgs fields as follows:
\beqn\label{eq:SU8_Higgs}
 \{ H \}_{ {\rm SU}(8)}^{n_g=3} &=& \repb{8_H}_{ \,, \omega}  \oplus \repb{28_H}_{ \,, \dot \omega}  \oplus \rep{70_H}  \oplus \underline{ \rep{63_H} } \,,~ {\rm dim}_{ \mathbf{H}}= 547 \,,
\eeqn
where the adjoint Higgs field of $\rep{63_H}$ is real, while all others are complex.
Accordingly, the non-anomalous global $\widetilde{ {\rm U}}(1)_T$ charges and the anomalous global $\widetilde{ {\rm U}}(1)_{\rm PQ}$ charges for all fermions and Higgs fields are assigned in Table~\ref{tab:U1TU1PQ}.

\subsection{The SWW symmetry breaking pattern}

\para
The SWW symmetry breaking pattern of the ${\rm SU}(8)$ theory follows the sequence of
\beqn\label{eq:Pattern-A} 
&& {\rm SU}(8) \xrightarrow{ v_U } \Gc_{441} \xrightarrow{ v_{441} } \Gc_{341} \xrightarrow{v_{341} } \Gc_{331} \xrightarrow{ v_{331} } \Gc_{\rm SM} \xrightarrow{ v_{\rm EW} } {\rm SU}(3)_{c}  \otimes  {\rm U}(1)_{\rm EM} \,, \non
&&\Gc_{441} \equiv {\rm SU}(4)_{s} \otimes {\rm SU}(4)_W \otimes  {\rm U}(1)_{X_0 } \,, ~ \Gc_{341} \equiv {\rm SU}(3)_{c} \otimes {\rm SU}(4)_W \otimes  {\rm U}(1)_{X_1 } \,,\non
&&\Gc_{331} \equiv {\rm SU}(3)_{c} \otimes {\rm SU}(3)_W \otimes  {\rm U}(1)_{X_2 } \,,~ \Gc_{\rm SM} \equiv  {\rm SU}(3)_{c} \otimes {\rm SU}(2)_W \otimes  {\rm U}(1)_{Y } \,,\non
&&  \textrm{with}~  v_U\gg v_{441}  \gg v_{341} \gg v_{331} \gg v_{\rm EW} \,,
\eeqn
which was previously studied in Refs.~\cite{Chen:2023qxi,Chen:2024cht,Chen:2024deo}.
Once one assumes that the extended gauge symmetries in the strong sector break first, there is no ambiguity of the sequential symmetry breaking patterns.
Sequentially, the ${\rm U}(1)_{X_1}$, ${\rm U}(1)_{X_2}$, and ${\rm U}(1)_{Y}$ charges are defined according to the ${\rm SU}(4)_s$ and the ${\rm SU}(4)_W$ fundamental representations as follows:
\beqs\label{eqs:SWW_U1charges_fund}
\beqn
\hat X_1(\rep{4_s}) &\equiv&  {\rm diag} \, \Big( \underbrace{  (- \frac{1}{12}+ \Xc_0 ) \mathbb{I}_{3\times 3} }_{ \rep{3_c} } \,, \frac{1}{4}+ \Xc_0 \Big) \,,\label{eq:SWW_X1charge_4sfund}\\[1mm]
\hat X_2 ( \rep{4_W} )&\equiv& {\rm diag} \, \Big( \underbrace{  ( \frac{1}{12} + \Xc_1 ) \mathbb{I}_{3\times 3} }_{ \rep{3_W} } \,, -\frac{1}{4} + \Xc_1 \Big)  \,, \label{eq:SWW_X2charge_4Wfund} \\[1mm]
\hat Y ( \rep{4_W} )&\equiv&  {\rm diag} \, \Big(  ( \frac{1}{6}+ \Xc_2 ) \mathbb{I}_{2\times 2} \,,- \frac{1}{3}+ \Xc_2 \,, \Xc_2 \Big) \non
&=& {\rm diag} \, \Big(  \underbrace{ ( \frac{1}{4} + \Xc_0 ) \mathbb{I}_{2\times 2} }_{ \rep{2_W} } \,,  ( - \frac{1}{4} + \Xc_0  ) \mathbb{I}_{2\times 2} \Big) \,, \label{eq:SWW_Ycharge_4Wfund} \\[1mm]
\hat Q_e ( \rep{4_W} )&\equiv& T_{ {\rm SU}(4) }^3 +  \hat Y ( \rep{4_W} ) = {\rm diag} \, \Big( \frac{3}{4} + \Xc_1  \,, ( - \frac{1}{4} + \Xc_1  ) \mathbb{I}_{3\times 3}  \Big) \,. \label{eq:SWW_Qcharge_4Wfund}
\eeqn
\eeqs
The nonanomalous global $\widetilde{ {\rm U}}(1)_T$ symmetry becomes the global $\widetilde{ {\rm U}}(1)_{B-L}$ at the EW scale according to the following sequence~\cite{Chen:2023qxi}
\beqn\label{eq:U1T_defA}
&& \Gc_{441}~:~ \Tc^\prime \equiv \Tc - 4t \Xc_0 \,, \quad  \Gc_{341}~:~ \Tc^{ \prime \prime} \equiv \Tc^\prime + 8t\Xc_1 \,, \non
&& \Gc_{331}~:~   \Tc^{ \prime \prime \prime} \equiv  \Tc^{ \prime \prime}  \,, \quad  \Gc_{\rm SM}~:~  \Bc- \Lc \equiv  \Tc^{ \prime \prime \prime}  \,.
\eeqn

\subsection{The WSW symmetry breaking pattern}

\para
The WSW symmetry breaking pattern of the ${\rm SU}(8)$ theory follows the sequence of
\beqn\label{eq:Pattern-B}
&& {\rm SU}(8) \xrightarrow{ v_U } \Gc_{441} \xrightarrow{ v_{441} } \Gc_{431} \xrightarrow{v_{431} } \Gc_{331} \xrightarrow{ v_{331} } \Gc_{\rm SM} \xrightarrow{ v_{\rm EW} } {\rm SU}(3)_{c}  \otimes  {\rm U}(1)_{\rm EM}  \,, \non
&&\Gc_{441} \equiv {\rm SU}(4)_{s} \otimes {\rm SU}(4)_W \otimes  {\rm U}(1)_{X_0 } \,, ~ \Gc_{431} \equiv {\rm SU}(4)_{c} \otimes {\rm SU}(3)_W \otimes  {\rm U}(1)_{X_1 } \,,\non
&&\Gc_{331} \equiv {\rm SU}(3)_{c} \otimes {\rm SU}(3)_W \otimes  {\rm U}(1)_{X_2 } \,,~ \Gc_{\rm SM} \equiv  {\rm SU}(3)_{c} \otimes {\rm SU}(2)_W \otimes  {\rm U}(1)_{Y } \,,\non
&&  \textrm{with}~  v_U\gg v_{441}  \gg v_{431} \gg v_{331} \gg v_{\rm EW} \,.
\eeqn
The ${\rm U}(1)_{X_1}$, ${\rm U}(1)_{X_2}$, and ${\rm U}(1)_{Y}$ charges along the WSW sequence are defined according to the ${\rm SU}(4)_s$ and the ${\rm SU}(4)_W$ fundamental representations as follows:
\beqs\label{eqs:WSW_U1charges_fund}
\beqn
\hat X_1 ( \rep{4_W} )&\equiv& {\rm diag} \, \Big( \underbrace{  ( \frac{1}{12} + \Xc_0 ) \mathbb{I}_{3\times 3} }_{ \rep{3_W} } \,, -\frac{1}{4} + \Xc_0 \Big)  \,, \label{eq:WSW_X2charge_4Wfund} \\[1mm]
\hat X_2(\rep{4_s}) &\equiv&  {\rm diag} \, \Big( \underbrace{  (- \frac{1}{12}+ \Xc_1 ) \mathbb{I}_{3\times 3} }_{ \rep{3_c} } \,, \frac{1}{4}+ \Xc_1 \Big) \,,\label{eq:WSW_X1charge_4sfund}\\[1mm]
\hat Y ( \rep{4_W} )&\equiv&  {\rm diag} \, \Big(  ( \frac{1}{6}+ \Xc_2 ) \mathbb{I}_{2\times 2} \,,- \frac{1}{3}+ \Xc_2 \,, \Xc_2 \Big) \non
&=& {\rm diag} \, \Big(  \underbrace{ ( \frac{1}{4} + \Xc_0 ) \mathbb{I}_{2\times 2} }_{ \rep{2_W} } \,,  ( - \frac{1}{4} + \Xc_0  ) \mathbb{I}_{2\times 2} \Big) \,, \label{eq:WSW_Ycharge_4Wfund} \\[1mm]
\hat Q_e ( \rep{4_W} )&\equiv& T_{ {\rm SU}(4) }^3 +  \hat Y ( \rep{4_W} ) = {\rm diag} \, \Big( \frac{3}{4} + \Xc_0  \,, ( - \frac{1}{4} + \Xc_0  ) \mathbb{I}_{3\times 3}  \Big) \,. \label{eq:WSW_Qcharge_4Wfund}
\eeqn
\eeqs
The nonanomalous global $\widetilde{ {\rm U}}(1)_T$ symmetry becomes the global $\widetilde{ {\rm U}}(1)_{B-L}$ at the EW scale according to the following sequence~\cite{Chen:2023qxi}:
\beqn\label{eq:U1T_defB}
&& \Gc_{441}~:~ \Tc^\prime \equiv \Tc + 4t \Xc_0 \,, \quad  \Gc_{431}~:~ \Tc^{ \prime \prime} \equiv \Tc^\prime -8t\Xc_1 \,, \non
&& \Gc_{331}~:~   \Tc^{ \prime \prime \prime} \equiv  \Tc^{ \prime \prime} + 8 t \Xc_2 \,, \quad  \Gc_{\rm SM}~:~  \Bc- \Lc \equiv  \Tc^{ \prime \prime \prime}  \,.
\eeqn
%
%

\begin{table}[htp] {\small
\begin{center}
\begin{tabular}{c|c|c|c|c}
\hline \hline
   $\gSU(8)$   &  $\Gc_{441}$  & $\Gc_{431}$  & $\Gc_{331}$  &  $\Gc_{\rm SM}$  \\
\hline \hline
 $\repb{ 8_F}^\Omega$   
 & $( \repb{4} \,, \rep{1}\,,  +\frac{1}{4} )_{ \mathbf{F} }^\Omega$  
 & $(\repb{4} \,, \rep{1} \,, +\frac{1}{4} )_{ \mathbf{F} }^\Omega$  
 & $(\repb{3} \,, \rep{1} \,, +\frac{1}{3} )_{ \mathbf{F} }^\Omega$  
 &  $( \repb{3} \,, \rep{ 1}  \,, +\frac{1}{3} )_{ \mathbf{F} }^{\Omega}~:~ { \Dc_R^\Omega}^c$  \\
 &  &   &  $( \rep{1} \,, \rep{1} \,, 0)_{ \mathbf{F} }^{\Omega}$
 &  $( \rep{1} \,, \rep{1} \,, 0)_{ \mathbf{F} }^{\Omega} ~:~ \check \Nc_L^{\Omega }$  \\[1mm]
 & $(\rep{1}\,, \repb{4}  \,,  -\frac{1}{4})_{ \mathbf{F} }^\Omega$  
 &  $(\rep{1}\,, \repb{3}  \,,  -\frac{1}{3})_{ \mathbf{F} }^\Omega$  
 &  $( \rep{1} \,, \repb{3} \,,  -\frac{1}{3})_{ \mathbf{F} }^{\Omega}$  
 &  $( \rep{1} \,, \repb{2} \,,  -\frac{1}{2})_{ \mathbf{F} }^{\Omega } ~:~\Lc_L^\Omega =( \Ec_L^\Omega \,, - \Nc_L^\Omega )^T$   \\
 &   &  &  &  $( \rep{1} \,, \rep{1} \,,  0)_{ \mathbf{F} }^{\Omega^\prime} ~:~ \check \Nc_L^{\Omega^\prime }$  \\
  &   & $( \rep{1} \,, \rep{1} \,, 0)_{ \mathbf{F} }^{\Omega^{\prime\prime} }$  
  &   $( \rep{1} \,, \rep{1} \,, 0)_{ \mathbf{F} }^{\Omega^{\prime\prime} }$
  &   $( \rep{1} \,, \rep{1} \,, 0)_{ \mathbf{F} }^{\Omega^{\prime\prime} } ~:~ \check \Nc_L^{\Omega^{\prime \prime} }$   \\
\hline\hline
\end{tabular}
\caption{The $\gSU(8)$ fermion representation of $\repb{8_F}^\Omega$ under the $\Gc_{441}\,,\Gc_{431}\,, \Gc_{331}\,, \Gc_{\rm SM}$ subgroups of the WSW pattern, with $\Omega\equiv(\omega \,, \dot \omega)$.
}
\label{tab:SU8_8ferm_B}
\end{center}
}
\end{table}%

\begin{table}[htp] {\small
\begin{center}
\begin{tabular}{c|c|c|c|c}
\hline \hline
   $\gSU(8)$   &  $\Gc_{441}$  & $\Gc_{431}$  & $\Gc_{331}$  &  $\Gc_{\rm SM}$  \\
\hline \hline
 $\rep{28_F}$   
 & $( \rep{6}\,, \rep{ 1} \,, - \frac{1}{2})_{ \mathbf{F}}$
 & $( \rep{6}\,, \rep{ 1} \,, - \frac{1}{2})_{ \mathbf{F}}$
 & $( \rep{3}\,, \rep{ 1} \,, - \frac{1}{3})_{ \mathbf{F}}$
 & $( \rep{3}\,, \rep{ 1} \,, - \frac{1}{3})_{ \mathbf{F}} ~:~\DG_L$  \\
&   &  & $( \repb{3}\,, \rep{ 1} \,, - \frac{2}{3})_{ \mathbf{F}}$  & $\underline{( \repb{3}\,, \rep{ 1} \,, - \frac{2}{3})_{ \mathbf{F}}~:~ {t_R }^c }$   \\[1mm]
& $( \rep{1}\,, \rep{ 6} \,, +\frac{1}{2})_{ \mathbf{F}}$
&$( \rep{1}\,, \rep{ 3} \,, +\frac{1}{3})_{ \mathbf{F}}$
&  $( \rep{1}\,, \rep{ 3} \,, +\frac{1}{3})_{ \mathbf{F}}$
& $( \rep{1}\,, \rep{2} \,, +\frac{1}{2})_{ \mathbf{F}} ~:~( {\eG_R }^c \,, { \nG_R }^c)^T$  \\
&   &   &   & $( \rep{1}\,, \rep{1} \,, 0 )_{ \mathbf{F}} ~:~ \check \nG_R^c $ \\
&   &  $( \rep{1}\,, \repb{ 3} \,, +\frac{2}{3})_{ \mathbf{F}}$
& $( \rep{1}\,, \repb{ 3} \,, +\frac{2}{3})_{ \mathbf{F}}$  
& $( \rep{1}\,, \repb{2} \,, +\frac{1}{2})_{ \mathbf{F}}^\prime ~:~( { \nG_R^{\prime} }^c\,, - {\eG_R^{\prime} }^c  )^T$   \\
&   &   &   & $\underline{ ( \rep{1}\,, \rep{1} \,, +1 )_{ \mathbf{F}} ~:~ {\tau_R}^c }$ \\[1mm]
& $( \rep{4}\,, \rep{4} \,,  0)_{ \mathbf{F}}$
&  $( \rep{4}\,, \rep{3} \,,  +\frac{1}{12})_{ \mathbf{F}}$   
& $( \rep{3}\,, \rep{3} \,,  0)_{ \mathbf{F}}$  & $\underline{ ( \rep{3}\,, \rep{2} \,,  +\frac{1}{6})_{ \mathbf{F}}~:~ (t_L\,, b_L)^T}$  \\
&   &   &   & $( \rep{3}\,, \rep{1} \,,  -\frac{1}{3})_{ \mathbf{F}}^{\prime} ~:~\DG_L^\prime$  \\
 &   &  & $( \rep{1}\,, \rep{3} \,,  +\frac{1}{3} )_{ \mathbf{F}}^{\prime\prime}$  
 & $( \rep{1}\,, \rep{2} \,,  +\frac{1}{2} )_{ \mathbf{F}}^{\prime\prime} ~:~( {\eG_R^{\prime\prime} }^c \,, { \nG_R^{\prime\prime}}^c )^T$  \\
 &   &   &   & $( \rep{1}\,, \rep{1}\,, 0)_{ \mathbf{F}}^{\prime} ~:~ \check \nG_R^{\prime\,c}$ \\
 &   &  $( \rep{4}\,, \rep{1} \,,  -\frac{1}{4} )_{ \mathbf{F}}$ & $( \rep{3}\,, \rep{1} \,,  -\frac{1}{3})_{ \mathbf{F}}^{\prime\prime}$  & $( \rep{3}\,, \rep{1} \,,  -\frac{1}{3})_{ \mathbf{F}}^{\prime\prime} ~:~\DG_L^{\prime \prime}$ \\
&   &   & $( \rep{1}\,, \rep{1}\,, 0)_{ \mathbf{F}}^{\prime\prime}$ & $( \rep{1}\,, \rep{1}\,, 0)_{ \mathbf{F}}^{\prime\prime} ~:~\check \nG_R^{\prime \prime \,c}$ \\[1mm]  
\hline\hline
\end{tabular}
\caption{
The $\gSU(8)$ fermion representation of $\rep{28_F}$ under the $\Gc_{441}\,,\Gc_{431}\,, \Gc_{331}\,, \Gc_{\rm SM}$ subgroups of the WSW pattern.
All IRs for SM fermions are marked with underlines.}
\label{tab:SU8_28ferm_B}
\end{center}
}
\end{table}

\begin{table}[htp] {\small} 
\begin{center}
\begin{tabular}{c|c|c|c|c}
\hline \hline
   $\gSU(8)$   &  $\Gc_{441}$  & $\Gc_{431}$  & $\Gc_{331}$  &  $\Gc_{\rm SM}$  \\
\hline \hline
     $\rep{56_F}$   
     & $( \rep{ 1}\,, \repb{4} \,, +\frac{3}{4})_{ \mathbf{F}}$  
     &  $( \rep{ 1}\,, \repb{3} \,, +\frac{2}{3})_{ \mathbf{F}}^\prime$
     & $( \rep{ 1}\,, \repb{3} \,, +\frac{2}{3})_{ \mathbf{F}}^\prime$   
     &  $( \rep{ 1}\,, \repb{2} \,, +\frac{1}{2})_{ \mathbf{F}}^{\prime\prime\prime} ~:~( {\nG_R^{\prime\prime\prime }}^c \,, -{\eG_R^{\prime\prime\prime } }^c )^T$  \\
                        &   &   &   & $\underline{ ( \rep{ 1}\,, \rep{1} \,, +1)_{ \mathbf{F}}^{\prime} ~:~ { e_R}^c}$ \\
                        &   & $( \rep{ 1}\,, \rep{1} \,, +1)_{ \mathbf{F}}^{\prime\prime}$   
                        & $( \rep{ 1}\,, \rep{1} \,, +1)_{ \mathbf{F}}^{\prime\prime}$  
                        & $( \rep{ 1}\,, \rep{1} \,, +1)_{ \mathbf{F}}^{\prime \prime} ~:~{\EG_R}^c$   \\[1mm]
                       & $( \repb{ 4}\,, \rep{1} \,, -\frac{3}{4})_{ \mathbf{F}}$  
                       &  $( \repb{ 4}\,, \rep{1} \,, -\frac{3}{4})_{ \mathbf{F}}$
                       & $( \repb{3}\,, \rep{1} \,, -\frac{2}{3})_{ \mathbf{F}}^\prime$  
                       & $\underline{ ( \repb{3}\,, \rep{1} \,, -\frac{2}{3})_{ \mathbf{F}}^{\prime} ~:~ {c_R}^c }$ \\
                       &   &  &$( \rep{1}\,, \rep{1} \,, -1)_{ \mathbf{F}}$
                       &  $( \rep{1}\,, \rep{1} \,, -1)_{ \mathbf{F}} ~:~\EG_L$  \\[1mm]
                       & $( \rep{ 4}\,, \rep{6} \,, +\frac{1}{4})_{ \mathbf{F}}$  
                       &  $( \rep{4}\,, \rep{3} \,, +\frac{1}{12})_{ \mathbf{F}}$
                       & $( \rep{3}\,, \rep{3} \,, 0 )_{ \mathbf{F}}^\prime$ & $\underline{ ( \rep{3}\,, \rep{2} \,, +\frac{1}{6} )_{ \mathbf{F}}^{\prime} ~:~ ( u_L\,,  d_L)^T }$  \\
                       &   &   &   & $( \rep{3}\,, \rep{1} \,, -\frac{1}{3})_{ \mathbf{F}}^{\prime\prime \prime } ~:~\DG_L^{\prime \prime \prime}$ \\
                       &   &  & $( \rep{1}\,, \rep{3} \,, +\frac{1}{3})_{ \mathbf{F}}^\prime $
                       & $( \rep{1}\,, \rep{2} \,, +\frac{1}{2})_{ \mathbf{F}}^{\prime\prime \prime \prime} ~:~ ( {\eG_R^{\prime\prime\prime\prime }}^c \,, {\nG_R^{\prime\prime\prime\prime } }^c )^T$ \\
                       &   &   &   & $( \rep{1}\,, \rep{1} \,, 0 )_{ \mathbf{F}}^{\prime\prime \prime} ~:~ {\check \nG_R}^{\prime \prime\prime \,c}$ \\
                        &   &  $( \rep{4}\,, \repb{3} \,, +\frac{5}{12})_{ \mathbf{F}}$  & $( \rep{3}\,, \repb{3} \,, +\frac{1}{3})_{ \mathbf{F}}$ & $( \rep{3}\,, \repb{2} \,, +\frac{1}{6})_{ \mathbf{F}}^{\prime\prime} ~:~ (\dG_L \,, - \uG_L )^T$   \\
                       &   &   &   & $( \rep{3}\,, \rep{1} \,, +\frac{2}{3})_{ \mathbf{F}} ~:~\UG_L$  \\
                       &   &   & $( \rep{1}\,, \repb{3} \,, +\frac{2}{3})_{ \mathbf{F}}^{\prime\prime}$  & $( \rep{1}\,, \repb{2} \,, +\frac{1}{2})_{ \mathbf{F}}^{\prime\prime \prime \prime \prime} ~:~( {\nG_R^{\prime\prime\prime\prime\prime }}^c \,, -{\eG_R^{\prime\prime\prime\prime\prime } }^c )^T$  \\
                       &   &   &   & $\underline{ ( \rep{1}\,, \rep{1} \,, +1 )_{ \mathbf{F}}^{\prime\prime \prime } ~:~ { \mu_R}^c}$ \\[1mm]
                       & $( \rep{ 6}\,, \rep{4} \,, -\frac{1}{4})_{ \mathbf{F}}$  
                       & $( \rep{6}\,, \rep{3} \,, -\frac{1}{6})_{ \mathbf{F}}$
                       & $( \rep{3}\,, \rep{3} \,, 0)_{ \mathbf{F}}^{\prime\prime}$
                       & $\underline{ ( \rep{3}\,, \rep{2} \,, +\frac{1}{6})_{ \mathbf{F}}^{\prime\prime \prime } ~:~ ( c_L\,, s_L)^T} $ \\
                       &   &   &   &  $( \rep{3}\,, \rep{1} \,, -\frac{1}{3})_{ \mathbf{F}}^{\prime \prime \prime \prime} ~:~\DG_L^{\prime \prime \prime\prime}$ \\
                       &   & & $( \repb{3}\,, \rep{3} \,, -\frac{1}{3})_{ \mathbf{F}}$ & $( \repb{3}\,, \rep{2} \,, -\frac{1}{6})_{ \mathbf{F}} ~:~ ( {\dG_R}^c \,,{\uG_R}^c )^T$  \\
                       &   &   &   & $( \repb{3}\,, \rep{1} \,, -\frac{2}{3})_{ \mathbf{F}}^{\prime \prime} ~:~{\UG_R}^c$  \\
                       &   &   $( \rep{6}\,, \rep{1} \,, -\frac{1}{2})_{ \mathbf{F}}^{\prime}$ &  $( \rep{3}\,, \rep{1} \,, -\frac{1}{3})_{ \mathbf{F}}^{\prime \prime \prime\prime \prime}$ & $( \rep{3}\,, \rep{1} \,, -\frac{1}{3})_{ \mathbf{F}}^{\prime \prime \prime \prime \prime} ~:~ \DG_L^{\prime \prime \prime\prime \prime}$ \\
                       &   &   & $( \repb{3}\,, \rep{1} \,, -\frac{2}{3})_{ \mathbf{F}}^{\prime \prime \prime}$ & $\underline{ ( \repb{3}\,, \rep{1} \,, -\frac{2}{3})_{ \mathbf{F}}^{\prime \prime \prime} ~:~ { u_R}^c  }$  \\[1mm]
\hline\hline
\end{tabular}
\caption{
The $\gSU(8)$ fermion representation of $\rep{56_F}$ under the $\Gc_{441}\,,\Gc_{431}\,, \Gc_{331}\,, \Gc_{\rm SM}$ subgroups of the WSW pattern.
All IRs for SM fermions are marked with underlines.
}
\label{tab:SU8_56ferm_B}
\end{center}
\end{table}

\para
By following the symmetry breaking pattern in Eq.~\eqref{eq:Pattern-B}, we tabulate the fermion representations at various stages of the ${\rm SU}(8)$ theory in Tables~\ref{tab:SU8_8ferm_B}-\ref{tab:SU8_56ferm_B}.
For all heavy right-handed down-type quarks of ${\Dc_R^\Omega}^c$, they are named as follows~:\footnote{All SM right-handed down-type quarks are named according to the generational indices as $(  {\Dc_R^{ \dot 1 } }^c \,,  {\Dc_R^{ \dot 2 } }^c \,,  {\Dc_R^{ \dot 3 } }^c )\equiv ( {d_R}^c \,,  {s_R}^c \,, {b_R}^c )$.}
\beqn\label{eq:B-DR_names}
&&  {\Dc_R^{\rm IV } }^c \equiv {\DG_R^{\prime\prime }}^c \,, ~  {\Dc_R^{\rm V } }^c \equiv {\DG_R}^c  \,,~ {\Dc_R^{\rm VI } }^c \equiv {\DG_R^{\prime }}^c  \,, \non
&& {\Dc_R^{\dot {\rm VII} } }^c \equiv {\DG_R^{\prime\prime\prime\prime \prime }}^c \,, ~  {\Dc_R^{\dot {\rm VIII} } }^c \equiv {\DG_R^{\prime\prime \prime }}^c  \,,~ {\Dc_R^{  \dot {\rm IX} } }^c \equiv {\DG_R^{\prime\prime\prime \prime }}^c \,.
\eeqn
For the left-handed ${\rm SU}(2)_W$ lepton doublets of $(\Ec_L^\Omega \,, - \Nc_L^\Omega )$, they are named as follows~:\footnote{All SM left-handed ${\rm SU}(2)_W$ lepton doublets are named according to the generational indices as $ ( \Ec_L^{ \dot 1} \,,  - \Nc_L^{\dot 1})  \equiv (e_L\,, - \nu_{e\,L} )$, $( \Ec_L^{  \dot 2} \,,   - \Nc_L^{\dot 2})  \equiv( \mu_L \,, - \nu_{\mu\,L} )$, and $( \Ec_L^{ 3} \,, - \Nc_L^{3}) \equiv ( \tau_L \,, - \nu_{\tau\,L})$.}
\beqn\label{eq:B-ELNL_names}
&&  ( \Ec_L^{ \dot {\rm VII} } \,,  - \Nc_L^{ \dot {\rm VII} })  \equiv ( \eG_L^{ \prime\prime \prime \prime} \,, - \nG_L^{\prime\prime \prime \prime } )  \,, ~ ( \Ec_L^{ \dot {\rm VIII} } \,, - \Nc_L^{ \dot {\rm VIII} }  )  \equiv ( \eG_L^{ \prime\prime  \prime} \,, - \nG_L^{\prime\prime \prime} ) \,,~  ( \Ec_L^{\dot {\rm IX} } \,,  - \Nc_L^{ \dot {\rm IX} } ) \equiv ( \eG_L^{ \prime\prime \prime \prime \prime } \,,  - \nG_L^{\prime\prime \prime\prime \prime} )  \,,\non
&&    ( \Ec_L^{\rm IV } \,, - \Nc_L^{\rm IV }) \equiv ( \eG_L \,, - \nG_L )\,, ~ ( \Ec_L^{\rm V } \,, -  \Nc_L^{\rm V }) \equiv ( \eG_L^{\prime\prime}\,, - \nG_L^{\prime\prime} )  \,,~ ( \Ec_L^{\rm VI } \,, - \Nc_L^{\rm VI } ) \equiv  ( \eG_L^\prime\,, - \nG_L^\prime ) \,.
\eeqn
Through the analysis in Sec.~\ref{section:WSW_pattern}, we shall see that all heavy $(\Dc^\Omega\,, \Ec^\Omega\,, \Nc^\Omega)$ (with $\Omega={\rm IV}\,, \ldots \,,\dot {\rm IX}$) acquire vectorlike masses during the intermediate symmetry breaking stages.
For the remaining left-handed sterile neutrinos of $(  \check \Nc_L^\Omega \,, \check \Nc_L^{\Omega^\prime } \,, \check \Nc_L^{\Omega^{ \prime\prime} } )$, several of them are massive and they are named as follows
\beqn\label{eq:stNL_names_B}
&& \check \Nc_L^{{\rm IV} }  \equiv \check \nG_L^{\prime \prime }\,,~   \check \Nc_L^{{\rm IV}^{\prime } }  \equiv \check \nG_L^{} \,, ~ \check \Nc_L^{{\rm V}^{\prime } }  \equiv \check  \nG_L^\prime \,,~  \check \Nc_L^{\dot {\rm VII}^{\prime} }  \equiv \check \nG_L^{\prime \prime \prime } \,.
\eeqn
Notice that the first- and second-generational SM fermions are named differently in Table~\ref{tab:SU8_56ferm_B} as compared to the names following the SWW symmetry breaking pattern.
This will be elaborated in our derivation of the SM quark/lepton mass matrices in Sec.~\ref{section:WSW_pattern}.

\para
We decompose the Higgs fields in the Yukawa term into components that can be responsible for the sequential symmetry breaking pattern in Eq.~\eqref{eq:Pattern-B}.
All possible Higgs components that are likely to develop VEVs for the corresponding symmetry breaking stages are marked by $\langle ... \rangle$, while their UV origins are denoted by underlines.
For Higgs fields of $\repb{ 8_H}_{\,, \omega}$ they read
\begin{eqnarray}\label{eq:SU8B_Higgs_Br01}
\repb{8_H}_{\,,\omega }  &\supset&  \underline{  ( \repb{4} \,, \rep{1} \,, +\frac{1}{4} )_{\mathbf{H}\,, \omega }  } \oplus  \langle ( \rep{1} \,, \repb{4} \,, -\frac{1}{4} )_{\mathbf{H}\,, \omega } \rangle \non
&\supset&  \langle ( \repb{4} \,, \rep{1} \,, +\frac{1}{4 } )_{\mathbf{H}\,, \omega }  \rangle \oplus  \underline{ ( \rep{1} \,, \repb{3} \,, -\frac{1}{3 } )_{\mathbf{H}\,, \omega } }\non
&\supset&  \langle  ( \rep{1} \,, \repb{3} \,, -\frac{1}{3} )_{\mathbf{H}\,, \omega } \rangle \supset \langle ( \rep{1} \,, \repb{2} \,, -\frac{1}{2} )_{\mathbf{H}\,, \omega } \rangle   \,.
\end{eqnarray}
For Higgs fields of $\repb{28_H}_{\,,\dot \omega } $, they read
\begin{eqnarray}\label{eq:SU8B_Higgs_Br02}
\repb{28_H}_{\,,\dot \omega } &\supset& ( \rep{6} \,, \rep{1} \,,  +\frac{1}{2} )_{\mathbf{H}\,, \dot\omega }  \oplus  \underline{ ( \rep{1} \,, \rep{6} \,, -\frac{1}{2} )_{\mathbf{H}\,, \dot\omega } } \oplus \underline{ ( \repb{4} \,, \repb{4} \,, 0 )_{\mathbf{H}\,, \dot\omega } }  \non
&\supset & \left[ \underline{  ( \rep{1} \,, \repb{3} \,, -\frac{1}{3} )_{\mathbf{H}\,, \dot\omega }^\prime  } \oplus \underline{ ( \rep{1} \,, \rep{3} \,, -\frac{2}{3} )_{\mathbf{H}\,, \dot\omega }  } \right] \oplus \left [ \underline{  ( \repb{4} \,, \repb{3} \,, -\frac{1}{12} )_{\mathbf{H}\,, \dot\omega }  } \oplus \langle ( \repb{4} \,, \rep{1} \,, +\frac{1}{4} )_{\mathbf{H}\,, \dot\omega }  \rangle \right ] \non
&\supset&  \left [ \langle ( \rep{1} \,, \repb{3} \,, -\frac{1}{3} )_{\mathbf{H}\,, \dot\omega }^\prime \rangle \oplus \underline{ ( \rep{1} \,, \rep{3} \,, -\frac{2}{3} )_{\mathbf{H}\,, \dot\omega }  } \right ] \oplus  \langle ( \rep{1} \,, \repb{3} \,, -\frac{1}{3} )_{\mathbf{H}\,, \dot\omega }  \rangle    \non
&\supset&  \left [ \langle ( \rep{1} \,, \repb{2} \,, -\frac{1}{2} )_{\mathbf{H}\,, \dot\omega }^\prime  \rangle \oplus \langle ( \rep{1} \,, \rep{2} \,, -\frac{1}{2} )_{\mathbf{H}\,, \dot\omega }  \rangle \right ] \oplus  \langle ( \rep{1} \,, \repb{2} \,, -\frac{1}{2} )_{\mathbf{H}\,, \dot\omega } \rangle \,.
\end{eqnarray}
For Higgs fields of $\rep{70_H}$, they read
\begin{eqnarray}\label{eq:SU8B_Higgs_Br05}
\rep{70_H} &\supset& ( \rep{1} \,, \rep{1 } \,, -1 )_{\mathbf{H}}^{ \prime \prime } \oplus ( \rep{1} \,, \rep{1 } \,, +1 )_{\mathbf{H}}^{\prime \prime  \prime \prime }   \oplus  ( \repb{4} \,, \rep{4} \,, -\frac{1}{2} )_{\mathbf{H}}  \oplus ( \rep{6 } \,, \rep{6 } \,, 0 )_{\mathbf{H}} \oplus \underline{ ( \rep{4} \,, \repb{4} \,, +\frac{1}{2} )_{\mathbf{H}} }   \non
&\supset& \underline{ ( \rep{4} \,, \repb{3} \,, +\frac{5}{12} )_{\mathbf{H}} } \supset  \underline{ ( \rep{1} \,, \repb{3} \,, +\frac{2}{3} )_{\mathbf{H}}^{\prime\prime\prime} }  \supset  \langle ( \rep{1} \,, \repb{2} \,, +\frac{1}{2} )_{\mathbf{H}}^{\prime \prime\prime } \rangle \,.
\end{eqnarray}
\para
Schematically, we assign the Higgs VEVs according to the decompositions in Eqs.~\eqref{eq:SU8B_Higgs_Br01}-\eqref{eq:SU8B_Higgs_Br05} as follows:
\beqs\label{eqs:SU8_WSW_Higgs_VEVs}
\beqn
\Gc_{441} \to \Gc_{431} ~&:&~ \langle ( \rep{1} \,, \repb{4} \,, -\frac{1}{4} )_{\mathbf{H}\,, {\rm IV}} \rangle \equiv \frac{1}{\sqrt{2}}W_{ \repb{4}\,, {\rm IV}}\,, \label{eq:SU8_WSW_Higgs_VEVs_mini01}\\[1mm]
\Gc_{431} \to \Gc_{331} ~&:&~ \langle ( \repb{4} \,, \rep{1} \,, +\frac{1}{4} )_{\mathbf{H}\,, {\rm V} } \rangle \equiv \frac{1}{\sqrt{2}} w_{\repb{4}\,, {\rm V} }\,, ~ \langle ( \repb{4} \,, \rep{1} \,, +\frac{1}{4} )_{\mathbf{H}\,,\dot{1},  \dot {\rm VII}  } \rangle \equiv \frac{1}{\sqrt{2}}  w_{\repb{4}\,, \dot{1},\dot {\rm VII} } \,,\label{eq:SU8_WSW_Higgs_VEVs_mini02} \\[1mm]
\Gc_{331} \to \Gc_{\rm SM} ~&:&~  \langle ( \rep{1} \,, \repb{3} \,, -\frac{1}{3} )_{\mathbf{H}\,, 3,\rm{VI} } \rangle \equiv \frac{1}{\sqrt{2}} V_{ \repb{3}\,, 3,\rm{VI} } \,, \non
&&  \langle ( \rep{1} \,, \repb{3} \,, -\frac{1}{3} )_{\mathbf{H}\, ,\dot {\rm IX} } \rangle \equiv \frac{1}{\sqrt{2}} V_{ \repb{3}\,,\dot {\rm IX} }  \,, ~ \langle ( \rep{1} \,, \repb{3} \,, -\frac{1}{3} )_{\mathbf{H}\,,\dot{2}, \dot {\rm VIII}  }^\prime  \rangle \equiv \frac{1}{\sqrt{2}} V_{ \repb{3}\,, \dot{2},\dot {\rm VIII} }^\prime \,,  \label{eq:SU8_WSW_Higgs_VEVs_mini03}\\[1mm]
 {\rm EWSB} ~&:&~   \langle ( \rep{1} \,, \repb{2} \,, +\frac{1}{2} )_{\mathbf{H} }^{ \prime\prime \prime} \rangle \equiv \frac{1}{\sqrt{2}} v_{\rm EW}   \,.\label{eq:SU8_WSW_Higgs_VEVs_mini04}
\eeqn
\eeqs
For our later convenience, we also parametrize different symmetry breaking VEVs,
\begin{eqnarray}\label{eq:SU8_WSW_param}
    && \zeta_0 \equiv \frac{ v_U }{ M_{\rm pl} } \,, \quad \zeta_1 \equiv \frac{ W_{ \repb{4}\,, {\rm IV} } }{ M_{\rm pl} } \,, \quad \zeta_2 \equiv \frac{ w_{ \repb{4}\,, {\rm V} }  }{ M_{\rm pl} } \,,  \quad \dot \zeta_2 \equiv \frac{ w_{ \repb{4}\,,\dot{1},\dot {\rm VII} }  }{ M_{\rm pl} } \,,   \non
&&  \zeta_3 \equiv \frac{ V_{ \repb{3}\,,3,{\rm VI} } }{ M_{\rm pl} } \,, \quad \dot \zeta_3^\prime \equiv \frac{ V_{ \repb{3}\,, \dot{2},\dot {\rm VIII} }^\prime  }{ M_{\rm pl} }  \,, \quad \dot \zeta_3 \equiv \frac{ V_{ \repb{3}\,, \dot {\rm IX} } }{ M_{\rm pl} }     \,, \non
&& \zeta_0 \gg \zeta_1 \gg  \zeta_2 \sim \dot \zeta_2  \gg \zeta_3 \sim \dot \zeta_3^\prime \sim  \dot \zeta_3  \,, \quad \zeta_{i j} \equiv \frac{ \zeta_j }{ \zeta_i } \,, \quad ( i < j ) \,,
\end{eqnarray}
in terms of dimensionless quantities.
Here, we adopt the following conventions:
\begin{itemize}

 \item[(i)] The notations of $(W\,, w\,, V)$ and the dimensionless quantities of $( \zeta_1\,, \zeta_2 \sim \dot \zeta_2\,, \zeta_3 \sim \dot \zeta_3^\prime \sim  \dot \zeta_3)$ are used for the Higgs VEVs at the first, second, and third symmetry breaking stages, regardless of the specific symmetry breaking patterns.
 
 \item[(ii)] $W_{\repb{4}}$ and $w_{\repb{4}}$ represent the VEVs for the $\Gc_{441} \to \Gc_{431}$ and the $\Gc_{431} \to \Gc_{331}$ stages in Eqs.~\eqref{eqs:SU8_WSW_Higgs_VEVs}.
 In the SWW sequence~\cite{Chen:2024cht}, $W_{\repb{4}}$ and $w_{\repb{4}}$ represent the VEVs for the $\Gc_{441} \to \Gc_{341}$ and the $\Gc_{341} \to \Gc_{331}$ stages instead.
 
\end{itemize}

\subsection{The WWS symmetry breaking pattern}

\para
The WWS symmetry breaking pattern of the ${\rm SU}(8)$ theory follows the sequence of
\beqn\label{eq:Pattern-C}
&& {\rm SU}(8) \xrightarrow{ v_U } \Gc_{441} \xrightarrow{ v_{441} } \Gc_{431} \xrightarrow{v_{431} } \Gc_{421} \xrightarrow{ v_{421} } \Gc_{\rm SM} \xrightarrow{ v_{\rm EW} } {\rm SU}(3)_{c}  \otimes  {\rm U}(1)_{\rm EM}  \,, \non
&&\Gc_{441} \equiv {\rm SU}(4)_{s} \otimes {\rm SU}(4)_W \otimes  {\rm U}(1)_{X_0 } \,, ~ \Gc_{431} \equiv {\rm SU}(4)_{s} \otimes {\rm SU}(3)_W \otimes  {\rm U}(1)_{X_1 } \,,\non
&&\Gc_{421} \equiv {\rm SU}(4)_{s} \otimes {\rm SU}(2)_W \otimes  {\rm U}(1)_{X_2 } \,,~ \Gc_{\rm SM} \equiv  {\rm SU}(3)_{c} \otimes {\rm SU}(2)_W \otimes  {\rm U}(1)_{Y } \,,\non
&& \textrm{with}~  v_U\gg v_{441}  \gg v_{431} \gg v_{331} \gg v_{\rm EW} \,.
\eeqn
The ${\rm U}(1)_{X_1}$, ${\rm U}(1)_{X_2}$, and ${\rm U}(1)_{Y}$ charges along the WWS sequence are defined according to the ${\rm SU}(4)_s$ and the ${\rm SU}(4)_W$ fundamental representations as follows:
\beqs\label{eqs:WWS_U1charges_fund}
\beqn
\hat X_1 ( \rep{4_W} )&\equiv& {\rm diag} \, \Big( \underbrace{  ( \frac{1}{12} + \Xc_0 ) \mathbb{I}_{3\times 3} }_{ \rep{3_W} } \,, -\frac{1}{4} + \Xc_0 \Big)  \,, \label{eq:WWS_X2charge_4Wfund} \\[1mm]
\hat X_2(\rep{3_W}) &\equiv&  {\rm diag} \, \Big( \underbrace{  ( \frac{1}{6} + \Xc_1 ) \mathbb{I}_{2 \times 2} }_{ \rep{2_W } } \,, - \frac{1}{3}+ \Xc_1 \Big) \,,\label{eq:WWS_X1charge_4sfund}\\[1mm]
\hat Y ( \rep{4_s} )&\equiv&  {\rm diag} \, \Big(  \underbrace{  (- \frac{1}{12}+ \Xc_2 ) \mathbb{I}_{3\times 3} }_{ \rep{3_c} } \,, \frac{1}{4}+ \Xc_2 \Big) \,.\label{eq:WWS_Ycharge_4Wfund} 
\eeqn
\eeqs
The nonanomalous global $\widetilde{ {\rm U}}(1)_T$ symmetry becomes the global $\widetilde{ {\rm U}}(1)_{B-L}$ at the EW scale according to the following sequence~\cite{Chen:2023qxi}:
\beqn\label{eq:U1T_defC}
&& \Gc_{441}~:~ \Tc^\prime \equiv \Tc + 4t \Xc_0 \,, \quad  \Gc_{431}~:~ \Tc^{ \prime \prime} \equiv \Tc^\prime  \,, \non
&& \Gc_{421}~:~   \Tc^{ \prime \prime \prime} \equiv  \Tc^{ \prime \prime} - 8 t \Xc_2 \,, \quad  \Gc_{\rm SM}~:~  \Bc- \Lc \equiv  \Tc^{ \prime \prime \prime} + 8 t \Yc \,.
\eeqn
\begin{table}[htp] {\small
\begin{center}
\begin{tabular}{c|c|c|c|c}
\hline \hline
   $\gSU(8)$   &  $\Gc_{441}$  & $\Gc_{431}$  & $\Gc_{421}$  &  $\Gc_{\rm SM}$  \\
\hline \hline
 $\repb{ 8_F}^\Omega$   
 & $( \repb{4} \,, \rep{1}\,,  +\frac{1}{4} )_{ \mathbf{F} }^\Omega$  
 & $(\repb{4} \,, \rep{1} \,, +\frac{1}{4} )_{ \mathbf{F} }^\Omega$  
 & $(\repb{4} \,, \rep{1} \,, +\frac{1}{4} )_{ \mathbf{F} }^\Omega$  
 &  $( \repb{3} \,, \rep{ 1}  \,, +\frac{1}{3} )_{ \mathbf{F} }^{\Omega}~:~ { \Dc_R^\Omega}^c$  \\
 &  &   &  
 &  $( \rep{1} \,, \rep{1} \,, 0)_{ \mathbf{F} }^{\Omega} ~:~ \check \Nc_L^{\Omega }$  \\[1mm]
 & $(\rep{1}\,, \repb{4}  \,,  -\frac{1}{4})_{ \mathbf{F} }^\Omega$  
 &  $(\rep{1}\,, \repb{3}  \,,  -\frac{1}{3})_{ \mathbf{F} }^\Omega$  
 &  $( \rep{1} \,, \repb{2} \,,  -\frac{1}{2})_{ \mathbf{F} }^{\Omega}$  
 &  $( \rep{1} \,, \repb{2} \,,  -\frac{1}{2})_{ \mathbf{F} }^{\Omega } ~:~\Lc_L^\Omega =( \Ec_L^\Omega \,, - \Nc_L^\Omega )^T$   \\
 &   &  &  $( \rep{1} \,, \rep{1} \,,  0)_{ \mathbf{F} }^{\Omega^\prime}$  &  $( \rep{1} \,, \rep{1} \,,  0)_{ \mathbf{F} }^{\Omega^\prime} ~:~ \check \Nc_L^{\Omega^\prime }$  \\
  &   & $( \rep{1} \,, \rep{1} \,, 0)_{ \mathbf{F} }^{\Omega^{\prime\prime} }$  
  &   $( \rep{1} \,, \rep{1} \,, 0)_{ \mathbf{F} }^{\Omega^{\prime\prime} }$
  &   $( \rep{1} \,, \rep{1} \,, 0)_{ \mathbf{F} }^{\Omega^{\prime\prime} } ~:~ \check \Nc_L^{\Omega^{\prime \prime} }$   \\
\hline\hline
\end{tabular}
\caption{The $\gSU(8)$ fermion representation of $\repb{8_F}^\Omega$ under the $\Gc_{441}\,,\Gc_{431}\,, \Gc_{421}\,, \Gc_{\rm SM}$ subgroups of the WWS pattern, with $\Omega\equiv (\omega \,, \dot \omega)$.}
\label{tab:SU8_8ferm_C}
\end{center}
}
\end{table}%

\begin{table}[htp] {\small
\begin{center}
\begin{tabular}{c|c|c|c|c}
\hline \hline
   $\gSU(8)$   &  $\Gc_{441}$  & $\Gc_{431}$  & $\Gc_{421}$  &  $\Gc_{\rm SM}$  \\
\hline \hline
 $\rep{28_F}$   
 & $( \rep{6}\,, \rep{ 1} \,, - \frac{1}{2})_{ \mathbf{F}}$
 & $( \rep{6}\,, \rep{ 1} \,, - \frac{1}{2})_{ \mathbf{F}}$
 & $( \rep{6}\,, \rep{ 1} \,, - \frac{1}{2})_{ \mathbf{F}}$
 & $( \rep{3}\,, \rep{ 1} \,, - \frac{1}{3})_{ \mathbf{F}} ~:~\DG_L$  \\
&   &  & & $\underline{( \repb{3}\,, \rep{ 1} \,, - \frac{2}{3})_{ \mathbf{F}}~:~ {t_R }^c }$   \\[1mm]
& $( \rep{1}\,, \rep{ 6} \,, +\frac{1}{2})_{ \mathbf{F}}$
&$( \rep{1}\,, \rep{ 3} \,, +\frac{1}{3})_{ \mathbf{F}}$
&  $( \rep{1}\,, \rep{ 2} \,, +\frac{1}{2})_{ \mathbf{F}}$
& $( \rep{1}\,, \rep{2} \,, +\frac{1}{2})_{ \mathbf{F}} ~:~( {\eG_R }^c \,, { \nG_R }^c)^T$  \\
&   &   &    $( \rep{1}\,, \rep{1} \,, 0 )_{ \mathbf{F}}$
& $( \rep{1}\,, \rep{1} \,, 0 )_{ \mathbf{F}} ~:~ \check \nG_R^c $ \\
&   &  $( \rep{1}\,, \repb{ 3} \,, +\frac{2}{3})_{ \mathbf{F}}$
& $( \rep{1}\,, \repb{ 2} \,, +\frac{1}{2})_{ \mathbf{F}}^{\prime}$  
& $( \rep{1}\,, \repb{2} \,, +\frac{1}{2})_{ \mathbf{F}}^\prime ~:~( { \nG_R^{\prime} }^c\,, - {\eG_R^{\prime} }^c  )^T$   \\
&   &   & $  ( \rep{1}\,, \rep{1} \,, +1 )_{ \mathbf{F}} $
& $\underline{ ( \rep{1}\,, \rep{1} \,, +1 )_{ \mathbf{F}} ~:~ {\tau_R}^c} $ \\[1mm]
& $( \rep{4}\,, \rep{4} \,,  0)_{ \mathbf{F}}$
&  $( \rep{4}\,, \rep{3} \,,  +\frac{1}{12})_{ \mathbf{F}}$   
& $( \rep{4}\,, \rep{2} \,,  +\frac{1}{4})_{ \mathbf{F}}$  & $\underline{ ( \rep{3}\,, \rep{2} \,,  +\frac{1}{6})_{ \mathbf{F}}~:~ (t_L\,, b_L)^T}$  \\
 &   &  &
 & $( \rep{1}\,, \rep{2} \,,  +\frac{1}{2} )_{ \mathbf{F}}^{\prime\prime} ~:~( {\eG_R^{\prime\prime} }^c \,, { \nG_R^{\prime\prime}}^c )^T$  \\
&   &   & $( \rep{4}\,, \rep{1} \,,  -\frac{1}{4})_{ \mathbf{F}}$   & $( \rep{3}\,, \rep{1} \,,  -\frac{1}{3})_{ \mathbf{F}}^{\prime} ~:~\DG_L^\prime$  \\
 &   &   &   & $( \rep{1}\,, \rep{1}\,, 0)_{ \mathbf{F}}^{\prime} ~:~ \check \nG_R^{\prime\,c}$ \\
 &   &  $( \rep{4}\,, \rep{1} \,,  -\frac{1}{4} )_{ \mathbf{F}}^{\prime}$
 & $( \rep{4}\,, \rep{1} \,,  -\frac{1}{4})_{ \mathbf{F}}^{\prime}$  
 & $( \rep{3}\,, \rep{1} \,,  -\frac{1}{3})_{ \mathbf{F}}^{\prime\prime} ~:~\DG_L^{\prime \prime}$ \\
&   &   &  & $ ( \rep{1}\,, \rep{1}\,, 0)_{ \mathbf{F}}^{\prime\prime} ~:~\check \nG_R^{\prime \prime \,c} $\\[1mm]  
\hline\hline
\end{tabular}
\caption{The $\gSU(8)$ fermion representation of $\rep{28_F}$ under the $\Gc_{441}\,,\Gc_{431}\,, \Gc_{421}\,, \Gc_{\rm SM}$ subgroups of the WWS pattern.
All IRs for SM fermions are marked with underlines.}
\label{tab:SU8_28ferm_C}
\end{center}
}
\end{table}

\begin{table}[htp] {\small}
\begin{center}
\begin{tabular}{c|c|c|c|c}
\hline \hline
   $\gSU(8)$   &  $\Gc_{441}$  & $\Gc_{431}$  & $\Gc_{421}$  &  $\Gc_{\rm SM}$  \\
\hline \hline
     $\rep{56_F}$   
     & $( \rep{ 1}\,, \repb{4} \,, +\frac{3}{4})_{ \mathbf{F}}$  
     &  $( \rep{ 1}\,, \repb{3} \,, +\frac{2}{3})_{ \mathbf{F}}^\prime$
     & $( \rep{ 1}\,, \repb{2} \,, +\frac{1}{2})_{ \mathbf{F}}^{\prime\prime\prime}$   
     &  $( \rep{ 1}\,, \repb{2} \,, +\frac{1}{2})_{ \mathbf{F}}^{\prime\prime\prime} ~:~( {\nG_R^{\prime\prime\prime }}^c \,, -{\eG_R^{\prime\prime\prime } }^c )^T$  \\
                        &   &   & $( \rep{ 1}\,, \rep{1} \,, +1)_{ \mathbf{F}}^{\prime} $  & $\underline{( \rep{ 1}\,, \rep{1} \,, +1)_{ \mathbf{F}}^{\prime} ~:~ {e_R}^c }$ \\
                        &   & $( \rep{ 1}\,, \rep{1} \,, +1)_{ \mathbf{F}}^{\prime\prime}$   
                        & $( \rep{ 1}\,, \rep{1} \,, +1)_{ \mathbf{F}}^{\prime\prime}$  
                        & $( \rep{ 1}\,, \rep{1} \,, +1)_{ \mathbf{F}}^{\prime \prime} ~:~{\EG_R}^c$   \\[1mm]
                       & $( \repb{ 4}\,, \rep{1} \,, -\frac{3}{4})_{ \mathbf{F}}$  
                       &  $( \repb{ 4}\,, \rep{1} \,, -\frac{3}{4})_{ \mathbf{F}}$
                       & $( \repb{ 4}\,, \rep{1} \,, -\frac{3}{4})_{ \mathbf{F}}$
                       & $\underline{ ( \repb{3}\,, \rep{1} \,, -\frac{2}{3})_{ \mathbf{F}}^{\prime} ~:~ {c_R}^c }$ \\
                       &   &   &   &  $( \rep{1}\,, \rep{1} \,, -1)_{ \mathbf{F}} ~:~\EG_L$  \\[1mm]
                       & $( \rep{ 4}\,, \rep{6} \,, +\frac{1}{4})_{ \mathbf{F}}$  
                       &  $( \rep{4}\,, \rep{3} \,, +\frac{1}{12})_{ \mathbf{F}}^{\prime}$
                       & $( \rep{4}\,, \rep{2} \,, +\frac{1}{4} )_{ \mathbf{F}}^\prime$
                       & $\underline{ ( \rep{3}\,, \rep{2} \,, +\frac{1}{6} )_{ \mathbf{F}}^{\prime} ~:~  (u_L\,, d_L)^T} $ \\
                       &   &   &   & $( \rep{1}\,, \rep{2} \,, +\frac{1}{2})_{ \mathbf{F}}^{\prime\prime \prime \prime} ~:~ ( {\eG_R^{\prime\prime\prime\prime }}^c \,, {\nG_R^{\prime\prime\prime\prime } }^c )^T$  \\
                       &   &   &  $( \rep{4}\,, \rep{1} \,, -\frac{1}{4})_{ \mathbf{F}}^{\prime\prime  } $ & $( \rep{3}\,, \rep{1} \,, -\frac{1}{3})_{ \mathbf{F}}^{\prime\prime \prime } ~:~\DG_L^{\prime \prime \prime}$ \\
                       &   &   &   & $( \rep{1}\,, \rep{1} \,, 0 )_{ \mathbf{F}}^{\prime\prime \prime} ~:~ {\check \nG_R}^{\prime \prime\prime \,c}$ \\
                        &   &  $( \rep{4}\,, \repb{3} \,, +\frac{5}{12})_{ \mathbf{F}}$  
                        & $( \rep{4}\,, \repb{2} \,, +\frac{1}{4})_{ \mathbf{F}}$
                        & $( \rep{3}\,, \repb{2} \,, +\frac{1}{6})_{ \mathbf{F}}^{\prime\prime} ~:~ (\dG_L \,, - \uG_L )^T$   \\
                       &   &   &  & $( \rep{1}\,, \repb{2} \,, +\frac{1}{2})_{ \mathbf{F}}^{\prime\prime \prime \prime \prime} ~:~( {\nG_R^{\prime\prime\prime\prime\prime }}^c \,, -{\eG_R^{\prime\prime\prime\prime\prime } }^c )^T$  \\
                        &   &   &  $( \rep{4}\,, \rep{1} \,, +\frac{3}{4})_{ \mathbf{F}}$ & $( \rep{3}\,, \rep{1} \,, +\frac{2}{3})_{ \mathbf{F}} ~:~\UG_L$  \\
                       &   &   &   & $\underline{ ( \rep{1}\,, \rep{1} \,, +1 )_{ \mathbf{F}}^{\prime\prime \prime } ~:~ {\mu_R}^c  }$ \\[1mm]
                       & $( \rep{ 6}\,, \rep{4} \,, -\frac{1}{4})_{ \mathbf{F}}$  
                       & $( \rep{6}\,, \rep{3} \,, -\frac{1}{6})_{ \mathbf{F}}$
                       & $( \rep{6}\,, \rep{2} \,, 0)_{ \mathbf{F}}$
                       & $\underline{ ( \rep{3}\,, \rep{2} \,, +\frac{1}{6})_{ \mathbf{F}}^{\prime\prime \prime } ~:~  (c_L\,, s_L)^T}$  \\
                       &   & & & $( \repb{3}\,, \rep{2} \,, -\frac{1}{6})_{ \mathbf{F}} ~:~ ( {\dG_R}^c \,,{\uG_R}^c )^T$  \\
                       &   &   &  $( \rep{6}\,, \rep{1} \,, -\frac{1}{2})_{ \mathbf{F}}^{\prime}$
                       &  $( \rep{3}\,, \rep{1} \,, -\frac{1}{3})_{ \mathbf{F}}^{\prime \prime \prime \prime} ~:~\DG_L^{\prime \prime \prime\prime}$ \\
                       &   &   &  & $( \repb{3}\,, \rep{1} \,, -\frac{2}{3})_{ \mathbf{F}}^{\prime \prime} ~:~{\UG_R}^c$  \\
                       &   &   $( \rep{6}\,, \rep{1} \,, -\frac{1}{2})_{ \mathbf{F}}^{\prime\prime}$
                       & $( \rep{6}\,, \rep{1} \,, -\frac{1}{2})_{ \mathbf{F}}^{\prime\prime}$
                       & $( \rep{3}\,, \rep{1} \,, -\frac{1}{3})_{ \mathbf{F}}^{\prime \prime \prime \prime \prime} ~:~ \DG_L^{\prime \prime \prime\prime \prime}$ \\
                       &   &   &  & $\underline{ ( \repb{3}\,, \rep{1} \,, -\frac{2}{3})_{ \mathbf{F}}^{\prime \prime \prime} ~:~ {u_R}^c }$  \\[1mm]
\hline\hline
\end{tabular}
\caption{The $\gSU(8)$ fermion representation of $\rep{56_F}$ under the $\Gc_{441}\,,\Gc_{431}\,, \Gc_{421}\,, \Gc_{\rm SM}$ subgroups of the WWS pattern.
All IRs for SM fermions are marked with underlines.
}
\label{tab:SU8_56ferm_C}
\end{center}
\end{table}%

\para
By following the symmetry breaking pattern in Eq.~\eqref{eq:Pattern-C}, we tabulate the fermion representations at various stages of the ${\rm SU}(8)$ theory in Tables.~\ref{tab:SU8_8ferm_C}-\ref{tab:SU8_56ferm_C}.
For all heavy right-handed down-type quarks of ${\Dc_R^\Omega}^c$, they are named as follows:
\beqn\label{eq:DR_names_C}
&&   {\Dc_R^{\rm IV } }^c \equiv {\DG_R^{\prime\prime }}^c \,, ~  {\Dc_R^{\rm V } }^c \equiv {\DG_R^{ }}^c  \,,~ {\Dc_R^{\rm VI } }^c \equiv {\DG_R^{\prime }}^c  \,, \non
&&  {\Dc_R^{\dot {\rm VII} } }^c \equiv {\DG_R^{\prime\prime\prime\prime \prime }}^c \,, ~  {\Dc_R^{\dot {\rm VIII} } }^c \equiv {\DG_R^{\prime\prime \prime }}^c  \,,~ {\Dc_R^{  \dot {\rm IX} } }^c \equiv {\DG_R^{\prime\prime\prime \prime }}^c  \,.
\eeqn
For all heavy left-handed ${\rm SU}(2)_W$ lepton doublets of $(\Ec_L^\Omega \,, - \Nc_L^\Omega )$, they are named as follows:
\beqn\label{eq:ELNL_names_C}
&&    ( \Ec_L^{\rm IV } \,, - \Nc_L^{\rm IV }) \equiv ( \eG_L \,, - \nG_L )\,, ~  ( \Ec_L^{\rm V } \,, -  \Nc_L^{\rm V }) \equiv ( \eG_L^{\prime\prime}\,, - \nG_L^{\prime\prime} )  \,,~ ( \Ec_L^{\rm VI } \,, - \Nc_L^{\rm VI } ) \equiv  ( \eG_L^\prime\,, - \nG_L^\prime ) \,, \non
&&  ( \Ec_L^{ \dot {\rm VII} } \,,  - \Nc_L^{ \dot {\rm VII} })  \equiv ( \eG_L^{ \prime\prime \prime \prime} \,, - \nG_L^{\prime\prime \prime \prime } )  \,, ~ ( \Ec_L^{ \dot {\rm VIII} } \,, - \Nc_L^{ \dot {\rm VIII} }  )  \equiv ( \eG_L^{ \prime\prime  \prime} \,, - \nG_L^{\prime\prime \prime} ) \,,~  ( \Ec_L^{\dot {\rm IX} } \,,  - \Nc_L^{ \dot {\rm IX} } ) \equiv ( \eG_L^{ \prime\prime \prime \prime \prime } \,,  - \nG_L^{\prime\prime \prime\prime \prime} )  \,.
\eeqn
Through the analysis in Sec.~\ref{section:WWS_pattern}, we shall see that all heavy $(\Dc^\Omega\,, \Ec^\Omega\,, \Nc^\Omega)$ (with $\Omega={\rm IV}\,, \ldots \,,\dot {\rm IX}$) acquire vectorlike masses during the intermediate symmetry breaking stages.
For the remaining left-handed sterile neutrinos of $(  \check \Nc_L^\Omega \,, \check \Nc_L^{\Omega^\prime } \,, \check \Nc_L^{\Omega^{ \prime\prime} } )$, several of them are massive and they are named as follows:
\beqn\label{eq:stNL_names_C}
&& \check \Nc_L^{{\rm IV} }  \equiv \check \nG_L^{\prime \prime }\,,~   \check \Nc_L^{{\rm IV}^{\prime } }  \equiv \check \nG_L^{} \,, ~ \check \Nc_L^{{\rm V}^{\prime } }  \equiv \check  \nG_L^\prime \,,~  \check \Nc_L^{\dot {\rm VII}^{\prime} }  \equiv \check \nG_L^{\prime \prime \prime } \,.
\eeqn
Similar to the situation in the WSW pattern, the SM fermion names in Table~\ref{tab:SU8_56ferm_C} are exchanged from what we have obtained in the SWW pattern.
The detailed derivation of the corresponding SM quark/lepton mass matrices will be elaborated in Sec.~\ref{section:WWS_pattern}.

\para
We decompose the Higgs fields in the Yukawa term into components that can be responsible for the sequential symmetry breaking pattern in Eq.~\eqref{eq:Pattern-C}.
For Higgs fields of $\repb{ 8_H}_{\,, \omega}$ they read
\begin{eqnarray}\label{eq:SU8C_Higgs_Br01}
\repb{8_H}_{\,,\omega }  &\supset&  \underline{  ( \repb{4} \,, \rep{1} \,, +\frac{1}{4} )_{\mathbf{H}\,, \omega }  } \oplus  \langle ( \rep{1} \,, \repb{4} \,, -\frac{1}{4} )_{\mathbf{H}\,, \omega } \rangle \non
&\supset&  \underline{ ( \repb{4} \,, \rep{1} \,, +\frac{1}{4} )_{\mathbf{H}\,, \omega }  } \oplus  \langle ( \rep{1} \,, \repb{3} \,, -\frac{1}{3} )_{\mathbf{H}\,, \omega }  \rangle \non
&\supset&  \langle ( \repb{4} \,, \rep{1} \,, +\frac{1}{4} )_{\mathbf{H}\,, \omega }  \rangle \oplus  \underline{ ( \rep{1} \,, \repb{2} \,, -\frac{1}{2} )_{\mathbf{H}\,, \omega } } \supset  \langle ( \rep{1} \,, \repb{2} \,, -\frac{1}{2} )_{\mathbf{H}\,, \omega } \rangle   \,.
\end{eqnarray}
For Higgs fields of $\repb{28_H}_{\,,\dot \omega }$, they read
\begin{eqnarray}\label{eq:SU8C_Higgs_Br02}
\repb{28_H}_{\,,\dot \omega } &\supset& ( \rep{6} \,, \rep{1} \,,  +\frac{1}{2} )_{\mathbf{H}\,, \dot\omega }  \oplus  \underline{ ( \rep{1} \,, \rep{6} \,, -\frac{1}{2} )_{\mathbf{H}\,, \dot\omega } } \oplus \underline{ ( \repb{4} \,, \repb{4} \,, 0 )_{\mathbf{H}\,, \dot\omega } }  \non
&\supset & \[ \langle  ( \rep{1} \,, \repb{3} \,, -\frac{1}{3} )_{\mathbf{H}\,, \dot\omega }^\prime  \rangle \oplus \underline{ ( \rep{1} \,, \rep{3} \,, -\frac{2}{3} )_{\mathbf{H}\,, \dot\omega }  } \] \oplus \[ \underline{  ( \repb{4} \,, \repb{3} \,, -\frac{1}{12} )_{\mathbf{H}\,, \dot\omega }  } \oplus \underline{ ( \repb{4} \,, \rep{1} \,, +\frac{1}{4} )_{\mathbf{H}\,, \dot\omega }^{\prime }  }  \] \non
&\supset&  \left [ \underline{ ( \rep{1} \,, \repb{2} \,, -\frac{1}{2} )_{\mathbf{H}\,, \dot\omega }^\prime } \oplus \underline{ ( \rep{1} \,, \rep{2} \,, -\frac{1}{2} )_{\mathbf{H}\,, \dot\omega }  } \right ] \non
&& \oplus \[  \underline{ ( \repb{4} \,, \repb{2} \,, -\frac{1}{4} )_{\mathbf{H}\,, \dot\omega } }  \oplus  \langle ( \repb{4} \,, \rep{1} \,, +\frac{1}{4} )_{\mathbf{H}\,, \dot\omega } \rangle  \oplus  \langle ( \repb{4} \,, \rep{1} \,, +\frac{1}{4} )_{\mathbf{H}\,, \dot\omega }^{\prime }  \rangle \] \non
&\supset&  \[ \langle ( \rep{1} \,, \repb{2} \,, -\frac{1}{2} )_{\mathbf{H}\,, \dot\omega }^\prime \rangle \oplus \langle ( \rep{1} \,, \rep{2} \,, -\frac{1}{2} )_{\mathbf{H}\,, \dot\omega }  \rangle \] \oplus \[  \langle ( \rep{1} \,, \repb{2} \,, +\frac{1}{2} )_{\mathbf{H}\,, \dot\omega } \rangle  \] \,.
\end{eqnarray}
For Higgs fields of $\rep{70_H}$, they read
\begin{eqnarray}\label{eq:SU8C_Higgs_Br05}
\rep{70_H} &\supset& ( \rep{1} \,, \rep{1 } \,, -1 )_{\mathbf{H}}^{ \prime\prime }  \oplus ( \rep{1} \,, \rep{1 } \,, +1 )_{\mathbf{H}}^{ \prime\prime \prime \prime }  \oplus \underline{ ( \rep{4} \,, \repb{4} \,, +\frac{1}{2} )_{\mathbf{H}} }  \oplus  ( \repb{4} \,, \rep{4} \,, -\frac{1}{2} )_{\mathbf{H}}  \oplus ( \rep{6 } \,, \rep{6 } \,, 0 )_{\mathbf{H}}  \non
&\supset& \underline{ ( \rep{4} \,, \repb{3} \,, +\frac{5}{12} )_{\mathbf{H}} } \supset  \underline{ ( \rep{4} \,, \repb{2} \,, +\frac{1}{4} )_{\mathbf{H}}^{} }  \supset  \langle ( \rep{1} \,, \repb{2} \,, +\frac{1}{2} )_{\mathbf{H}}^{\prime \prime\prime } \rangle \,.
\end{eqnarray}

\para
Schematically, we assign the Higgs VEVs according to the symmetry breaking pattern as follows
\beqs\label{eqs:SU8_WWS_Higgs_VEVs}
\beqn
\Gc_{441} \to \Gc_{431} ~&:&~ \langle ( \rep{1} \,, \repb{4} \,, -\frac{1}{4} )_{\mathbf{H}\,, {\rm IV}} \rangle \equiv \frac{1}{\sqrt{2}}W_{ \repb{4}\,, {\rm IV}}\,, \label{eq:SU8_WWS_Higgs_VEVs_mini01}\\[1mm]
\Gc_{431} \to \Gc_{421} ~&:&~ \langle ( \rep{1} \,, \repb{3} \,, -\frac{1}{3} )_{\mathbf{H}\,,{\rm V} } \rangle \equiv \frac{1}{\sqrt{2}} w_{\repb{3}\,, {\rm V} }\,, ~ \langle ( \rep{1} \,, \repb{3} \,, -\frac{1}{3} )_{\mathbf{H}\,, \dot{1}, \dot {\rm VII}  }^\prime \rangle \equiv \frac{1}{\sqrt{2}}  w_{\repb{3}\,, \dot{1},\dot {\rm VII} } \,,\label{eq:SU8_WWS_Higgs_VEVs_mini02} \\[1mm]
\Gc_{421} \to \Gc_{\rm SM} ~&:&~  \langle ( \repb{4} \,, \rep{1} \,, +\frac{1}{4} )_{\mathbf{H}\,,3, \rm{VI} } \rangle \equiv \frac{1}{\sqrt{2}} V_{ \repb{4}\,,3, \rm{VI} } \,, \non
&&  \langle ( \repb{4} \,, \rep{1} \,, +\frac{1}{4} )_{\mathbf{H}\,, \dot {\rm IX} } \rangle \equiv \frac{1}{\sqrt{2}} V_{ \repb{4}\,,\dot {\rm IX} }  \,, ~ \langle ( \repb{4} \,, \rep{1} \,, +\frac{1}{4} )_{\mathbf{H} \,,\dot{2} \,, \dot {\rm VIII}  }^\prime \rangle \equiv \frac{1}{\sqrt{2}} V_{ \repb{4} \,,\dot{2} \,, \dot {\rm VIII} }^\prime \,,  \label{eq:SU8_WWS_Higgs_VEVs_mini03}\\[1mm]
 {\rm EWSB} ~&:&~   \langle ( \rep{1} \,, \repb{2} \,, +\frac{1}{2} )_{\mathbf{H} }^{ \prime\prime \prime} \rangle \equiv \frac{1}{\sqrt{2}} v_{\rm EW}   \,.\label{eq:SU8_WWS_Higgs_VEVs_mini04}
\eeqn
\eeqs
At the third stage, the Higgs VEVs from the rank-three sector are due to the components of $( \repb{4} \,, \rep{1} \,, +\frac{1}{4} )_{\mathbf{H}\,, \dot {\rm IX} } \subset ( \repb{4} \,, \repb{4} \,, 0 )_{\mathbf{H}\,, \dot {\rm IX} }$ and $( \repb{4} \,, \rep{1} \,, +\frac{1}{4} )_{\mathbf{H} \,,\dot{2} \,, \dot {\rm VIII}  }^\prime \subset ( \repb{4} \,, \repb{4} \,, 0 )_{\mathbf{H}\,, \dot 2\,, \dot {\rm VIII} }$.
This is different from the third stage of the WSW pattern, where the Higgs VEVs from the rank-three sector are from different IRs of $( \rep{1} \,, \rep{6} \,, -\frac{1}{2} )_{\mathbf{H}\,, \dot 2\,, \dot {\rm VIII} }$ and $( \repb{4} \,, \repb{4} \,, 0 )_{\mathbf{H}\,,  \dot {\rm IX} }$ in Eq.~\eqref{eq:SU8_WSW_Higgs_VEVs_mini03}, respectively.
For our later convenience, we also parametrize different symmetry breaking VEVs,
\begin{eqnarray}\label{eq:SU8_WWS_param}
    && \zeta_0 \equiv \frac{ v_U }{ M_{\rm pl} } \,, \quad \zeta_1 \equiv \frac{ W_{ \repb{4}\,, {\rm IV} } }{ M_{\rm pl} } \,, \quad \zeta_2 \equiv \frac{ w_{ \repb{3}\,, 3,{\rm V} }  }{ M_{\rm pl} } \,,  \quad \dot \zeta_2 \equiv \frac{ w_{ \repb{3}\,,\dot{1},\dot {\rm VII} }  }{ M_{\rm pl} } \,,   \non
&&  \zeta_3 \equiv \frac{ V_{ \repb{4}\,,{\rm VI} } }{ M_{\rm pl} } \,, \quad \dot \zeta_3^\prime \equiv \frac{ V_{ \repb{4} \,,\dot{2} \,, \dot {\rm VIII} }^\prime  }{ M_{\rm pl} }  \,, \quad \dot \zeta_3 \equiv \frac{ V_{ \repb{4}\,, \dot {\rm IX} } }{ M_{\rm pl} }     \,, \non
&& \zeta_0 \gg \zeta_1 \gg  \zeta_2 \sim \dot \zeta_2  \gg \zeta_3 \sim \dot \zeta_3^\prime \sim  \dot \zeta_3  \,, \quad \zeta_{i j} \equiv \frac{ \zeta_j }{ \zeta_i } \,, \quad ( i < j ) \,,
\end{eqnarray}
in terms of dimensionless quantities.
Notice that we use the same notations as in Eq.~\eqref{eq:SU8_WSW_param} to manifest the intrinsic hierarchies, while their symmetry breaking patterns should be distinguishable.

\subsection{The $d=5$ operators for the SM quark and lepton masses}

\begin{figure}
\begin{center}
\begin{tikzpicture}[scale=1.2]
\coordinate[] (S1);
    \coordinate [above left=1.8 and 0.4 of S1 ] (T1);
    \coordinate [above right=1.8 and 0.4 of S1 ] (T2);
    \coordinate [above left=0.6 and 1.2 of S1 ] (T3);
    \coordinate [above right=0.6 and 1.2 of S1 ] (T4);
    \draw[scalar] (T1)--node[above left=0.9 and 0.1]{$\langle\Phi_2 \rangle$} (S1);
    \draw[scalar] (T2)--node[above right=0.9 and 0.1]{$\langle\Phi_3 \rangle$} (S1);   
    \draw[scalar] (T3)--node[above left=0.2 and 0.4]{$\langle\Phi_1 \rangle$} (S1);
    \draw[scalar] (T4)--node[above right=0.2 and 0.4]{$\langle\Phi_4 \rangle$} (S1);
    \coordinate [ below=1.0 of S1 ] (S2);
    \draw[scalar] (S2)--node[ right=0.2]{$\Phi^\prime$} (S1);
    \coordinate [below left=1.0 and 1.2 of S2 ] (R1);
    \coordinate [below right=1.0 and 1.2 of S2 ] (R2);
    \draw[fermion] (R1)--node[below left=0.5]{$\Fc_{L}$} (S2);
    \draw[fermion] (R2)--node[below right=0.5]{$\Fc_{R}^{\prime\, c}$} (S2);
\end{tikzpicture}
\end{center}
\caption{The indirect Yukawa couplings through the VEV insertions to the $d=5$ Higgs mixing operators.}\label{fig:indirectY_treed5}
\end{figure}
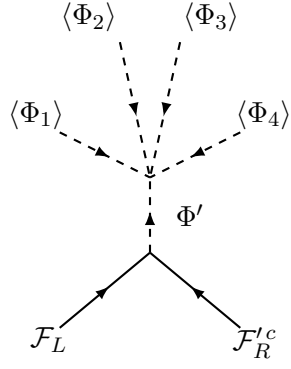

\para
In Ref.~\cite{Chen:2023qxi}, we found that only the top quark obtains the natural tree-level mass from the $Y_\Tc \rep{28_F} \rep{28_F} \rep{70_H} + {\rm H.c.}$ term with $Y_\Tc \sim \Oc(1)$ in Eq.~\eqref{eq:Yukawa_SU8}.
All other lighter SM quark/lepton masses are due to the $d=5$ operators~\cite{Chen:2024cht}, which all explicitly break the emergent global symmetries in Eqs.~\eqref{eq:DRS_SU8} and \eqref{eq:PQ_SU8}.
Among them, we have the $d=5$ direct Yukawa coupling terms of
\beqs\label{eqs:d5_ffHH}
\beqn
c_3\, \Oc_\Fc^{ (3\,,2)} &=& c_3\, \repb{8_F}^{ \dot \omega } \rep{56_F} \cdot \repb{28_H}_{\,, \dot \kappa }^\dag  \cdot  \rep{70_H}^\dag \,,\label{eq:d5_ffHH_32} \\[1mm]
c_4\, \Oc_\Fc^{ (4\,,1)}  &=& c_4\, \rep{56_F}  \rep{56_F} \cdot \repb{28_H}_{\,, \dot \omega }  \cdot  \rep{70_H}  \,,\label{eq:d5_ffHH_41} \\[1mm]
c_5\, \Oc_\Fc^{ (5\,,1)}  &=& c_5\, \rep{28_F}  \rep{56_F} \cdot  \repb{8_H}_{\,,\omega }  \cdot  \rep{70_H} \,.\label{eq:d5_ffHH_51}
\eeqn
\eeqs
The operators of $\Oc_\Fc^{ (4\,,1)}$ and $\Oc_\Fc^{ (5\,,1)}$ were found to generate the hierarchical up-type quark masses.
To generate the masses for all down-type quarks and charged leptons, we conjecture two sets of Higgs mixing terms
\beqs\label{eqs:d5_Hmixings}
\beqn
d_{\mathscr A}\, \Oc_{\mathscr A}^{d=5} &\equiv&d_{\mathscr A}\,  \epsilon_{ \omega_1 \omega_2 \omega_3  \omega_4 } \repb{8_{H}}_{\,, \omega_1}^\dag   \repb{8_{H}}_{\,, \omega_2}^\dag \repb{8_{H}}_{\,, \omega_3}^\dag  \repb{8_{H}}_{\,, \omega_4 }^\dag  \rep{70_H}^\dag  \,, \quad \Pc\Qc = 2 ( 2p + 3 q_2 ) \neq 0 \,, \label{eq:d5_Hmixing_A}\\[1mm]
d_{\mathscr B}\, \Oc_{\mathscr B}^{d=5} &\equiv&d_{\mathscr B}\,   ( \repb{28_H}_{\,,\dot \kappa_1 }^\dag \repb{28_H}_{\,,\dot \kappa_2 } ) \cdot  \repb{28_H}_{\,,\dot \omega_1}^\dag \repb{28_H}_{\,,\dot \omega_2}^\dag   \rep{70_H}^\dag  \,, \quad  \Pc\Qc =  2 ( p + q_2 + q_3) \,, \non
&& {\rm with} ~~ \dot \kappa_2 \neq ( \dot \kappa_1 \,, \dot \omega_1 \,, \dot \omega_2 )  \,, \label{eq:d5_Hmixing_B}
\eeqn
\eeqs
where one of the Higgs fields will mix with two renormalizable Yukawa couplings of $Y_\Bc  \repb{8_F}^\omega  \rep{28_F}  \repb{8_{H}}_{\,,\omega }+ Y_\Dc \repb{8_F}^{\dot \omega  } \rep{56_F}  \repb{28_{H}}_{\,,\dot \omega }   +{\rm H.c.}$ in Eq.~\eqref{eq:Yukawa_SU8}, as were depicted in Fig.~\ref{fig:indirectY_treed5}.
The SM quark/lepton mass terms along two separate WSW and WWS symmetry breaking patterns will be rederived in Appendixes~\ref{section:WSW_process} and \ref{section:WWS_process}.

\section{The WSW symmetry breaking pattern}
\label{section:WSW_pattern}

\subsection{A summary of the vectorlike fermion masses}

%
%
 \begin{table}[htp] {\small
\begin{center}
\begin{tabular}{c|c|c|c|c}
\hline \hline
   Stages &  $Q_e=- \frac{1}{3}$  & $Q_e=+ \frac{2}{3}$   & $Q_e= -1$ & $Q_e= 0$  \\
\hline \hline
  $v_{441}$   & $\DG^{\prime\prime}$   &  $...$  &  $( \eG\,, \nG )$  &  $\{ \check \nG\,, \check \nG^{\prime\prime} \}$  \\
  $\{ \Omega \}$  &  ${\rm IV}$  &    & ${\rm IV}$ &  $\{ {\rm IV}^\prime \,, {\rm IV} \}$  \\[1mm]
    \hline
  $v_{431}$   &  $\dG\,, \{  \DG\,,  \DG^{\prime\prime\prime\prime\prime} \}$  & $\uG\,, \UG$   &  $\EG\,,( \eG^{\prime\prime}\,, \nG^{\prime\prime} )\,,( \eG^{ \prime\prime\prime\prime}\,, \nG^{\prime\prime\prime\prime} )$  & $\{ \check \nG^{\prime}\,, \check \nG^{\prime\prime\prime} \}$   \\
   $\{ \Omega \}$   &  $\{ {\rm V} \,, \dot {\rm VII}  \}$  &    &  $\{ {\rm V} \,, \dot {\rm VII}  \}$  &  $\{ {\rm V}^\prime \,, \dot {\rm VII}^{\prime} \}$  \\[1mm]
    \hline
   $v_{331}$   &  $\{  \DG^{ \prime}\,,  \DG^{ \prime\prime\prime}\,, \DG^{\prime\prime\prime\prime} \}$  &  $...$  &  $( \eG^\prime \,, \nG^\prime )\,,( \eG^{ \prime\prime\prime}\,, \nG^{\prime\prime\prime} )\,, ( \eG^{ \prime\prime\prime\prime\prime}\,, \nG^{\prime\prime\prime\prime\prime} )$  &  $...$  \\
    $\{ \Omega\}$  &  $\{ {\rm VI} \,, \dot {\rm VIII}\,, \dot {\rm IX} \}$  &   &  $\{ {\rm VI} \,, \dot {\rm VIII}\,, \dot {\rm IX} \}$  &    \\[1mm]
\hline\hline
\end{tabular}
\caption{
The vectorlike fermions at different intermediate symmetry breaking scales in the ${\rm SU}(8)$ theory.}
\label{tab:SU8_vectorferm_B}
\end{center}
}
\end{table}%

 \para
Based on our results in Appendix~\ref{section:WSW_process}, we suggest the different intermediate symmetry breaking scales at which the vectorlike fermions obtain masses in Table~\ref{tab:SU8_vectorferm_B}.

\subsection{The CKM matrix and the benchmark}

\para
The biunitary transformations of
\begin{eqnarray} \label{eq:SU8_B_diag}
    && \hat \Fc_L \Mc_\Fc  \hat \Fc_R^\dag = \Mc_\Fc^{\rm diag}\,,\quad \Mc_\Fc \Mc_\Fc^\dag = \hat \Fc_L^\dag ( \Mc_\Fc^{\rm diag} )^2  \hat \Fc_L  \,,\quad \Fc = (\Uc \,, \Dc) \,,
\end{eqnarray}
diagonalize the unhatted flavor eigenstates into their hatted mass eigenstates.
To obtain the CKM matrix of the quark sector, we derive the left-handed mixing matrices of $(\hat \Uc_L \,, \hat \Dc_L)$ of
\begin{eqnarray} \label{eq:SU8_B_Quarks_Rotation}
    &&  \left( \ba{c}  \hat u_L \\  \hat c_L  \\ \hat t_L  \\  \ea  \right)  = \hat \Uc_L \cdot \left( \ba{c}  u_L \\  c_L  \\  t_L  \\  \ea  \right)  \,, \quad  \left( \ba{c}  \hat d_L \\  \hat s_L  \\ \hat b_L  \\  \ea  \right)  = \hat \Dc_L \cdot \left( \ba{c}  d_L \\  s_L  \\  b_L  \\  \ea  \right) \,,
\end{eqnarray}
through their perturbative expansions in Eqs.~\eqref{eqs:WSW_Uquark_masses} and \eqref{eqs:WSW_SMdmasses}.
Explicitly, we find that
\begin{eqnarray} \label{eq:SU8_B_UL}
&& \hat \Uc_L = \hat \Uc_L^{ (12)} \cdot \hat \Uc_L^{ (23)} \cdot \hat \Uc_L^{ (13)} \cdot \hat \Uc_L^{ID}\,,\non
&&\hat \Uc_L^{ ID} = \left( \ba{ccc}
0  &   1 &  0    \\
 -1 & 0   &   0    \\
0  & 0   &  1     \\   \ea  \right)  \,, \quad
\hat \Uc_L^{ (12)} = \left( \ba{ccc}
\cos{\epsilon_1}  &   -\sin{\epsilon_1} &  0    \\
 \sin{\epsilon_1} & \cos{\epsilon_1}   &   0    \\
 0  & 0   &  1     \\   \ea  \right)  \,, \quad
 \sin{\epsilon_1} \simeq\frac{m_u}{m_c}\frac{\zeta_1}{\zeta_2}-\frac{\zeta_2}{\zeta_1} \sim \Oc (10^{-2})  , \non
   &&  \hat \Uc_L^{(13)} \approx   \left( \ba{ccc}
1  &  0  &  - \frac{c_5 \zeta_2}{\sqrt{2}Y_\Tc}    \\
 0 &  1  &  0   \\
 \frac{c_5 \zeta_2}{\sqrt{2}Y_\Tc} &  0  &  1  \\   \ea  \right)  \,, \quad
\hat \Uc_L^{ (23)} \approx   \left( \ba{ccc}
1  & 0   & 0   \\
 0 &  1  &   \frac{c_5 \zeta_1}{\sqrt{2}Y_\Tc}    \\
 0 &  -\frac{c_5 \zeta_1}{\sqrt{2}Y_\Tc} &  1   \\    \ea  \right)  \,,
\end{eqnarray}
and
\begin{eqnarray} \label{eq:SU8_B_DL}
    && \hat \Dc_L= \left( \ba{ccc}
\sin \lambda  &  \cos \lambda  &       \\
- \cos \lambda &  \sin \lambda &       \\
  &    & 1   \\   \ea  \right)  \approx
  \left( \ba{ccc}
 \lambda  &  1-\lambda^2/2   &       \\
- 1 + \lambda^2/2  &  \lambda &       \\
  &    & 1   \\   \ea  \right) \,.
\end{eqnarray}
The CKM matrix can be approximated as the Wolfenstein parametrization
\begin{eqnarray}\label{eq:WSW_CKM}
\hat V_{\rm CKM} \Big|_{ {\rm SU}(8)\,, {\rm WSW} }&=& \hat \Uc_L \hat \Dc_L^\dag  \approx \left( \ba{ccc}
1-\lambda^2/2  & \lambda   & -c_5 \zeta_2 /Y_\Tc \\
-\lambda  &  1-\lambda^2/2  &   c_5 \zeta_1 /Y_\Tc \\
  c_5  ( \lambda \zeta_1 + \zeta_2 ) /Y_\Tc &  - c_5 \zeta_1/Y_\Tc  & 1 \\  \ea \right) \,,
\end{eqnarray}
which is identical to what we have obtained in the SWW sequence~\cite{Chen:2024cht}.
For completeness, we tabulate the benchmark point of the ${\rm SU}(8)$ along the WSW sequence, as well as the predicted SM quantities in Table~\ref{tab:SU8_WSW_benchmark}.
Three dimensionless parameters of $(\zeta_1\,, \zeta_2\,, \zeta_3)$ can be translated into three intermediate symmetry breaking scales in Eq.~\eqref{eq:Pattern-B} as follows:
\beqn\label{eq:WSW_benchmark}
&&  v_{441}\simeq 1.4 \times 10^{17 }\,{\rm GeV}  \,, \quad v_{431} \simeq  4.8\times 10^{15} \,{\rm GeV} \,, \quad v_{331} \simeq 4.8\times 10^{13} \,{\rm GeV} \,.
\eeqn

\begin{table}[htp]
\begin{center}
\begin{tabular}{cccccc}
\hline\hline
$\zeta_1$  &  $\zeta_2$ &  $\zeta_3$ & $Y_\Dc$  & $Y_\Bc$  &   $Y_\Tc$  \\
$ 6.0 \times10^{-2}$  &  $2.0\times10^{-3}$  & $2.0\times10^{-5}$  &  $0.5$ & $0.5$ & $0.8$  \\
\hline
  $c_3$&$c_4$&$c_5$&$d_{\mathscr A}$&  $d_{\mathscr B}$  &$\lambda$    \\
  $1.0$& $ 0.2$ & $1.0$ & $ 0.01$ & $ 0.01$ & $0.22$   \\
\hline
$m_u$  &  $m_c$ &  $m_t$  &  $m_d=m_e$  &  $m_s=m_\mu$  &  $m_b=m_\tau$       \\
$1.6 \times 10^{-3}$  &  $0.6$ &  $139.2$  &  $ 0.5\times 10^{-3}$  &  $6.4 \times 10^{-2}$  &  $1.5$      \\
\hline
$|V_{ud}|$  &  $|V_{us}|$ &  $|V_{ub}|$  &   &   &        \\
$0.98$  &  $0.22$ &  $2.1 \times10^{-3}$  &   &  &       \\
\hline
  $|V_{cd}|$  &  $|V_{cs}|$  &  $|V_{cb}|$  &   &   &  $$    \\
  $0.22$  &  $0.98$  &  $ 5.3\times 10^{-2}$  &   &   &  $$    \\
\hline
$|V_{td}|$  &  $|V_{ts}|$ &   $|V_{tb}|$  &    &   &        \\
$0.013$  &  $ 5.3 \times 10^{-2}$ &  $1$  &   &    &      \\
\hline\hline
\end{tabular}
\end{center}
\caption{The parameters of the $\gSU(8)$ benchmark point along the WSW sequence and the predicted SM quark/lepton masses (in unit of ${\rm GeV}$) as well as the CKM mixings.}
\label{tab:SU8_WSW_benchmark}
\end{table}%

\section{The WWS symmetry breaking pattern}
\label{section:WWS_pattern}

 \subsection{A summary of the vectorlike fermion masses}

 \begin{table}[htp] {\small
\begin{center}
\begin{tabular}{c|c|c|c|c}
\hline \hline
   Stages &  $Q_e=- \frac{1}{3}$  & $Q_e=+ \frac{2}{3}$   & $Q_e= -1$ & $Q_e= 0$  \\
\hline \hline
  $v_{441}$   & $\DG^{\prime\prime}$   &  $...$  &  $( \eG\,, \nG )$  &  $\{ \check \nG\,, \check \nG^{\prime\prime} \}$  \\
  $\{ \Omega \}$  &  ${\rm IV}$  &    & ${\rm IV}$ &  $\{ {\rm IV}^\prime \,, {\rm IV} \}$  \\[1mm]
    \hline
  $v_{431}$   &  $ \{  \DG^{\prime}\,,  \DG^{\prime\prime\prime} \}$  &  $...$  &  $ ( \eG^{\prime}\,, \nG^{\prime} )\,,( \eG^{ \prime\prime\prime}\,, \nG^{\prime\prime\prime} )$  & $\{ \check \nG^{\prime}\,, \check \nG^{\prime\prime\prime} \}$   \\
   $\{ \Omega \}$   &  $\{ {\rm V} \,, \dot {\rm VII}  \}$  &    &  $\{ {\rm V} \,, \dot {\rm VII}  \}$  &  $\{ {\rm V} \,, \dot {\rm VII}^{\prime} \}$  \\[1mm]
    \hline
   $v_{421}$   & $\dG \,,  \{  \DG^{}\,,  \DG^{ \prime\prime\prime\prime}\,, \DG^{\prime\prime\prime\prime\prime} \}$  &  $\uG\,, \UG$  &  $\EG\,,  ( \eG^{\prime\prime} \,, \nG^{\prime\prime} )\,,( \eG^{ \prime\prime\prime\prime\prime}\,, \nG^{\prime\prime\prime\prime\prime} )\,, ( \eG^{ \prime\prime\prime\prime}\,, \nG^{\prime\prime\prime\prime} )$  &  $...$  \\
    $\{ \Omega\}$  &  $\{ {\rm VI} \,, \dot {\rm VIII}\,, \dot {\rm IX} \}$  &   &  $\{ {\rm VI} \,, \dot {\rm VIII}\,, \dot {\rm IX} \}$  &    \\[1mm]
\hline\hline
\end{tabular}
\caption{
The vectorlike fermions at different intermediate symmetry breaking scales in the ${\rm SU}(8)$ theory.}
\label{tab:SU8_vectorferm_C}
\end{center}
}
\end{table}%

Based on our results in Appendix~\ref{section:WWS_process}, we suggest the different intermediate symmetry breaking scales at which the vectorlike fermions obtain masses in Table~\ref{tab:SU8_vectorferm_C}.

\subsection{The CKM matrix and the benchmark}

\para
The biunitary transformations of
\begin{eqnarray}  \label{eq:SU8_C_diag}
    && \hat \Fc_L \Mc_\Fc  \hat \Fc_R^\dag = \Mc_\Fc^{\rm diag}\,,\quad \Mc_\Fc \Mc_\Fc^\dag = \hat \Fc_L^\dag ( \Mc_\Fc^{\rm diag} )^2  \hat \Fc_L  \,,\quad \Fc = (\Uc \,, \Dc) \,,
\end{eqnarray}
diagonalize the unhatted flavor eigenstates into their hatted mass eigenstates.
To obtain the CKM matrix of the quark sector, we derive the left-handed mixing matrices of $(\hat \Uc_L \,, \hat \Dc_L)$ of
\begin{eqnarray} \label{eq:SU8_C_Quarks_Rotation}
    &&  \left( \ba{c}  \hat u_L \\  \hat c_L  \\ \hat t_L  \\  \ea  \right)  = \hat \Uc_L \cdot \left( \ba{c}  u_L \\  c_L  \\  t_L  \\  \ea  \right)  \,, \quad  \left( \ba{c}  \hat d_L \\  \hat s_L  \\ \hat b_L  \\  \ea  \right)  = \hat \Dc_L \cdot \left( \ba{c}  d_L \\  s_L  \\  b_L  \\  \ea  \right) \,,
\end{eqnarray}
through their perturbative expansions in Eqs.~\eqref{eqs:WSW_Uquark_masses} and \eqref{eqs:WSW_SMdmasses}.
Explicitly, we find that
\begin{eqnarray} \label{eq:SU8_C_UL}
&& \hat \Uc_L = \hat \Uc_L^{ (12)} \cdot \hat \Uc_L^{ (23)} \cdot \hat \Uc_L^{ (13)} \cdot \hat \Uc_L^{ID}\,,\non
&&\hat \Uc_L^{ ID} = \left( \ba{ccc}
0  &   1 &  0    \\
-1 & 0   &   0    \\
0  & 0   &  1     \\   \ea  \right)  \,, \quad
\hat \Uc_L^{ (12)} = \left( \ba{ccc}
\cos{\epsilon_2}  &   -\sin{\epsilon_2} &  0    \\
\sin{\epsilon_2} & \cos{\epsilon_2}   &   0    \\
0  & 0   &  1     \\   \ea  \right)  \,, \quad
\sin{\epsilon_2} \simeq\frac{m_u}{m_c}\frac{\zeta_1}{\zeta_3}-\frac{\zeta_3}{\zeta_1} \sim \Oc (\zeta_{13})  , \non
&&  \hat \Uc_L^{(13)} \approx   \left( \ba{ccc}
1  &  0  &  - \frac{c_5 \zeta_3}{\sqrt{2}Y_\Tc}    \\
0 &  1  &  0   \\
\frac{c_5 \zeta_3}{\sqrt{2}Y_\Tc} &  0  &  1  \\   \ea  \right)  \,, \quad
\hat \Uc_L^{ (23)} \approx   \left( \ba{ccc}
1  & 0   & 0   \\
0 &  1  &   \frac{c_5 \zeta_1}{\sqrt{2}Y_\Tc}    \\
0 &  -\frac{c_5 \zeta_1}{\sqrt{2}Y_\Tc} &  1   \\    \ea  \right)  \,,
\end{eqnarray}
and
\begin{eqnarray} \label{eq:SU8_C_DL}
    && \hat \Dc_L 
\approx
   \left( \ba{ccc}
   \zeta_{ 23 }  &  1- \zeta_{ 23 }^2/2  &    0   \\
    -1+\zeta_{ 23 }^2/2 &  \zeta_{ 23 } &    0   \\
 0  & 0   & 1   \\   \ea  \right) \,.
\end{eqnarray}
The CKM matrix can be approximated as the Wolfenstein parametrization
\begin{eqnarray}\label{eq:WWS_CKM}
\hat V_{\rm CKM} \Big|_{ {\rm SU}(8)\,, {\rm WWS} }&=& \hat \Uc_L \hat \Dc_L^\dag  \approx \left( \ba{ccc}
 1-\zeta_{ 23 }^2/2&  \zeta_{ 23 }   & -\frac{c_5 \zeta_3}{\sqrt{2}Y_\Tc}\\
- \zeta_{ 23 }   &  1- \zeta_{ 23 }^2/2  &   \frac{c_5 \zeta_1}{\sqrt{2}Y_\Tc} \\
   \frac{c_5\left(\zeta_{ 23 } \zeta_1+ \zeta_3  \right)}{\sqrt{2}Y_\Tc} &  -\frac{c_5 \zeta_1 }{\sqrt{2}Y_\Tc}  & 1 \\  \ea \right)  \,,
\end{eqnarray}
where the Cabibbo mixing parameter is interpreted as the ratio of $\lambda = | V_{us} | =\zeta_{ 23 }$.
The observed mixing hierarchy of $|V_{cb}| \gg | V_{ub}|$ is due to the hierarchy of $\zeta_1 \gg \zeta_3$, while this was due to the hierarchy of $\zeta_1 \gg \zeta_2$ in the SWW and the WSW patterns.
To obtain the reasonable CKM mixing parameters, one has to choose a suppressed $c_3=0.01$ for the reasonable $(d\,,e)$ masses, and a relatively enhanced Higgs mixing parameter of $d_{\mathscr A}=0.2$ for the reasonable $(b\,, \tau)$ masses.

\begin{table}[htp]
	\begin{center}
		\begin{tabular}{cccccc}
			\hline\hline
			$\zeta_1$  &  $\zeta_2$ &  $\zeta_3$ & $Y_\Dc$  & $Y_\Bc$  &   $Y_\Tc$  \\
			$ 6.0 \times10^{-2}$  &  $2.0\times10^{-3}$  &  $4.4\times10^{-4}$  &  $1.0$ & $0.4$ & $0.8$  \\
			\hline
			$c_3$&$c_4$&$c_5$&$d_{\mathscr A}$&  $d_{\mathscr B}$  &$\zeta_{ 23 }$    \\
			$0.01$  & $ 1.5$ & $1.0$ & $ 0.2$ & $ 0.01$ & $0.22$   \\
			\hline
			$m_u$  &  $m_c$ &  $m_t$  &  $m_d=m_e$  &  $m_s=m_\mu$  &  $m_b=m_\tau$       \\
			$0.6 \times 10^{-3}$  &  $0.4$ &  $139.2$  &  $ 0.5\times 10^{-3}$  &  $2.5 \times 10^{-2}$  &  $1.3$      \\
			\hline
			$|V_{ud}|$  &  $|V_{us}|$ &  $|V_{ub}|$  &   &   &        \\
			$0.98$  &  $0.22$ &  $0.4 \times10^{-3}$  &   &  &       \\
			\hline
			$|V_{cd}|$  &  $|V_{cs}|$  &  $|V_{cb}|$  &   &   &  $$    \\
			$0.22$  &  $0.98$  &  $ 5.3\times 10^{-2}$  &   &   &  $$    \\
			\hline
			$|V_{td}|$  &  $|V_{ts}|$ &   $|V_{tb}|$  &    &   &        \\
			$0.012 $  &  $ 5.3 \times 10^{-2}$ &  $1$  &   &    &      \\
			\hline\hline
		\end{tabular}
	\end{center}
	\caption{The parameters of the $\gSU(8)$ benchmark point along the WWS sequence and the predicted SM quark/lepton masses (in unit of ${\rm GeV}$) as well as the CKM mixings.}
	\label{tab:SU8_WWS_benchmark}
\end{table}%

\para
Based on the predicted quark masses in Eqs.~\eqref{eq:WWS_SMumasses} and \eqref{eqs:WWS_SMdmasses} and the CKM matrix in Eq.~\eqref{eq:WSW_CKM}, we suggest a set of benchmark point of the ${\rm SU}(8)$ input parameters, as well as the predicted SM quantities in Table~\ref{tab:SU8_WWS_benchmark}.
Three dimensionless parameters of $(\zeta_1\,, \zeta_2\,, \zeta_3)$ can be translated into three intermediate symmetry breaking scales in Eq.~\eqref{eq:Pattern-C} as follows:
\beqn\label{eq:WWS_benchmark}
&& v_{441}\simeq 1.4 \times 10^{17 }\,{\rm GeV}  \,, \quad v_{431} \simeq  4.8\times 10^{15} \,{\rm GeV}\,, \quad v_{421} \simeq 1.1\times 10^{15} \,{\rm GeV} \,.
\eeqn
Obviously, the intermediate scale at the third symmetry breaking stage is much higher than what we have obtained in the SWW and the WSW symmetry breaking patterns~\cite{Chen:2024cht}.

\section{ The gauge coupling evolutions in the SU(8) theory}
\label{section:RGE}
%

\para
The gauge coupling evolutions rely on the RGEs of the ${\rm SU}(8)$ theory.
The two-loop RGE of a gauge coupling of $\alpha_\Upsilon$ in the $\overline{\rm MS}$ scheme is given by~\cite{Machacek:1983tz}
\beqn \label{eq:SU8_2loop_RGE}
&&  \frac{d \alpha_\Upsilon ( \mu ) }{ d\log \mu} = \frac{ b_\Upsilon^{ (1) } }{2\pi } \alpha_\Upsilon^2 ( \mu ) +  \Big( \sum_{ \Upsilon^\prime } \frac{b_{  \Upsilon \Upsilon^\prime }^{ ( 2 ) } }{ 8 \pi^2  } \alpha_{\Upsilon^\prime } ( \mu ) + \sum_{ \Fc = \Tc\,, \Bc \,, \Dc } f_\Upsilon ( Y_\Fc ) \Big) \cdot  \alpha_\Upsilon^2 ( \mu ) \,,
\eeqn
where the one- and two-loop $\beta$ coefficients for the non-Abelian groups are
\beqs\label{eqs:SU8_1_2_loop_beta}
\begin{align}
b^{\(1\)}_{\Upsilon}=&-\frac{11}{3}C_2\(\Gc_\Upsilon\)+\frac{2}{3}\sum_{F}T\(\Rc^{F}_{\Upsilon}\)+\frac{1}{3}\sum_{S}T\(\Rc^{S}_{\Upsilon}\)\,, \label{eq:SU8_1_loop_beta} \\
b^{\(2\)}_{\Upsilon\Upsilon^\prime}=&-\frac{34}{3}C_2\(\Gc_\Upsilon\)^2  + \sum_F\[2   C_2\(\Rc^{F}_{\Upsilon^\prime}\) + \frac{10}{3}C_2\(\Gc_\Upsilon\)   \]T\(\Rc^F_\Upsilon\) \non
&+\sum_S\[4  C_2\(\Rc^{S}_{\Upsilon^\prime}\)  +  \frac{2}{3}C_2\(\Gc_\Upsilon\)   \]T\(\Rc^S_\Upsilon\)\,,\label{eq:SU8_2_loop_beta}
\end{align}
\eeqs
and the $f_\Upsilon(Y_\Fc)$ are the polynomials to account for the renormalizable Yukawa couplings at the two-loop level.
Here, $C_2( \Gc_\Upsilon )$ is the quadratic Casimir of the gauge group $\Gc_\Upsilon$, and $T(  \Rc^F_\Upsilon ) $ and $T(  \Rc^S_\Upsilon ) $ are trace invariants of the chiral fermions in the IR of $\Rc^F_\Upsilon$ and complex scalars in the IR of $\Rc^S_\Upsilon$.
For the ${\rm U}(1)$ Abelian groups with charges denoted as $\Xc^{F/S}$, the one- and two-loop $\beta$ coefficients read
\beqs
\beqn
&& b^{\(1\)}_{\Upsilon}= \frac{2}{3}\sum_{F}\(\Xc^{F}_{\Upsilon}\)^2+\frac{1}{3}\sum_{S}\(\Xc^{S}_{\Upsilon}\)^2\,, \label{eq:SU8_U1_beta} \\
&& b_{\Upsilon \Upsilon^\prime }^{ (2 ) } =   2    \sum_{ F } (  \Xc^F_{\Upsilon^\prime } )^2   \cdot (  \Xc^F_\Upsilon )^2  + 4    \sum_{ S  }  (  \Xc^S_{\Upsilon^\prime } )^2  \cdot (  \Xc^S_\Upsilon )^2 \,.
\eeqn
\eeqs
Below, we first list the minimal set of massless Higgs fields according to the survival hypothesis~\cite{Georgi:1979md,Glashow:1979nm,Barbieri:1979ag,Barbieri:1980vc,Barbieri:1980tz,delAguila:1980qag} and the massless fermions between different symmetry breaking stages according to the minimal Higgs VEVs in Eqs.~\eqref{eqs:SU8_WSW_Higgs_VEVs} and \eqref{eqs:SU8_WWS_Higgs_VEVs}.
The two-loop RGEs within the minimal ${\rm SU}(8)$ setup are therefore derived by using the PyR@TE code~\cite{Sartore:2020gou}.

\subsection{RGEs in the WSW symmetry breaking pattern}

\para
Between the $v_{441}\leq \mu\leq v_U$, almost all massless Higgs fields are in the first lines of Eqs.~\eqref{eq:SU8B_Higgs_Br01}-\eqref{eq:SU8B_Higgs_Br05}, which are
\beqn \label{eq:SU8_WSW_441_massless_H}
&& ( \repb{4}\,, \rep{1}\,, +\frac{1}{4})_{ \mathbf{H}\,, \omega} \oplus
( \rep{1}\,, \repb{4}\,, -\frac{1}{4})_{ \mathbf{H}\,, \omega}
\subset \repb{8_H}_{, \omega }\,, \non
&& ( \rep{6}\,, \rep{1}\,, +\hf)_{\mathbf{H}\,,\dot \omega} \oplus
(\rep{1}\,,\rep{6}\,,-\hf)_{\mathbf{H\,,\dot\omega}} \oplus
(\repb{4}\,,\repb{4}\,,0)_{\mathbf{H}\,,\dot\omega} \subset \repb{28_H}_{,\dot\omega}\,, \non
&&(\rep{4}\,,\repb{4}\,,\hf)_\mathbf{H} \oplus
(\repb{4}\,,\rep{4}\,,\hf)_\mathbf{H} \oplus
(\rep{6}\,,\rep{6}\,,0)_\mathbf{H} \subset \rep{70_H}\,.
\eeqn
All ${\rm SU}(8)$ fermions in Eq.~\eqref{eq:SU8_3gen_fermions} remain massless after the decomposition into the $\Gc_{441}$ IRs.
Correspondingly, we have the $\Gc_{441}$ $\beta$ coefficients of
\beqn \label{eq:WSW_441_beta}
&& (b^{(1)}_{{\rm SU}(4)_{s}}\,,b^{(1)}_{{\rm SU}(4)_{W}}\,,b^{(1)}_{{\rm U}(1)_{X_0}})=(+\frac{13}{3}\,,+\frac{13}{3}\,,+\frac{55}{3})\,, \non
&& b^{(2)}_{\Gc_{441}}=
\begin{pmatrix}
2299/6&405/2&12\\
405/2&2299/6&12\\
180&180&31
\end{pmatrix}\,, \non
&& f_{ {\rm SU}(4)_s } = f_{ {\rm SU}(4)_W } = -24 Y_\Tc^2 - 38 Y_\Bc^2 - 165 Y_\Dc^2 \,, \non
&& f_{ {\rm U}(1)_{X_0 } } = -54 Y_\Tc^2 - 38 Y_\Bc^2 - 165 Y_\Dc^2 \,.
\eeqn
When the RGEs evolve down to $v_{441}$, the gauge couplings should match according to the following relations:
\beqn\label{eq:WSW_441_coupMatch}
&& \alpha_{3W }^{-1}(v_{441} ) = \alpha_{4W }^{-1} (v_{441} ) \,,~ \alpha_{X_1}^{-1} (v_{441} ) = \frac{1}{6} \alpha_{4W }^{-1}(v_{441} ) +  \alpha_{X_0}^{-1} (v_{441} ) \,.
\eeqn

\para
Between the $v_{431}\leq \mu\leq v_{441}$, the massless Higgs fields are
\beqn \label{eq:SU8_WSW_431_massless_H}
&& ( \repb{4}\,, \rep{1}\,, +\frac{1}{4})_{ \mathbf{H}\,, \rm{V}} \subset \repb{8_H}_{, \rm{V} }\,, \non
&& ( \rep{1}\,, \repb{3}\,, -\frac{1}{3})_{ \mathbf{H}\,, 3\,,\rm{VI}} \subset ( \rep{1}\,, \repb{4}\,, -\frac{1}{4})_{ \mathbf{H}\,, 3\,,\rm{VI}} \subset      \repb{8_H}_{, 3\,,\rm{VI} }\,,\non
&& (\repb{4}\,,\repb{3}\,,-\frac{1}{12})_{\mathbf{H}\,,\dot{\rm{IX}}} \subset (\repb{4}\,,\repb{4}\,,    0        )_{\mathbf{H}\,,\dot{\rm{IX}}} \subset \repb{28_H}_{,\dot{\rm{IX}}}\,, \non
&&(\repb{4}\,,\rep{1}\,,+\frac{1}{4})_{\mathbf{H}\,,\dot{1}\,,\dot{\rm{VII}}} \subset (\repb{4}\,,\rep{4}\,,   0 )_{\mathbf{H}\,,\dot{1}\,,\dot{\rm{VII}}} \subset \repb{28_H}_{,\dot{1}\,,\dot{\rm{VII}}}\,, \non
&& (\rep{1}\,,\repb{3}\,,-\frac{1}{3})^{\prime}_{\mathbf{H}\,,\dot{2}\,,\dot{\rm{VIII}}} \subset (\rep{1}\,,\rep{6}\,,-\frac{1}{2})_{\mathbf{H}\,,\dot{2}\,,\dot{\rm{VIII}}} \subset \repb{28_H}_{,\dot{2}\,,\dot{\rm{VIII}}}\,, \non
&& (\rep{4}\,,\repb{3}\,,+\frac{5}{12})_\mathbf{H} \subset (\rep{4}\,,\repb{4}\,,+\frac{1}{ 2})_\mathbf{H} \subset \rep{70_H}\,.
\eeqn
The massless $\Gc_{431}$ fermions are listed in Eq.~\eqref{eq:431B_fermions}.
Correspondingly, we have the $\Gc_{431}$ $\beta$ coefficients of
\beqn \label{eq:WSW_431_beta}
&& (b^{(1)}_{{\rm SU}(4)_{s}}\,,b^{(1)}_{{\rm SU}(3)_{W}}\,,b^{(1)}_{{\rm U}(1)_{X_1}})=(-\frac{23}{6}\,,+\frac{1}{3}\,,+\frac{443}{36})\,, \non
&& b^{(2)}_{\Gc_{431}}=
\begin{pmatrix}
659/6&36&17/4\\
135/2&358/3&181/36\\
255/4&362/9&2641/216
\end{pmatrix}\,, \non
&& f_{ {\rm SU}(4)_s }  = - \frac{ 21}{ 2} Y_\Tc^2 - 11 Y_\Bc^2 - 31 Y_\Dc^2 \,, \non
&& f_{ {\rm SU}(3)_W }  = - 6 Y_\Tc^2 - 12 Y_\Bc^2 - 32 Y_\Dc^2 \,, \non
&& f_{ {\rm U}(1)_{X_1 } } = - \frac{37 }{4}Y_\Tc^2 -  \frac{27 }{2} Y_\Bc^2 -  \frac{ 173 }{6} Y_\Dc^2 \,.
\eeqn
When the RGEs evolve down to $v_{431}$, the gauge couplings should match according to the following relations:
\beqn\label{eq:WSW_431_coupMatch}
&& \alpha_{3c }^{-1}(v_{431} ) = \alpha_{4s }^{-1} (v_{431} ) \,,~ \alpha_{X_2}^{-1} (v_{431} ) = \frac{1}{6} \alpha_{4s }^{-1}(v_{431} ) +  \alpha_{X_1}^{-1} (v_{431} ) \,.
\eeqn

\para
Between the $v_{331}\leq \mu\leq v_{431}$, the massless Higgs fields are
\beqn \label{eq:SU8_WSW_331_massless_H}
&&( \rep{1}\,, \repb{3}\,, -\frac{1}{3})_{ \mathbf{H}\,, 3\,,\rm{VI}} \subset  ( \rep{1}\,, \repb{4}\,, -\frac{1}{4})_{ \mathbf{H}\,, 3\,,\rm{VI}} \subset\repb{8_H}_{, 3\,,\rm{VI} }\,,\non
&& (\rep{1}\,,\repb{3}\,,-\frac{1}{3})_{\mathbf{H}\,,\dot{\rm{IX}}}  \subset(\repb{4}\,,\repb{3}\,,-\frac{1}{12})_{\mathbf{H}\,,\dot{\rm{IX}}} \subset (\repb{4}\,,\repb{4}\,,    0        )_{\mathbf{H}\,,\dot{\rm{IX}}} \subset \repb{28_H}_{,\dot{\rm{IX}}}\,, \non
&& (\rep{1}\,,\repb{3}\,,-\frac{1}{3})^{\prime}_{\mathbf{H}\,,\dot{2}\,,\dot{\rm{VIII}}} \subset (\rep{1}\,,\rep{6}\,,-\frac{1}{2})_{\mathbf{H}\,,\dot{2}\,,\dot{\rm{VIII}}} \subset \repb{28_H}_{,\dot{2}\,,\dot{\rm{VIII}}}\,, \non
&& (\rep{1}\,,\repb{3}\,,+\frac{2}{3})_\mathbf{H}^{\prime \prime \prime} \subset (\rep{4}\,,\repb{3}\,,+\frac{5}{12})_\mathbf{H} \subset (\rep{4}\,,\repb{4}\,,+\frac{1}{ 2})_\mathbf{H} \subset \rep{70_H}\,. 
\eeqn
The massless $\Gc_{331}$ fermions are listed in Eq.~\eqref{eq:331B_fermions}.
Correspondingly, we have the $\Gc_{331}$ $\beta$ coefficients of
\beqn\label{eq:WSW_331_beta}
&&(b^{(1)}_{{\rm SU}(3)_{c}}\,,b^{(1)}_{{\rm SU}(3)_{W}}\,,b^{(1)}_{{\rm U}(1)_{X_2}})=(-5\,,-4\,,+9)\,,\non
&&b^{(2)}_{\Gc_{331}}=
\begin{pmatrix}
12&12&2\\
12&34&4\\
16&32&100/9
\end{pmatrix}\,, \non
&& f_{ {\rm SU}(3)_c }  = -3 Y_\Tc^2 - 6 Y_\Bc^2 -  6 Y_\Dc^2 \,, \non
&& f_{ {\rm SU}(3)_W }  = - \frac{3 }{2} Y_\Tc^2 - 7 Y_\Bc^2 - 7 Y_\Dc^2 \,, \non
&& f_{ {\rm U}(1)_{X_2 } } = - 4 Y_\Tc^2 -  \frac{26 }{3} Y_\Bc^2 -  \frac{ 26 }{3} Y_\Dc^2 \,.
\eeqn
When the RGEs evolve down to $v_{331}$, the gauge couplings should match according to the following relations:
\beqn\label{eq:331_coupMatch}
&&  \alpha_{2W }^{-1} (v_{331} ) =  \alpha_{3W }^{-1} (v_{331} ) \,, ~ \alpha_Y^{-1} (v_{331} ) = \frac{1}{3} \alpha_{3W }^{-1} (v_{331} ) +  \alpha_{X_2}^{-1} ( v_{331} ) \,.
\eeqn

\para
Between the $v_{\rm EW}\leq \mu\leq v_{331}$, the massless Higgs fields are
\beqn \label{eq:SU8_WSW_321_massless_H}
(\rep{1}\,,\repb{2}\,,+\frac{1}{2})^{\prime \prime \prime}_\mathbf{H} \subset (\rep{1}\,,\repb{3}\,,+\frac{2}{3})_\mathbf{H}^{\prime \prime \prime} \subset (\rep{4}\,,\repb{3}\,,+\frac{5}{12})_\mathbf{H} \subset (\rep{4}\,,\repb{4}\,,+\frac{1}{ 2})_\mathbf{H} \subset \rep{70_H} \, .
\eeqn
The massless $\Gc_{\rm SM}$ fermions are listed at Eq.~\eqref{eq:321B_fermions}.
Correspondingly, we have the $\Gc_{\rm SM}$ $\beta$ coefficients of
\beqn \label{eq:WSW_SM_beta}
&& (b^{(1)}_{{\rm SU}(3)_{c}}\,,b^{(1)}_{{\rm SU}(2)_{W}}\,,b^{(1)}_{{\rm U}(1)_{Y}})=(-7\,,-\frac{19}{6}\,,+\frac{41}{6})\,, \non
&& b^{(2)}_{\Gc_{\rm{SM}}}=
\begin{pmatrix}
-26&9/2&11/6\\
12&35/6&3/2\\
44/3&9/2&199/18
\end{pmatrix}\,.
\eeqn
%
%

\begin{figure}[htb]
\centering
\includegraphics[height=6.5cm]{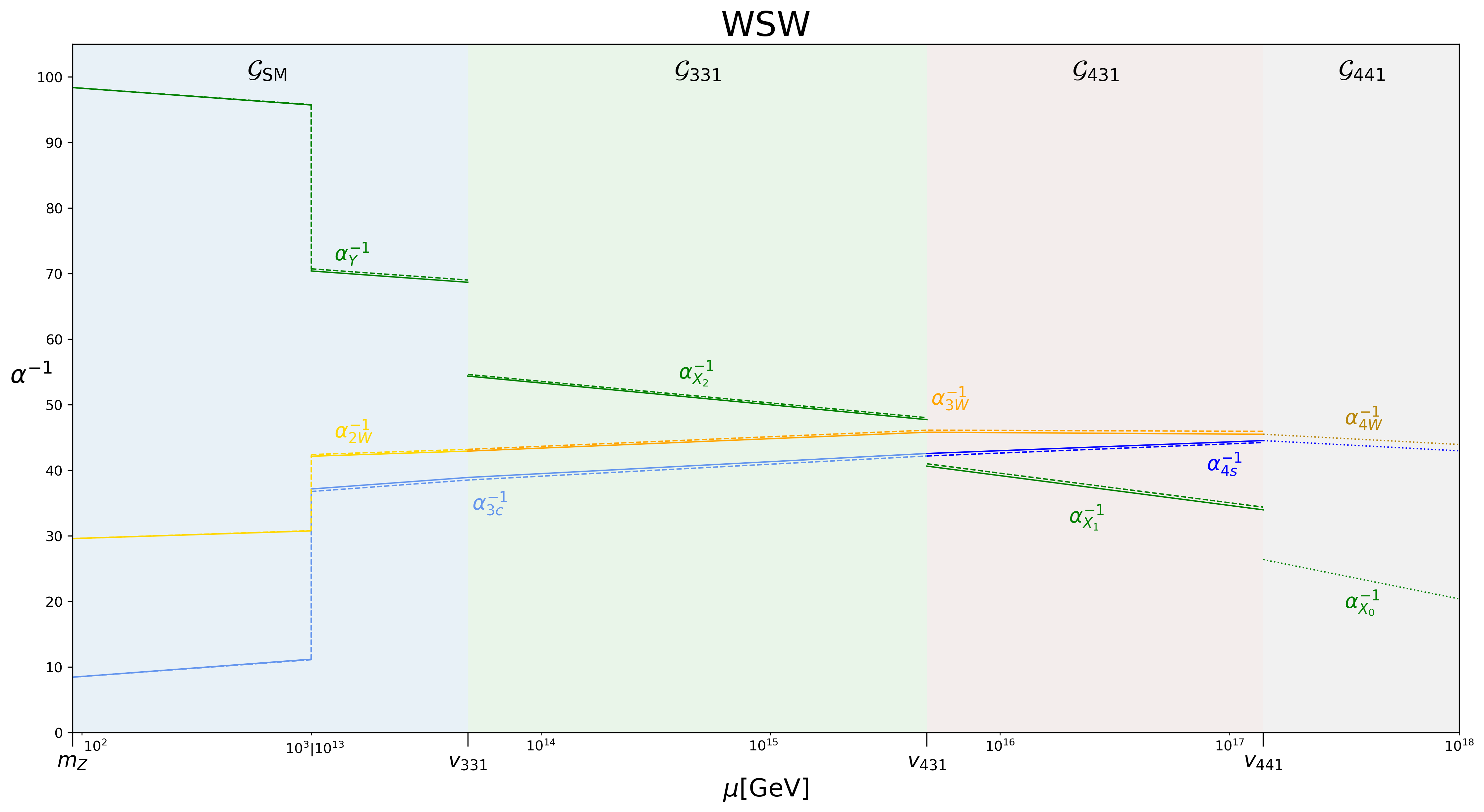}
\caption{The RGEs of the minimal ${\rm SU}(8)$ setup for the WSW symmetry breaking pattern.
The dashed lines and the solid lines represent the one- and two-loop RGEs, respectively.}
\label{fig:WSW_RGE_mini} 
\end{figure}

\para
With the one- and two-loop $\beta$ coefficients, we plot the RGEs of the minimal ${\rm SU}(8)$ setup according to the WSW symmetry breaking pattern in Fig.~\ref{fig:WSW_RGE_mini}.
Three intermediate symmetry breaking scales are given in Eq.~\eqref{eq:WSW_benchmark}, and three renormalizable Yukawa couplings follow from the benchmark point in Table~\ref{tab:SU8_WSW_benchmark}.
Obviously, the evolutions of three gauge couplings cannot achieve the unification.

\subsection{RGEs in the WWS symmetry breaking pattern}

\para
Between the $v_{441}\leq \mu\leq v_U$, the $\Gc_{441}$ $\beta$ coefficients are identical to Eq.~\eqref{eq:WSW_441_beta}, while the matching conditions at $v_{441}$ are identical to Eq.~\eqref{eq:WSW_441_coupMatch}.

\para
Between the $v_{431}\leq \mu\leq v_{441}$, the massless Higgs fields are
\beqn  \label{eq:SU8_WWS_431_massless_H}
	&& ( \repb{4}\,, \rep{1}\,, +\frac{1}{4})_{ \mathbf{H}\,, 3\,, \rm{VI}} \subset \repb{8_H}_{, 3\,,\rm{VI} }\,,\non
	&& ( \rep{1}\,, \repb{3}\,, -\frac{1}{3})_{ \mathbf{H}\,,\rm{V}} \subset ( \rep{1}\,, \repb{4}\,, -\frac{1}{4})_{ \mathbf{H}\,,\rm{V}} \subset      \repb{8_H}_{, \rm{V} }\,,\non
	&&	(\repb{4}\,,\repb{3}\,,-\frac{1}{12})_{\mathbf{H}\,,\dot{\rm{IX}}} \subset (\repb{4}\,,\repb{4}\,,    0        )_{\mathbf{H}\,,\dot{\rm{IX}}} \subset \repb{28_H}_{,\dot{\rm{IX}}}\,,\non
	&& (\repb{4}\,,\rep{1}\,,+\frac{1}{4})^{\prime}_{\mathbf{H}\,,\dot{2}\,,\dot{\rm{VIII}}} \subset (\repb{4}\,,\rep{4}\,,   0 )_{\mathbf{H}\,,\dot{2}\,,\dot{\rm{VIII}}} \subset \repb{28_H}_{,\dot{1}\,,\dot{\rm{VII}}}\,,\non
	&&	(\rep{1}\,,\repb{3}\,,-\frac{1}{3})^{\prime}_{\mathbf{H}\,,\dot{1}\,,\dot{\rm{VII}}} \subset (\rep{1}\,,\rep{6}\,,-\frac{1}{2})_{\mathbf{H}\,,\dot{1}\,,\dot{\rm{VII}}} \subset \repb{28_H}_{,\dot{2}\,,\dot{\rm{VIII}}}\, ,\non
	&& (\rep{4}\,,\repb{3}\,,+\frac{5}{12})_\mathbf{H} \subset (\rep{4}\,,\repb{4}\,,+\frac{1}{ 2})_\mathbf{H} \subset \rep{70_H}\,.
\eeqn
The massless $\Gc_{431}$ fermions are listed in Eq.~\eqref{eq:431B_fermions}.
Correspondingly, we have the $\Gc_{431}$ $\beta$ coefficients of
\beqn \label{eq:WWS_431_beta}
&& (b^{(1)}_{{\rm SU}(4)_{s}}\,,b^{(1)}_{{\rm SU}(3)_{W}}\,,b^{(1)}_{{\rm U}(1)_{X_1}})=(-\frac{10}{3}\,,+\frac{5}{6}\,,+\frac{110}{9})\,,\non
&& b^{(2)}_{\Gc_{431}}=
\begin{pmatrix}
1501/12&44&103/24\\
165/2&391/3&175/36\\
515/8&350/9&5219/432
\end{pmatrix}\,, \non
&& f_{ {\rm SU}(4)_s }  = -\frac{ 21}{ 2} Y_\Tc^2 - 13 Y_\Bc^2 -  31 Y_\Dc^2 \,, \non
&& f_{ {\rm SU}(3)_W }  = - 6 Y_\Tc^2 - 12 Y_\Bc^2 - 32 Y_\Dc^2 \,, \non
&& f_{ {\rm U}(1)_{X_1 } } = - \frac{37}{ 4} Y_\Tc^2 -  \frac{29 }{2 } Y_\Bc^2 -  \frac{ 173 }{6} Y_\Dc^2 \,.
\eeqn
When the RGEs evolve down to $v_{431}$, the gauge couplings should match according to the following relations:
\beqn\label{eq:WWS_431_coupMatch}
&&  \alpha_{2W }^{-1} (v_{431} ) =  \alpha_{3W }^{-1} (v_{431} ) \,, ~ \alpha_{X_2}^{-1} (v_{431} ) = \frac{1}{3} \alpha_{3W }^{-1} (v_{431} ) +  \alpha_{X_1}^{-1} ( v_{431} ) \,.
\eeqn

\para
Between the $v_{421}\leq \mu\leq v_{431}$, the massless Higgs fields are
\beqn \label{eq:SU8_WWS_421_massless_H}
&&( \repb{4}\,, \rep{1}\,, +\frac{1}{4})_{ \mathbf{H}\,, 3\,,\rm{VI}} \subset \repb{8_H}_{, 3\,,\rm{VI} }\,, \non
&& (\repb{4}\,,\rep{1}\,,+\frac{1}{4})_{\mathbf{H}\,,\dot{\rm{IX}}} \subset (\repb{4}\,,\repb{3}\,,-\frac{1}{12})_{\mathbf{H}\,,\dot{\rm{IX}}} \subset (\repb{4}\,,\repb{4}\,,    0        )_{\mathbf{H}\,,\dot{\rm{IX}}} \subset \repb{28_H}_{,\dot{\rm{IX}}}\,,\non
&&(\repb{4}\,,\rep{1}\,,+\frac{1}{4})^{\prime}_{\mathbf{H}\,,\dot{2}\,,\dot{\rm{VIII}}} \subset (\repb{4}\,,\rep{4}\,,   0 )_{\mathbf{H}\,,\dot{2}\,,\dot{\rm{VIII}}} \subset \repb{28_H}_{,\dot{1}\,,\dot{\rm{VII}}}\,,\non
&& (\rep{4}\,,\repb{2}\,,+\frac{1}{ 4})_\mathbf{H} \subset (\rep{4}\,,\repb{3}\,,+\frac{5}{12})_\mathbf{H} \subset (\rep{4}\,,\repb{4}\,,+\frac{1}{ 2})_\mathbf{H} \subset \rep{70_H}\,.
\eeqn
The massless $\Gc_{421}$ fermions are listed in Eq.~\eqref{eq:421C_fermions}.
Correspondingly, we have the $\Gc_{421}$ $\beta$ coefficients of
\beqn  \label{eq:WWS_421_beta}
&&(b^{(1)}_{{\rm SU}(4)_{s}}\,,b^{(1)}_{{\rm SU}(2)_{W}}\,,b^{(1)}_{{\rm U}(1)_{X_2}})=(-\frac{11}{2}\,,+\frac{4}{3}\,,+\frac{151}{12})\,, \non
&&b^{(2)}_{\Gc_{421}}=
\begin{pmatrix}
131/2&21/2&17/4\\
105/2&184/3&11/4\\
255/4&33/4&125/8
\end{pmatrix}\,, \non
&& f_{ {\rm SU}(4)_s }  = -7 Y_\Tc^2 - 9 Y_\Bc^2 -  \frac{ 27}{2 } Y_\Dc^2 \,, \non
&& f_{ {\rm SU}(2)_W }  = - 6 Y_\Tc^2 - 8 Y_\Bc^2 - 12 Y_\Dc^2 \,, \non
&& f_{ {\rm U}(1)_{X_2 } } = - \frac{ 15}{ 2} Y_\Tc^2 -  \frac{ 25 }{ 2 } Y_\Bc^2 -  \frac{ 75 }{ 4} Y_\Dc^2 \,.
\eeqn
When the RGEs evolve down to $v_{421}$, the gauge couplings should match according to the following relations:
\beqn\label{eq:WWS_421_coupMatch}
&& \alpha_{3c }^{-1}(v_{421} ) = \alpha_{4s }^{-1} (v_{421} ) \,,~ \alpha_{Y}^{-1} (v_{421} ) = \frac{1}{6} \alpha_{4s }^{-1}(v_{421} ) +  \alpha_{X_2}^{-1} (v_{421} ) \,.
\eeqn

\para
Between the $v_{\rm EW}\leq \mu\leq v_{421}$, the massless Higgs fields are
\beqn \label{eq:SU8_WWS_321_massless_H}
(\rep{1}\,,\repb{2}\,,+\frac{1}{2})^{\prime\prime\prime}_\mathbf{H} \subset (\rep{4}\,,\repb{2}\,,+\frac{1}{ 4})_\mathbf{H} \subset (\rep{4}\,,\repb{3}\,,+\frac{5}{12})_\mathbf{H} \subset (\rep{4}\,,\repb{4}\,,+\frac{1}{ 2})_\mathbf{H} \subset \rep{70_H}\,.
\eeqn
The massless $\Gc_{\rm SM}$ fermions are listed in Eq.~\eqref{eq:321C_fermions}.
Correspondingly, we have the $\Gc_{\rm SM}$ $\beta$ coefficients of
\beqn  \label{eq:WWS_321_beta}
&&(b^{(1)}_{{\rm SU}(3)_{c}}\,,b^{(1)}_{{\rm SU}(2)_{W}}\,,b^{(1)}_{{\rm U}(1)_{Y}})=(-7\,,-\frac{19}{6}\,,+\frac{41}{6})\,,\non
&& b^{(2)}_{\Gc_{\rm{SM}}}=
\begin{pmatrix}
-26&9/2&11/6\\
12&35/6&3/2\\
44/3&9/2&199/18
\end{pmatrix}\,.
\eeqn

\begin{figure}[htb]
\centering
\includegraphics[height=6.5cm]{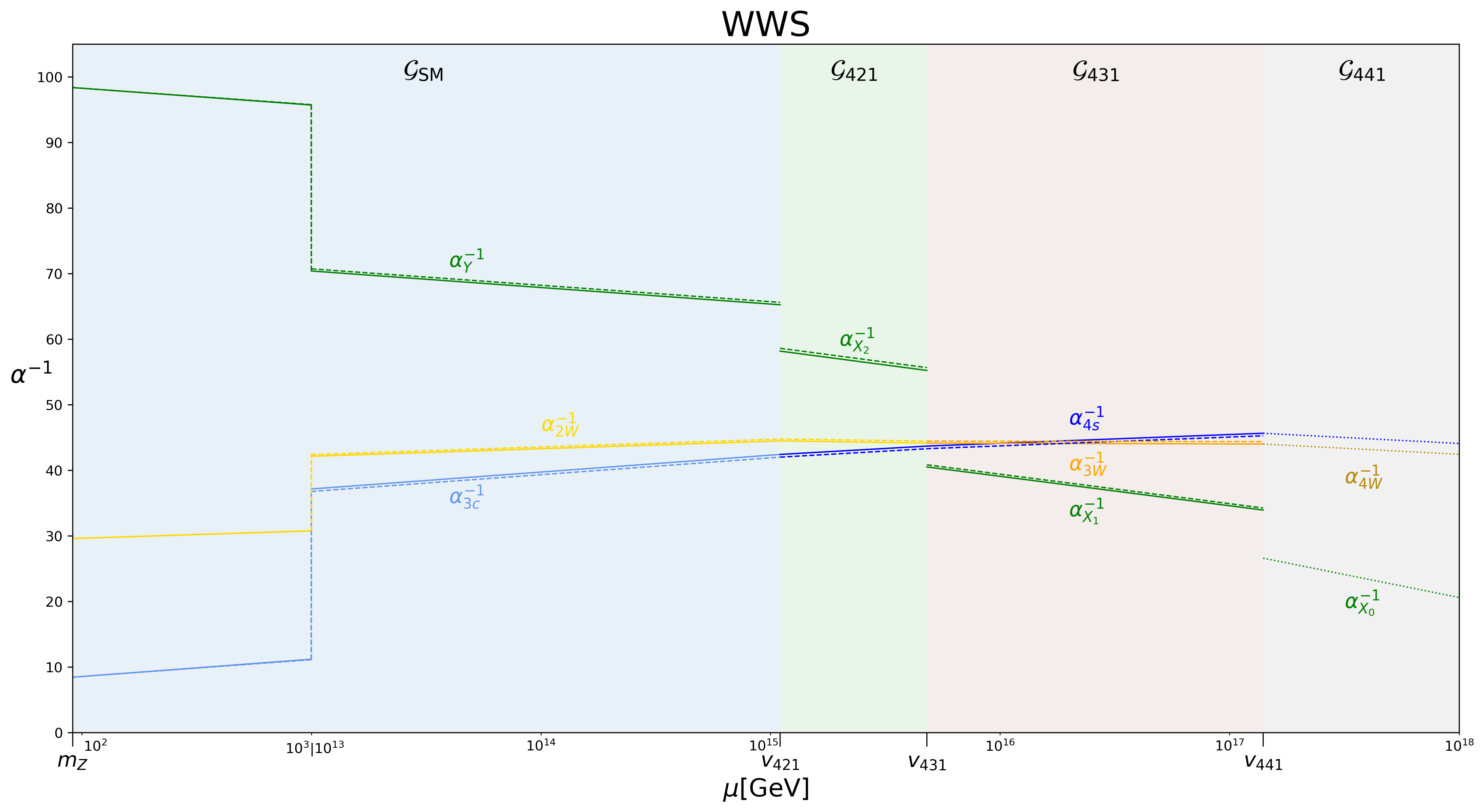}
\caption{The RGEs of the minimal ${\rm SU}(8)$ setup for the WWS symmetry breaking pattern.
The dashed lines and the solid lines represent the one- and two-loop RGEs, respectively.}
\label{fig:WWS_RGE_mini} 
\end{figure}

\para
With the one- and two-loop $\beta$ coefficients, we plot the RGEs of the minimal ${\rm SU}(8)$ setup according to the WWS symmetry breaking pattern in Fig.~\ref{fig:WWS_RGE_mini}.
Three intermediate symmetry breaking scales are given in Eq.~\eqref{eq:WWS_benchmark}, and three renormalizable Yukawa couplings follow from the benchmark point in Table~\ref{tab:SU8_WWS_benchmark}.
None of the evolutions of three gauge couplings can achieve the unification along the WWS sequence.

\section{Summary and outlook}
\label{section:conclusion}

\para
The WSW and the WWS symmetry breaking patterns following the maximally broken ${\rm SU}(8)$ theory have been analyzed in both the fermion spectra and the gauge coupling evolutions.
Throughout our analyses of both patterns, we compare our results from what we obtained previously in Refs.~\cite{Chen:2023qxi,Chen:2024cht,Chen:2024deo}.
\begin{itemize}

\item[(i)] Given the ${\rm SU}(8)$ chiral fermions in Eq.~\eqref{eq:SU8_3gen_fermions}, one finds six vectorlike $(\rep{5_F}\,, \repb{5_F})$ pairs and one $(\rep{10_F}\,, \repb{10_F})$ pair in addition to three-generational SM fermions through Georgi's decomposition rule~\cite{Georgi:1979md}.
Both symmetry breaking patterns are found to give rise to massive vectorlike fermions in the spectrum consistent to what we have found along the SWW pattern~\cite{Chen:2023qxi,Chen:2024cht}.

\item[(ii)] With the central motivation to address the SM quark/lepton masses through the nonuniversal Yukawa couplings to one unique SM Higgs boson in the spectrum, the SM fermion flavor identifications are elaborated along both the WSW and the SWW pattern.
In particular, we found that the flavor identifications between the first- and second-generational SM fermions are different from those in the SWW pattern, as one can compare the SM flavors in Tables~\ref{tab:SU8_56ferm_B} and \ref{tab:SU8_56ferm_C} versus the results in Table~\ref{tab:SU8_56ferm_A}.
The flavor identifications are determined by requiring the similar up-type quark mass matrices in all three symmetry breaking patterns, as one can find in Eqs.~\eqref{eqs:WSW_Uquark_masses}, \eqref{eqs:WWS_Uquark_masses}, and \eqref{eq:SWW_Uquark_masses}.
These flavor identifications, together with the down-type quarks, lead to both reasonable SM quark mass hierarchies as well as the observed CKM mixing pattern.
There is only one unique SM Higgs boson predicted in the current minimal model setup, regardless of the specific symmetry breaking pattern.
Hence we do not expect the discovery of any other beyond the SM particle(s) within few ${\rm TeV}$ scale at the LHC.

\item[(iii)] There are also observed differences in the SM quark/lepton mass spectra between the (SWW, WSW) and the WWS patterns.
The up-type quark mass matrix along the WWS pattern contains the dimensionless parameters of $(\zeta_1\,, \zeta_3\,, \dot \zeta_3^\prime)$ in Eq.~\eqref{eqs:WWS_Uquark_masses}, rather than dimensionless parameters of $(\zeta_1\,, \zeta_2\,, \dot \zeta_2)$.
For the down-type quark and charged lepton sectors along the WWS pattern, the mass matrices contain the dimensionless parameters of $(\dot \zeta_2\,, \dot \zeta_3)$.
Correspondingly, the Cabibbo angle is then interpreted as the ratio between two different symmetry breaking scales such that $\lambda=|V_{us}|=\zeta_{23}$.
In the SWW and the WSW patterns, the Cabibbo angle is interpreted as the relative ratio between two $\Gc_{331}$ breaking VEVs in Eq.~\eqref{eq:WSW_331VEVratio}.
These differences originate from the Higgs VEVs at the third symmetry breaking stage in the rank-three chiral IRAFFS sector since (i) along the SWW and the WSW patterns, the Higgs VEVs at the third symmetry breaking stage come from two different Higgs components of $( \rep{1} \,, \repb{3} \,, -\frac{1}{3} )_{\mathbf{H}\,, \dot\omega }^\prime \subset ( \rep{1} \,, \rep{6} \,, -\frac{1}{2} )_{\mathbf{H}\,, \dot\omega }$ and $( \rep{1} \,, \repb{3} \,, -\frac{1}{3} )_{\mathbf{H}\,, \dot\omega } \subset ( \repb{4} \,, \repb{4} \,, 0 )_{\mathbf{H}\,, \dot\omega }$ in Eq.~\eqref{eq:SU8B_Higgs_Br02}, and (ii) along the WWS pattern, the Higgs VEVs at the third symmetry breaking stage come from one single Higgs component of $( \repb{4} \,, \rep{1} \,, +\frac{1}{4} )_{\mathbf{H}\,, \dot\omega } \oplus ( \repb{4} \,, \rep{1} \,, +\frac{1}{4} )_{\mathbf{H}\,, \dot\omega }^\prime \subset ( \repb{4} \,, \repb{4} \,, 0 )_{\mathbf{H}\,, \dot\omega }$ in Eq.~\eqref{eq:SU8C_Higgs_Br02}. 
Another consequence of the above differences can be found in the suggested benchmark points in Tables~\ref{tab:SU8_WSW_benchmark} and \ref{tab:SU8_WWS_benchmark}.

\item[(iv)] The gauge coupling evolutions in both patterns are estimated according to the two-loop RGEs based on the suggested intermediate symmetry breaking scales in Eqs.~\eqref{eq:WSW_benchmark} and \eqref{eq:WWS_benchmark}.
Their behaviors are largely matching with what we have found in the SWW sequence~\cite{Chen:2024deo}, where large discrepancies of $| \alpha_{X_0}^{-1} - ( \alpha_{4S}^{-1} + \alpha_{ 4W}^{-1} )/2 | \sim 20$ above $\mu\sim 10^{17}\,{\rm GeV}$ can be observed in both Figs.~\ref{fig:WSW_RGE_mini} and \ref{fig:WWS_RGE_mini}.
Within the field theory context, such large discrepancy could not be reconciled with the one-loop threshold effects~\cite{Hall:1980kf,Weinberg:1980wa,Langacker:1992rq}.
Altogether, the gauge coupling unification should be achieved in the context of the affine Lie algebra such that
\beqn\label{eq:KM_unification}
&&k_s \alpha_{ 4S } ( v_U ) = k_W \alpha_{ 4 W} ( v_U ) = k_0 \alpha_{X_0} ( v_U ) = \alpha_U ( v_U )\,,
\eeqn
where $k_i$ represent the affine levels of each gauge coupling.
To achieve the string unification, it is necessary to consider the constraints of the unitarity and the conformal invariance to the affine $\widehat{ {\sG \uG}}(8)$ Lie algebra.
In a sequential work~\cite{Chen:2024wcj}, we have proved that the unification as in  Eq.~\eqref{eq:KM_unification} can be achieved with $k_0=\frac{1 }{4}$, as well as  the supersymmetric extension.
The unification scale and gauge coupling therein were suggested to be $v_U\approx 8.0\times 10^{17} \, \rm{GeV}$ and $\alpha_U^{-1} (v_U)\approx 30.9$.
By using the naive estimation~\cite{Langacker:1980js,Nath:2006ut} of 
\beqn
\tau [ p\to \pi^0 + e^+ ]&\approx&\frac{ v_U^4 }{ m_p^5} \,, \quad m_p = 938\,{\rm MeV} \,,
\eeqn
purely from the gauge sector contribution, we find a much enhanced lifetime of $\tau[ p\to \pi^0 + e^+] \sim 5.6 \times 10^{71}\,{\rm yrs}$.
Notice that the nontrivially flavor embedding structure could potentially further enhance this conventional proton decay lifetime.
This means several ongoing experimental probes of the proton decay modes are unlikely to discover such a signal, with the typical exclusion limit of $\sim {\cal O}(10^{34})\,{\rm yrs}$ from the Super-Kamiokande~\cite{Super-Kamiokande:2020wjk} or DUNE Collaboration~\cite{DUNE:2016hlj}.

\end{itemize}

\section*{ACKNOWLEDGEMENTS}
%
%
\para
We would like to thank Xiangpan Ji, Tianjun Li, Kaiwen Sun, Yuan Sun, Yinan Wang, Zhong-Zhi Xianyu, and Wenbin Yan for very enlightening discussions and communications. 
N.C. would like to thank Nanjing University, Central China Normal University, Sun Yat-Sen University, and South China Normal University for hospitality when preparing this work.
N.C. is partially supported by the National Natural Science Foundation of China (under Grants No. 12035008 and No. 12275140) and Nankai University.

\appendix

\section{Main results in the ${\rm SU}(8)$ SWW symmetry breaking pattern}
\label{section:SWW_results}

\begin{table}[htp] {\small
\begin{center}
\begin{tabular}{c|c|c|c|c}
\hline \hline
   $\gSU(8)$   &  $\Gc_{441}$  & $\Gc_{341}$  & $\Gc_{331}$  &  $\Gc_{\rm SM}$  \\
\hline \hline
 $\repb{ 8_F}^\Omega $   & $( \repb{4} \,, \rep{1}\,,  +\frac{1}{4} )_{ \mathbf{F} }^\Omega$  & $(\repb{3} \,, \rep{1} \,, +\frac{1}{3} )_{ \mathbf{F} }^\Omega $  & $(\repb{3} \,, \rep{1} \,, +\frac{1}{3} )_{ \mathbf{F} }^\Omega $  &  $( \repb{3} \,, \rep{ 1}  \,, +\frac{1}{3} )_{ \mathbf{F} }^{\Omega }~:~ { \Dc_R^\Omega }^c$  \\[1mm]
 &  &  $( \rep{1} \,, \rep{1} \,, 0)_{ \mathbf{F} }^{\Omega }$  &  $( \rep{1} \,, \rep{1} \,, 0)_{ \mathbf{F} }^{\Omega }$ &  $( \rep{1} \,, \rep{1} \,, 0)_{ \mathbf{F} }^{\Omega } ~:~ \check \Nc_L^{\Omega }$  \\[1.5mm]
 & $(\rep{1}\,, \repb{4}  \,,  -\frac{1}{4})_{ \mathbf{F} }^\Omega $  &  $(\rep{1}\,, \repb{4}  \,,  -\frac{1}{4})_{ \mathbf{F} }^\Omega$  &  $( \rep{1} \,, \repb{3} \,,  -\frac{1}{3})_{ \mathbf{F} }^{\Omega }$  &  $( \rep{1} \,, \repb{2} \,,  -\frac{1}{2})_{ \mathbf{F} }^{\Omega } ~:~\Lc_L^\Omega =( \Ec_L^\Omega \,, - \Nc_L^\Omega )^T$   \\[1mm]
 &   &   &   &  $( \rep{1} \,, \rep{1} \,,  0)_{ \mathbf{F} }^{\Omega^\prime} ~:~ \check \Nc_L^{\Omega^\prime }$  \\[1mm]
  &   &  &   $( \rep{1} \,, \rep{1} \,, 0)_{ \mathbf{F} }^{\Omega^{\prime\prime} }$ &   $( \rep{1} \,, \rep{1} \,, 0)_{ \mathbf{F} }^{\Omega^{\prime\prime} } ~:~ \check \Nc_L^{\Omega^{\prime \prime} }$   \\[1mm]   
\hline\hline
\end{tabular}
\caption{
The $\gSU(8)$ fermion representation of $\repb{8_F}^\Omega$ under the $\Gc_{441}\,,\Gc_{341}\,, \Gc_{331}\,, \Gc_{\rm SM}$ subgroups for the three-generational ${\rm SU}(8)$ theory, with $\Omega\equiv (\omega \,, \dot \omega )$.
Here, we denote $\underline{ {\Dc_R^\Omega}^c={d_R^\Omega}^c}$ for the SM right-handed down-type quarks, and ${\Dc_R^\Omega}^c={\DG_R^\Omega}^c$ for the right-handed down-type heavy partner quarks.
Similarly, we denote $\underline{ \Lc_L^\Omega = ( \ell_L^\Omega \,, - \nu_L^\Omega)^T}$ for the left-handed SM lepton doublets, and $\Lc_L^\Omega =( \eG_L^\Omega \,, - \nG_L^\Omega )^T$ for the left-handed heavy lepton doublets.
All left-handed neutrinos of $\check \Nc_L$ are sterile neutrinos, which are $\Gc_{\rm SM}$ singlets and do not couple to the EW gauge bosons.
}
\label{tab:SU8_8barferm_A}
\end{center}
}
\end{table}%

\begin{table}[htp] {\small
\begin{center}
\begin{tabular}{c|c|c|c|c}
\hline \hline
   $\gSU(8)$   &  $\Gc_{441}$  & $\Gc_{341}$  & $\Gc_{331}$  &  $\Gc_{\rm SM}$  \\
\hline \hline
 $\rep{28_F}$   & $( \rep{6}\,, \rep{ 1} \,, - \frac{1}{2})_{ \mathbf{F}}$ &  $ ( \rep{3}\,, \rep{ 1} \,, - \frac{1}{3})_{ \mathbf{F}}$   & $( \rep{3}\,, \rep{ 1} \,, - \frac{1}{3})_{ \mathbf{F}}$  & $( \rep{3}\,, \rep{ 1} \,, - \frac{1}{3})_{ \mathbf{F}} ~:~\DG_L$  \\[1mm]
                        &   & $( \repb{3}\,, \rep{ 1} \,, - \frac{2}{3})_{ \mathbf{F}}$  & $( \repb{3}\,, \rep{ 1} \,, - \frac{2}{3})_{ \mathbf{F}}$  & $\underline{( \repb{3}\,, \rep{ 1} \,, - \frac{2}{3})_{ \mathbf{F}}~:~ {t_R }^c }$   \\[1.5mm]
                        & $( \rep{1}\,, \rep{ 6} \,, +\frac{1}{2})_{ \mathbf{F}}$ & $( \rep{1}\,, \rep{ 6} \,, +\frac{1}{2})_{ \mathbf{F}}$   &  $( \rep{1}\,, \rep{ 3} \,, +\frac{1}{3})_{ \mathbf{F}}$ & $( \rep{1}\,, \rep{2} \,, +\frac{1}{2})_{ \mathbf{F}} ~:~( {\eG_R }^c \,, { \nG_R }^c)^T$  \\[1mm]
                       &   &   &   & $( \rep{1}\,, \rep{1} \,, 0 )_{ \mathbf{F}} ~:~ \check \nG_R^c $ \\[1mm]
                       &   &   & $( \rep{1}\,, \repb{ 3} \,, +\frac{2}{3})_{ \mathbf{F}}$  & $( \rep{1}\,, \repb{2} \,, +\frac{1}{2})_{ \mathbf{F}}^\prime ~:~( { \nG_R^{\prime} }^c\,, - {\eG_R^{\prime} }^c  )^T$   \\[1mm]
                       &   &   &   & $\underline{ ( \rep{1}\,, \rep{1} \,, +1 )_{ \mathbf{F}} ~:~ {\tau_R}^c}$ \\[1.5mm]
                        & $( \rep{4}\,, \rep{4} \,,  0)_{ \mathbf{F}}$ &  $( \rep{3}\,, \rep{4} \,,  -\frac{1}{12})_{ \mathbf{F}}$   & $( \rep{3}\,, \rep{3} \,,  0)_{ \mathbf{F}}$  & $\underline{ ( \rep{3}\,, \rep{2} \,,  +\frac{1}{6})_{ \mathbf{F}}~:~ (t_L\,, b_L)^T}$  \\[1mm]
                        &   &   &   & $( \rep{3}\,, \rep{1} \,,  -\frac{1}{3})_{ \mathbf{F}}^{\prime} ~:~\DG_L^\prime$  \\[1mm]
                        &   &   & $( \rep{3}\,, \rep{1} \,,  -\frac{1}{3})_{ \mathbf{F}}^{\prime\prime}$  & $( \rep{3}\,, \rep{1} \,,  -\frac{1}{3})_{ \mathbf{F}}^{\prime\prime} ~:~\DG_L^{\prime \prime}$ \\[1mm]
                        &   & $ ( \rep{1}\,, \rep{4} \,,  +\frac{1}{4} )_{ \mathbf{F}}$  & $( \rep{1}\,, \rep{3} \,,  +\frac{1}{3} )_{ \mathbf{F}}^{\prime\prime}$  & $( \rep{1}\,, \rep{2} \,,  +\frac{1}{2} )_{ \mathbf{F}}^{\prime\prime} ~:~( {\eG_R^{\prime\prime} }^c \,, { \nG_R^{\prime\prime}}^c )^T$  \\[1mm]
                        &   &   &   & $( \rep{1}\,, \rep{1}\,, 0)_{ \mathbf{F}}^{\prime} ~:~ \check \nG_R^{\prime\,c}$ \\[1mm]  
                        &   &   & $( \rep{1}\,, \rep{1}\,, 0)_{ \mathbf{F}}^{\prime\prime}$ & $( \rep{1}\,, \rep{1}\,, 0)_{ \mathbf{F}}^{\prime\prime} ~:~\check \nG_R^{\prime \prime \,c}$ \\[1mm]  
\hline\hline
\end{tabular}
\caption{
The $\gSU(8)$ fermion representation of $\rep{28_F}$ under the $\Gc_{441}\,,\Gc_{341}\,, \Gc_{331}\,, \Gc_{\rm SM}$ subgroups for the three-generational ${\rm SU}(8)$ theory.
All SM fermions are marked with underlines.}
\label{tab:SU8_28ferm_A}
\end{center}
}
\end{table}%

\begin{table}[htp] {\small
\begin{center}
\begin{tabular}{c|c|c|c|c}
\hline \hline
   $\gSU(8)$   &  $\Gc_{441}$  & $\Gc_{341}$  & $\Gc_{331}$  &  $\Gc_{\rm SM}$  \\
\hline \hline
     $\rep{56_F}$   & $( \rep{ 1}\,, \repb{4} \,, +\frac{3}{4})_{ \mathbf{F}}$  &  $( \rep{ 1}\,, \repb{4} \,, +\frac{3}{4})_{ \mathbf{F}}$ & $( \rep{ 1}\,, \repb{3} \,, +\frac{2}{3})_{ \mathbf{F}}^\prime$   &  $( \rep{ 1}\,, \repb{2} \,, +\frac{1}{2})_{ \mathbf{F}}^{\prime\prime\prime} ~:~( {\nG_R^{\prime\prime\prime }}^c \,, -{\eG_R^{\prime\prime\prime } }^c )^T$  \\[1mm]
                            &   &   &   & $\underline{( \rep{ 1}\,, \rep{1} \,, +1)_{ \mathbf{F}}^{\prime} ~:~ {\mu_R}^c}$ \\[1mm]
                            &   &   & $( \rep{ 1}\,, \rep{1} \,, +1)_{ \mathbf{F}}^{\prime\prime}$  & $( \rep{ 1}\,, \rep{1} \,, +1)_{ \mathbf{F}}^{\prime \prime} ~:~{\EG_R}^c$   \\[1.5mm]
                       & $( \repb{ 4}\,, \rep{1} \,, -\frac{3}{4})_{ \mathbf{F}}$  &  $( \repb{3}\,, \rep{1} \,, -\frac{2}{3})_{ \mathbf{F}}^{\prime}$ & $( \repb{3}\,, \rep{1} \,, -\frac{2}{3})_{ \mathbf{F}}^\prime$  & $\underline{ ( \repb{3}\,, \rep{1} \,, -\frac{2}{3})_{ \mathbf{F}}^{\prime} ~:~{u_R}^c }$ \\[1mm]
                       &   &  $( \rep{1}\,, \rep{1} \,, -1)_{ \mathbf{F}}$ & $( \rep{1}\,, \rep{1} \,, -1)_{ \mathbf{F}}$  &  $( \rep{1}\,, \rep{1} \,, -1)_{ \mathbf{F}} ~:~\EG_L$  \\[1.5mm]
                       & $( \rep{ 4}\,, \rep{6} \,, +\frac{1}{4})_{ \mathbf{F}}$  &  $( \rep{3}\,, \rep{6} \,, +\frac{1}{6})_{ \mathbf{F}}$ & $( \rep{3}\,, \rep{3} \,, 0 )_{ \mathbf{F}}^\prime$ & $\underline{ ( \rep{3}\,, \rep{2} \,, +\frac{1}{6} )_{ \mathbf{F}}^{\prime} ~:~ ( c_L \,, s_L)^T} $  \\[1mm]
                       &   &   &   & $( \rep{3}\,, \rep{1} \,, -\frac{1}{3})_{ \mathbf{F}}^{\prime\prime \prime } ~:~\DG_L^{\prime \prime \prime}$ \\[1mm]
                       &   &   & $( \rep{3}\,, \repb{3} \,, +\frac{1}{3})_{ \mathbf{F}}$ & $( \rep{3}\,, \repb{2} \,, +\frac{1}{6})_{ \mathbf{F}}^{\prime\prime} ~:~ (\dG_L \,, - \uG_L )^T$   \\[1mm]
                       &   &   &   & $( \rep{3}\,, \rep{1} \,, +\frac{2}{3})_{ \mathbf{F}} ~:~\UG_L$  \\[1mm]
                       &   & $( \rep{1}\,, \rep{6} \,, +\frac{1}{2})_{ \mathbf{F}}^\prime$ & $( \rep{1}\,, \rep{3} \,, +\frac{1}{3})_{ \mathbf{F}}^\prime $ & $( \rep{1}\,, \rep{2} \,, +\frac{1}{2})_{ \mathbf{F}}^{\prime\prime \prime \prime} ~:~ ( {\eG_R^{\prime\prime\prime\prime }}^c \,, {\nG_R^{\prime\prime\prime\prime } }^c )^T$ \\[1mm]
                       &   &   &   & $( \rep{1}\,, \rep{1} \,, 0 )_{ \mathbf{F}}^{\prime\prime \prime} ~:~ {\check \nG_R}^{\prime \prime\prime \,c}$ \\[1mm]
                       &   &   & $( \rep{1}\,, \repb{3} \,, +\frac{2}{3})_{ \mathbf{F}}^{\prime\prime}$  & $( \rep{1}\,, \repb{2} \,, +\frac{1}{2})_{ \mathbf{F}}^{\prime\prime \prime \prime \prime} ~:~( {\nG_R^{\prime\prime\prime\prime\prime }}^c \,, -{\eG_R^{\prime\prime\prime\prime\prime } }^c )^T$  \\[1mm]
                       &   &   &   & $\underline{ ( \rep{1}\,, \rep{1} \,, +1 )_{ \mathbf{F}}^{\prime\prime \prime } ~:~ {e_R}^c }$ \\[1.5mm]
                       & $( \rep{ 6}\,, \rep{4} \,, -\frac{1}{4})_{ \mathbf{F}}$  & $( \rep{3}\,, \rep{4} \,, -\frac{1}{12})_{ \mathbf{F}}^\prime$ & $( \rep{3}\,, \rep{3} \,, 0)_{ \mathbf{F}}^{\prime\prime}$ & $\underline{ ( \rep{3}\,, \rep{2} \,, +\frac{1}{6})_{ \mathbf{F}}^{\prime\prime \prime } ~:~  (u_L \,, d_L)^T} $ \\[1mm]
                       &   &   &   &  $( \rep{3}\,, \rep{1} \,, -\frac{1}{3})_{ \mathbf{F}}^{\prime \prime \prime \prime} ~:~\DG_L^{\prime \prime \prime\prime}$ \\[1mm]
                       &   &   &  $( \rep{3}\,, \rep{1} \,, -\frac{1}{3})_{ \mathbf{F}}^{\prime \prime \prime\prime \prime}$ & $( \rep{3}\,, \rep{1} \,, -\frac{1}{3})_{ \mathbf{F}}^{\prime \prime \prime \prime \prime} ~:~ \DG_L^{\prime \prime \prime\prime \prime}$ \\[1mm]
                       &   & $( \repb{3}\,, \rep{4} \,, -\frac{5}{12})_{ \mathbf{F}}$ & $( \repb{3}\,, \rep{3} \,, -\frac{1}{3})_{ \mathbf{F}}$ & $( \repb{3}\,, \rep{2} \,, -\frac{1}{6})_{ \mathbf{F}} ~:~ ( {\dG_R}^c \,,{\uG_R}^c )^T$  \\[1mm]
                       &   &   &   & $( \repb{3}\,, \rep{1} \,, -\frac{2}{3})_{ \mathbf{F}}^{\prime \prime} ~:~{\UG_R}^c$  \\[1mm]
                       &   &   & $( \repb{3}\,, \rep{1} \,, -\frac{2}{3})_{ \mathbf{F}}^{\prime \prime \prime}$ & $\underline{ ( \repb{3}\,, \rep{1} \,, -\frac{2}{3})_{ \mathbf{F}}^{\prime \prime \prime} ~:~{c_R}^c }$  \\[1mm]
\hline\hline
\end{tabular}
\caption{
The $\gSU(8)$ fermion representation of $\rep{56_F}$ under the $\Gc_{441}\,,\Gc_{341}\,, \Gc_{331}\,, \Gc_{\rm SM}$ subgroups for the three-generational ${\rm SU}(8)$ theory.
All SM fermions are marked with underlines.
}
\label{tab:SU8_56ferm_A}
\end{center}
}
\end{table}%

\para
By following the symmetry breaking pattern in Eq.~\eqref{eq:Pattern-A} and the ${\rm U}(1)$ charges defined in Eqs.~\eqref{eqs:SWW_U1charges_fund}, we tabulate the ${\rm SU}(8)$ fermion representations in Tables~\ref{tab:SU8_8barferm_A}-\ref{tab:SU8_56ferm_A}.
For the right-handed down-type quarks of ${\Dc_R^\Omega}^c$, they are named as follows:
\beqn\label{eq:DR_names_A}
&& {\Dc_R^{ \dot 1} }^c \equiv {d_R}^c \,, ~ {\Dc_R^{  \dot 2} }^c \equiv {s_R}^c \,,~ {\Dc_R^{\dot {\rm VII} } }^c \equiv {\DG_R^{\prime\prime\prime\prime \prime }}^c \,, ~  {\Dc_R^{\dot {\rm VIII} } }^c \equiv {\DG_R^{\prime\prime \prime }}^c  \,,~ {\Dc_R^{  \dot {\rm IX} } }^c \equiv {\DG_R^{\prime\prime\prime \prime }}^c \,, \non
&&{\Dc_R^{3} }^c \equiv {b_R}^c \,,~  {\Dc_R^{\rm IV } }^c \equiv {\DG_R}^c \,, ~  {\Dc_R^{\rm V } }^c \equiv {\DG_R^{\prime\prime }}^c  \,,~ {\Dc_R^{\rm VI } }^c \equiv {\DG_R^{\prime }}^c  \,.
\eeqn
For the left-handed ${\rm SU}(2)_W$ lepton doublets of $(\Ec_L^\Omega \,, - \Nc_L^\Omega )$, they are named as follows:
\beqn\label{eq:ELNL_names_A}
&&   ( \Ec_L^{ \dot 1} \,,  - \Nc_L^{\dot 1})  \equiv (e_L\,, - \nu_{e\,L} ) \,, ~( \Ec_L^{  \dot 2} \,,   - \Nc_L^{\dot 2})  \equiv( \mu_L \,, - \nu_{\mu\,L} )  \,, \non
&&  ( \Ec_L^{ \dot {\rm VII} } \,,  - \Nc_L^{ \dot {\rm VII} })  \equiv ( \eG_L^{ \prime\prime \prime \prime} \,, - \nG_L^{\prime\prime \prime \prime } )  \,, ~ ( \Ec_L^{ \dot {\rm VIII} } \,, - \Nc_L^{ \dot {\rm VIII} }  )  \equiv ( \eG_L^{ \prime\prime  \prime} \,, - \nG_L^{\prime\prime \prime} ) \,,~  ( \Ec_L^{\dot {\rm IX} } \,,  - \Nc_L^{ \dot {\rm IX} } ) \equiv ( \eG_L^{ \prime\prime \prime \prime \prime } \,,  - \nG_L^{\prime\prime \prime\prime \prime} )  \,,\non
&&  ( \Ec_L^{ 3} \,, - \Nc_L^{3}) \equiv ( \tau_L \,, - \nu_{\tau\,L})\,,~  ( \Ec_L^{\rm IV } \,, - \Nc_L^{\rm IV }) \equiv ( \eG_L^{\prime\prime} \,, - \nG_L^{\prime\prime} ) \,, \non
&& ( \Ec_L^{\rm V } \,, -  \Nc_L^{\rm V }) \equiv ( \eG_L \,, - \nG_L )  \,,~ ( \Ec_L^{\rm VI } \,, - \Nc_L^{\rm VI } ) \equiv  ( \eG_L^\prime\,, - \nG_L^\prime ) \,.
\eeqn
According to the flavor identifications in Tables~\ref{tab:SU8_8barferm_A}-\ref{tab:SU8_56ferm_A}, we find the following up-type quark mass matrix:
\beqn\label{eq:SWW_Uquark_masses}
\Mc_u   &=&   \frac{1}{\sqrt{2} }  \left( \ba{ccc}  
0 &  0  & c_5 \zeta_1 /\sqrt{2} \\
 c_4  \dot \zeta_2 /\sqrt{2}  & 0  & c_5 \zeta_2 / \sqrt{2}   \\
   c_5 \zeta_1 /\sqrt{2}  &  c_5 \zeta_2/\sqrt{2}  &   Y_\Tc \\  \ea  \right) v_{\rm EW} \,,
\eeqn
from the renormalizable Yukawa coupling term of $Y_\Tc \rep{28_F} \rep{28_F} \rep{70_H}+{\rm H.c.}$ and the $d=5$ direct Yukawa coupling terms in Eqs.~\eqref{eq:d5_ffHH_41} and \eqref{eq:d5_ffHH_51}.
The down-type quark and the charge lepton mass matrices are found to be
\beqn\label{eq:SWW_Dquark_massdd}
&&  \Big( \Mc_d \Big)_{3\times 3}  = \Big( \Mc_\ell^T \Big)_{3\times 3}  \non
&\approx&  \frac{1}{4} \left( \ba{ccc}
 ( 2 c_3 + Y_\Dc d_{\mathscr B}  )\dot \zeta_3^\prime  &  ( 2 c_3  +  Y_\Dc d_{\mathscr B}  \Delta_{ \dot 2} ) \dot \zeta_3^\prime  & 0   \\
  ( 2 c_3  +  Y_\Dc d_{\mathscr B}  \Delta_{ \dot 1}^\prime )  \dot \zeta_3  &  (  2 c_3  +  Y_\Dc d_{\mathscr B}   \zeta_{23 }^{-2 } ) \dot \zeta_3   &  0   \\
  0 & 0  &  Y_\Bc d_{\mathscr A}   \zeta_{23}^{-1} \zeta_1   \\  \ea  \right) v_{\rm EW}    \,,
\eeqn
from the $d=5$ indirect Yukawa coupling terms in Eqs.~\eqref{eqs:d5_Hmixings}.

\section{The intermediate stages along the WSW symmetry breaking pattern}\label{section:WSW_process}

\subsection{The first stage}
\label{section:WSW_stage1}

\begin{table}[htp]
	\begin{center}
		\begin{tabular}{c|cccc}
			\hline\hline
			$\repb{8_F}^\Omega$ & $( \repb{4}\,, \rep{1}\,, +\frac{1}{4} )_{\rep{F}}^\Omega$  &  $( \rep{1}\,, \repb{4}\,,  -\frac{1}{4} )_{\rep{F}}^\Omega$  &   &     \\[1mm]
			\hline
			$\Tc^\prime$  &  $ -2t$  &  $-4t$  &   &     \\[1mm]
			\hline
			$\rep{28_F}$ &  $( \rep{6}\,, \rep{1}\,,  -\frac{1}{2} )_{\rep{F}}$ &  $( \rep{1}\,, \rep{ 6}\,,  + \frac{1}{2} )_{\rep{F}}$ &  $( \rep{4 }\,, \rep{4}\,,  0 )_{\rep{F}}$   &    \\[1mm]
			\hline
			$\Tc^\prime$  &  $0$  & $+4t$ & $+2t$   &    \\[1mm]
			\hline
			$\rep{56_F}$  &  $( \rep{1}\,, \repb{4}\,,  +\frac{3}{4} )_{\rep{F}}$ &  $( \repb{4}\,,  \rep{1}\,,  -\frac{3}{4} )_{\rep{F}}$ & $( \rep{4}\,,  \rep{6}\,,  +\frac{1}{4} )_{\rep{F}}$  & $( \rep{6}\,,  \rep{4}\,,  -\frac{1}{4} )_{\rep{F}}$   \\[1mm]
			\hline
			$\Tc^\prime$  & $+4t$  &  $-2t$  & $+2t$  & $0$  \\[1mm]
			\hline\hline
			$\repb{8_H}_{\,, \omega }$  & $( \repb{4}\,, \rep{1}\,, +\frac{1}{4} )_{\rep{H}\,,\omega}$  &  $( \rep{1}\,, \repb{4}\,,  -\frac{1}{4} )_{\rep{H}\,,\omega }$  &   &      \\[1mm]
			\hline
			$\Tc^\prime$  &  $+2t$  &  $0$  &   &     \\[1mm]
			\hline
			$\repb{28_H}_{\,,\dot \omega}$ & $( \rep{1}\,, \rep{6}\,,  -\frac{1}{2} )_{\rep{H}\,, \dot \omega}$  &  $( \repb{4}\,, \repb{4}\,,  0 )_{\rep{H}\,, \dot \omega}$  &   &    \\[1mm]
			\hline
			$\Tc^\prime$  &  $0$  &  $+2t$  &      &   \\[1mm]
			\hline
			$\rep{70_H}$  & $( \rep{4}\,, \repb{4}\,,  +\frac{1}{2} )_{\rep{H}}$   &  $( \repb{4}\,, \rep{4}\,,  - \frac{1}{2} )_{\rep{H}}$  &   &       \\[1mm]
			\hline
			$\Tc^\prime$  & $-2t$  & $-6t$ &   &      \\[1mm]
			\hline\hline
		\end{tabular}
	\end{center}
	\caption{The $\widetilde{ {\rm U}}(1)_{T^\prime } $ charges for massless fermions and possible symmetry breaking Higgs components in the $\Gc_{441}$ theory.}
	\label{tab:BG441_Tcharges}
\end{table}

\para
The first symmetry breaking stage of $\Gc_{441} \to \Gc_{431}$ is achieved by $( \rep{1}\,, \repb{4}\,,  -\frac{1}{4} )_{\rep{H}\,,\rm{IV} }\subset \repb{8_H}_{\,, \rm{IV} }$ in the rank-two sector, according to their $\Gc_{431}$-invariant and $\widetilde {\rm U}(1)_{ T^{ \prime }}$-neutral components in Table~\ref{tab:BG441_Tcharges}.
Accordingly, the term of $ Y_\Bc \repb{8_F}^{\rm IV} \rep{28_F} \repb{8_H}_{\,,\rm IV} +{\rm H.c.}$ leads to the vectorial masses of  $(\DG^{\prime \prime} \,, \check \nG^{\prime \prime }  \,, \eG \,, \nG\,,  \check \nG)$.
After this stage, the remaining massless fermions expressed in terms of the $\Gc_{431}$ IRs are the following:
\begin{eqnarray}\label{eq:431B_fermions}
	&& ( \repb{4}\,, \rep{1}\,, +\frac{1}{4} )_{ \mathbf{F}}^\Omega \oplus \Big[ ( \rep{1}\,, \repb{3}\,, -\frac{1}{3} )_{ \mathbf{F}}^\Omega \oplus ( \rep{1}\,, \rep{1}\,, 0 )_{ \mathbf{F}}^{\Omega^{\prime\prime}} \Big]  \oplus   \subset \repb{8_F}^\Omega \,, \non
	&& \Omega = ( \omega\,, \dot \omega) \,,\quad \omega = (3\,, {\rm V}\,, {\rm VI} ) \,,\quad \dot \omega = ( \dot 1\,, \dot 2\,, \dot {\rm VII} \,,\dot {\rm VIII} \,, \dot {\rm IX}) \,,\non
	&& ( \rep{1}\,, \rep{1}\,, 0 )_{ \mathbf{F}}^{{\rm IV}^{\prime\prime} } \subset \repb{8_F}^{ {\rm IV}} \,,\non
	&&( \rep{6}\,, \rep{1}\,, -\frac{1}{2} )_{ \mathbf{F}}  \oplus \Big[ \cancel{ ( \rep{1}\,, \rep{3}\,, +\frac{1}{3} )_{ \mathbf{F}} } \oplus ( \rep{1}\,, \repb{3}\,, +\frac{2}{3} )_{ \mathbf{F}} \Big] \oplus  \Big[ ( \rep{4}\,, \rep{3}\,, +\frac{1}{12} )_{ \mathbf{F}} \oplus \cancel{ ( \rep{4}\,, \rep{1}\,, -\frac{1}{4} )_{ \mathbf{F}} } \Big] \subset \rep{28_F}\,, \non
	&& \Big[ ( \rep{1}\,, \repb{3}\,, +\frac{2}{3} )_{ \mathbf{F}}^\prime \oplus ( \rep{1}\,, \rep{1}\,, -1 )_{ \mathbf{F}}^{\prime\prime} \Big] \oplus ( \repb{4}\,, \rep{1}\,, -\frac{3}{4} )_{ \mathbf{F}} \oplus  \Big[ ( \rep{4}\,, \rep{3}\,, +\frac{1}{12} )_{ \mathbf{F}} \oplus ( \rep{4}\,, \repb{3}\,, +\frac{5}{12})_{ \mathbf{F}} \Big] \non
	&\oplus& \Big[ ( \rep{6}\,, \rep{3}\,, -\frac{1}{6} )_{ \mathbf{F}} \oplus ( \rep{6}\,, \rep{1}\,, -\frac{1}{2})_{ \mathbf{F}}^{\prime} \Big]   \subset \rep{56_F}\,.
\end{eqnarray}
Fermions that become massive at this stage are crossed out by slashes.
We find only one massive $\repb{8_F}^{\rm IV}$ is integrated out from the anomaly-free conditions of $[ {\rm SU}(4)_s]^2 \cdot {\rm U}(1)_{X_1}=0$, $\[ {\rm SU}(3)_W \]^2 \cdot {\rm U}(1)_{X_1}=0$, and $\[ {\rm U}(1)_{X_1} \]^3=0$, except for one left-handed sterile neutrino of $\check \Nc_L^{{\rm IV}^{ \prime \prime} } \equiv ( \rep{1}\,, \rep{1} \,, 0 )_{ \rep{F}}^{{\rm IV}^{ \prime \prime } } \subset \repb{8_F}^{\rm IV}$.

\subsection{The second stage}

\begin{table}[htp]
	\begin{center}
		\begin{tabular}{c|cccc}
			\hline\hline
			$\repb{8_F}^\Omega $ & $( \repb{4}\,, \rep{1}\,, +\frac{1}{4} )_{\rep{F}}^\Omega $ & $( \rep{1}\,, \repb{3}\,,  -\frac{1}{3} )_{\rep{F}}^\Omega $  &  $( \rep{1}\,, \rep{1}\,,  0 )_{\rep{F}}^{\Omega^{\prime\prime}}$   &   \\[1mm]
			\hline
			$\Tc^{\prime\prime}$  &  $ -4t$ & $ - \frac{4}{3} t$  &  $ - 4 t$   &    \\[1mm]
			\hline
			$\rep{28_F}$  &  $( \rep{6}\,, \rep{1}\,,  -\frac{1}{2 } )_{\rep{F}}$ &  $( \rep{1}\,, \repb{ 3}\,,  + \frac{2}{3} )_{\rep{F}}$ & $( \rep{4  }\,, \rep{3}\,, + \frac{1 }{12 } )_{\rep{F}}$  &    \\[1mm]
			\hline
			$\Tc^{\prime\prime}$   & $+4t$  &  $ - \frac{4}{3} t$  &  $+\frac{4 }{3 } t$  &    \\[1mm]
			\hline
			$\rep{56_F}$ & $( \rep{1}\,, \repb{3}\,,  + \frac{ 2}{3} )_{\rep{F}}^{\prime}$&$( \rep{1}\,, \rep{1}\,,  +1 )_{\rep{F}}^{\prime\prime}$ & $( \repb{4}\,, \rep{1}\,,  - \frac{ 3}{4} )_{\rep{F}}$  & $( \rep{4}\,,  \rep{3}\,,  + \frac{1 }{12} )_{\rep{F}}$    \\[1mm]
			\hline
			$\Tc^{\prime\prime}$  &  $- \frac{4 }{3} t$  &  $-4t$  &  $+4 t$  & $+ \frac{4 }{3} t$       \\[1mm]
			\hline
			&  $( \rep{4}\,,  \repb{3}\,,  + \frac{5}{12} )_{\rep{F}}$  &  $( \rep{6}\,,  \rep{3}\,,  -\frac{1}{6} )_{\rep{F}}$ & $( \rep{6}\,,  \rep{1}\,,  -\frac{1}{2} )_{\rep{F}}^{\prime}$ &    \\[1mm]
			\hline
			$\Tc^{\prime\prime}$ & $ -\frac{4}{3 } t$     & $ +\frac{4}{3 } t$ &   $ +4 t$& \\[1mm]
			\hline\hline
			$\repb{8_H}_{\,, \omega}$  &$( \repb{4}\,, \rep{1}\,,  +\frac{1}{4} )_{\rep{H}\,,\omega}$    & $( \rep{1}\,, \repb{3}\,,  -\frac{1}{3} )_{\rep{H}\,,\omega}$  &  &    \\[1mm]
			\hline
			$\Tc^{\prime\prime}$  &  $0$  & $+\frac{8}{3}t$  &   &     \\[1mm]
			\hline
			$\repb{28_H}_{\,, \dot \omega }$  & $( \rep{1}\,, \repb{3}\,,  -\frac{1}{3} )_{\rep{H}\,,\dot \omega }^{\prime}$  &  $( \repb{4}\,, \repb{3}\,,  -\frac{1}{12} )_{\rep{H}\,,\dot \omega }$  & $( \repb{4}\,, \rep{1}\,, + \frac{1}{4} )_{\rep{H}\,,\dot \omega }$  &     \\[1mm]
			\hline
			$\Tc^{\prime\prime}$  &  $ +\frac{8}{3}t$  &  $+\frac{8}{3}t $  & $0$ &     \\[1mm]
			\hline
			$\rep{70_H}$ &  $( \rep{4 }\,, \repb{3}\,,  +\frac{5}{12} )_{\rep{H}}$  & $( \repb{4}\,, \rep{3}\,,  - \frac{5 }{12} )_{\rep{H}}$   &  &    \\[1mm]
			\hline
			$\Tc^{\prime\prime}$  & $-\frac{16}{3}t$  &  $ -\frac{8}{3} t$  &   &     \\[1mm]
			\hline\hline
		\end{tabular}
	\end{center}
	\caption{The $\widetilde{ {\rm U}}(1)_{T^{\prime \prime} }$ charges for massless fermions and possible symmetry breaking Higgs components in the $\Gc_{431}$ theory.
	}
	\label{tab:BG431_Tcharges}
\end{table}

\para
The second symmetry breaking stage of $\Gc_{431}\to \Gc_{331}$ can be achieved by $( \repb{4}\,, \rep{1}\,,  +\frac{1}{4} )_{\rep{H}\,,\rm  V} \subset \repb{8_H}_{\,, \rm V}$ in the rank-two sector and $( \repb{4}\,, \rep{1}\,,  +\frac{1}{4} )_{\rep{H}\,, \dot{\rm VII}\,,\dot{1} } \subset \repb{28_H}_{\,, \dot{\rm VII}\,,\dot{1}}$ in the rank-three sector, according to their $\Gc_{331}$-invariant and $\widetilde {\rm U}(1)_{ T^{ \prime \prime}}$-neutral components in Table~\ref{tab:BG431_Tcharges}.
Accordingly, the term of $ Y_\Bc \repb{8_F}^{\rm V} \rep{28_F} \repb{8_H}_{\,,{\rm V }} + {\rm H.c.}$ leads to the vectorial masses of $( \DG \,, \eG^{\prime\prime} \,, \nG^{\prime\prime} \,, \check \nG^{\prime} )$,  the term of    $Y_\Dc \repb{8_F}^{\dot{\rm VII}\,, \dot{1}} \rep{56_F} \repb{28_H}_{\,,\dot{\rm VII}\,, \dot{1}} +{\rm H.c.}$ leads to the vectorial masses of $(\DG^{\prime\prime\prime\prime \prime } \,, \eG^{\prime \prime  \prime\prime }  \,,   \nG^{\prime \prime  \prime\prime } \,, \check \nG^{\prime\prime\prime }  )$, and the term of  $\frac{c_4}{M_{\rm{pl}}}\rep{56_F}\rep{56_F} \langle  \rep{63_H} \rangle \repb{28_H}_{\,,\dot{\rm VII}\,,\dot{1}}^{\dag} + {\rm H.c.}$ leads to the vectorial masses of  $( \EG\,,\dG\,, \uG\,,\UG   )$, respectively. 
After integrating out the massive fermions, the remaining massless fermions expressed in terms of the $\Gc_{331}$ IRs are the following:
\begin{eqnarray}\label{eq:331B_fermions}
	&& \Big[ ( \repb{3}\,, \rep{1}\,, +\frac{1}{3} )_{ \mathbf{F}}^\Omega \oplus ( \rep{1}\,, \rep{1}\,, 0 )_{ \mathbf{F}}^\Omega \Big]  \oplus \Big[ ( \rep{1}\,, \repb{3}\,, -\frac{1}{3} )_{ \mathbf{F}}^\Omega \oplus ( \rep{1}\,, \rep{1}\,, 0 )_{ \mathbf{F}}^{\Omega^{\prime \prime }}  \Big]  \subset \repb{8_F}^\Omega \,, \non
	&& \Omega = ( \omega\,, \dot \omega ) \,, \quad \omega = (3\,, {\rm VI})\,,  \quad \dot \omega = (\dot 1\,, \dot 2\,, \dot {\rm VIII}\,, \dot {\rm IX} )\,,\non
	&& ( \rep{1}\,, \rep{1}\,, 0)_{ \mathbf{F}}^{{\rm IV}^{\prime\prime} } \subset \repb{8_F}^{{\rm IV} } \,, \quad ( \rep{1}\,, \rep{1}\,, 0)_{ \mathbf{F}}^{{\rm V} } \oplus ( \rep{1}\,, \rep{1}\,, 0)_{ \mathbf{F}}^{ {\rm V}^{ \prime \prime} } \subset \repb{8_F}^{ \rm V } \,,  \non
	&&  ( \rep{1}\,, \rep{1}\,, 0)_{ \mathbf{F}}^{\dot {\rm VII} } \oplus ( \rep{1}\,, \rep{1}\,, 0)_{ \mathbf{F}}^{\dot {\rm VII}^{ \prime \prime} } \subset \repb{8_F}^{\dot {\rm VII} } \,,  \non
	&&\Big[ \bcancel{ ( \rep{3}\,, \rep{1}\,, -\frac{1}{3} )_{ \mathbf{F}} } \oplus ( \repb{3}\,, \rep{1}\,, -\frac{2}{3} )_{ \mathbf{F}} \Big] \oplus \Big[ \cancel{ ( \rep{1}\,, \rep{3}\,, +\frac{1}{3} )_{ \mathbf{F}} } \oplus ( \rep{1}\,, \repb{3}\,, +\frac{2}{3} )_{ \mathbf{F}} \Big] \non
	&\oplus& \Big[ ( \rep{3}\,, \rep{3}\,, 0 )_{ \mathbf{F}} \oplus \bcancel{ ( \rep{1}\,, \rep{3}\,, +\frac{1}{3} )_{ \mathbf{F}}^{\prime\prime} } \Big]  \oplus \cancel{ \Big[ ( \rep{3}\,, \rep{1}\,, -\frac{1}{3} )_{ \mathbf{F}}^{\prime\prime}  \oplus  ( \rep{1}\,, \rep{1}\,, 0 )_{ \mathbf{F}}^{\prime\prime}  \Big] }  \subset \rep{28_F}\,, \non
	&&\Big[ ( \rep{1}\,, \repb{3}\,, +\frac{2}{3} )_{ \mathbf{F}}^\prime \oplus \bcancel{ ( \rep{1}\,, \rep{1}\,, +1 )_{ \mathbf{F}}^{\prime\prime} } \Big] \oplus \Big[  ( \repb{3}\,, \rep{1}\,, -\frac{2}{3} )_{ \mathbf{F}}^\prime \oplus  \bcancel{ ( \rep{1}\,, \rep{1}\,, -1 )_{ \mathbf{F}} }  \Big] \non
	&\oplus&  \Big[ ( \rep{3}\,, \rep{3}\,, 0 )_{ \mathbf{F}}^\prime \oplus \bcancel{  ( \rep{1}\,, \rep{3}\,, +\frac{1}{3} )_{ \mathbf{F}}^\prime }  \oplus \bcancel{( \rep{3}\,, \repb{3}\,, +\frac{1}{3} )_{ \mathbf{F}} }  \oplus ( \rep{1}\,, \repb{3}\,, +\frac{2}{3})_{ \mathbf{F}}^{\prime \prime} \Big] \non
	&\oplus& \Big[ ( \rep{3}\,, \rep{3}\,, 0)_{ \mathbf{F}}^{\prime\prime} \oplus \bcancel{ ( \repb{3}\,, \rep{3}\,, -\frac{1}{3})_{ \mathbf{F}} }  \oplus  \bcancel{ ( \rep{3}\,, \rep{1}\,, -\frac{1}{3})_{ \mathbf{F}}^{\prime\prime\prime\prime\prime}  } \oplus ( \repb{3}\,, \rep{1}\,, -\frac{2}{3})_{ \mathbf{F}}^{\prime\prime\prime} \Big]   \subset \rep{56_F}\,.
\end{eqnarray}
We use the slashes and the back slashes to cross out massive fermions at the first and the second stages, respectively.
From the anomaly-free conditions of $[ {\rm SU}(3)_c]^2 \cdot {\rm U}(1)_{X_2}=0$, $\[ {\rm SU}(3)_W \]^2 \cdot {\rm U}(1)_{X_2}=0$, and $\[ {\rm U}(1)_{X_2} \]^3=0$, we find that one  $\repb{8_F}^{\rm V}$ and one  $\repb{8_F}^{\dot {\rm VII}}$ are integrated out.

\subsection{The third stage}

\begin{table}[htp]
	\begin{center}
		\begin{tabular}{c|cccc}
			\hline\hline
			$\repb{8_F}^\Omega$ & $( \repb{3}\,, \rep{1}\,, +\frac{1}{3} )_{\rep{F}}^\Omega$ & $( \rep{1}\,, \rep{1}\,,  0 )_{\rep{F}}^\Omega$  &  $( \rep{1}\,, \repb{3}\,,  -\frac{1}{3} )_{\rep{F}}^\Omega$  &  $( \rep{1}\,, \rep{1}\,,  0 )_{\rep{F}}^{\Omega^{ \prime\prime} }$   \\[1mm]
			\hline
			$\Tc^{ \prime \prime \prime }$  &  $- \frac{4}{3} t$ & $- 4 t$  &  $- 4 t$  & $ -4 t$     \\[1mm]
			\hline
			$\rep{28_F}$ & $( \repb{3}\,, \rep{1}\,, - \frac{2}{3} )_{\rep{F}}$  &  $( \rep{1}\,, \repb{3}\,,  + \frac{2}{3} )_{\rep{F}}$  &  $( \rep{3}\,, \rep{3 }\,,  0 )_{\rep{F}}$   &   \\[1mm]
			\hline
			$\Tc^{ \prime \prime \prime }$  & $ - \frac{4}{3} t$  &  $+ 4 t$  &  $ +  \frac{4}{3} t$ &  \\[1mm]
			\hline
			$\rep{56_F}$  & $( \rep{1}\,,  \repb{3}\,, +\frac{2}{3} )_{\rep{F}}^\prime$ &  $( \repb{3 }\,, \rep{1}\,,  -\frac{2}{3 } )_{\rep{F}}^\prime$  & $( \rep{3}\,,  \rep{3}\,, 0 )_{\rep{F}}^\prime$   &     \\[1mm]
			\hline
			$\Tc^{ \prime \prime \prime }$   &  $ +4 t$  &  $  -\frac{4}{3 } t$  &  $+\frac{4}{3} t$  &   \\[1mm]
			\hline
			& $(  \rep{1}\,, \repb{3 }\,,  +\frac{2}{3 } )_{\rep{F}}^{\prime\prime} $ &  $( \rep{3}\,,  \rep{3}\,, 0 )_{\rep{F}}^{\prime \prime} $  & $( \repb{3 }\,,   \rep{1}\,,  -\frac{2}{3 } )_{\rep{F}}^{\prime\prime \prime}$&  \\[1mm]
			\hline
			$\Tc^{ \prime \prime \prime }$  &  $ + 4 t$  &  $+\frac{4}{3} t$  &  $ -\frac{4}{3} t$ & \\[1mm]
			\hline\hline
			$\repb{8_H}_{ \,, \omega }$   &  $( \rep{1}\,, \repb{3}\,,  - \frac{1}{3} )_{\rep{H}\,,\omega }$  &   &   &   \\[1mm]
			\hline
			$\Tc^{ \prime \prime \prime }$  &  $0$  &   &  &  \\[1mm]
			\hline
			$\repb{28_H}_{ \,, \dot \omega }$  &  $( \rep{1}\,,  \repb{3}\,, -\frac{1}{ 3} )_{\rep{H}\,, \dot \omega }^\prime$  & $( \rep{1}\,,  \rep{3}\,, -\frac{2 }{ 3} )_{\rep{H}\,, \dot \omega }$   &  $( \rep{1}\,,  \repb{3}\,, -\frac{1 }{ 3} )_{\rep{H}\,, \dot \omega }$  &    \\[1mm]
			\hline
			$\Tc^{ \prime \prime \prime }$  & $0$  & $0$  & $0$ &     \\[1mm]
			\hline
			$\rep{70_H}$  &  $( \rep{1}\,, \repb{3}\,,  + \frac{2}{3} )_{\rep{H}}^{ \prime\prime \prime}$  &  $( \rep{1}\,, \rep{3}\,,  - \frac{2}{3} )_{\rep{H}}^{ \prime\prime \prime}$  &  &    \\[1mm]
			\hline
			$\Tc^{ \prime \prime \prime }$  &   $0$ &   $-8 t$ &  &   \\[1mm]
			\hline\hline
		\end{tabular}
	\end{center}
	\caption{The $\widetilde{ {\rm U}}(1)_{T^{\prime \prime \prime } }$ charges for massless fermions and possible symmetry breaking Higgs components in the $\Gc_{331}$ theory.
	}
	\label{tab:BG331_Tcharges}
\end{table}

\para
The third symmetry breaking stage of $\Gc_{331} \to \Gc_{\rm SM}$ can be achieved by Higgs fields of $( \rep{1}\,, \repb{3}\,,  - \frac{1}{3} )_{\rep{H}\,,\rm{VI} \,,3 }\subset \repb{8_H}_{ \,, \rm{VI} \,,3 }$ and $\[ ( \rep{1}\,,  \repb{3}\,, -\frac{1}{ 3} )_{\rep{H}\,, \dot{\rm VII} \,,\dot{2}  }^\prime  \oplus ( \rep{1}\,,  \repb{3}\,, -\frac{1}{ 3} )_{\rep{H}\,, \dot{\rm IX} } \] \subset \repb{28_H}_{ \,, \dot{\rm VII} \,,\dot{2}\,, \dot{\rm IX} }$, according to the decompositions in Eqs.~\eqref{eq:SU8B_Higgs_Br01} and \eqref{eq:SU8B_Higgs_Br02}, as well as their $\widetilde{ {\rm U}}(1)_{T^{\prime \prime \prime } }$-neutral components in Table~\ref{tab:BG331_Tcharges}.
Accordingly, the term of $ Y_\Bc \repb{8_F}^{\rm{VI} \,,3} \rep{28_F} \repb{8_H}_{\,,\rm{VI} \,,3 } + {\rm H.c.}$  leads to the vectorial masses of   $(\DG_L^\prime \,, \eG^{\prime} \,, \nG^{\prime})   $,  the term of $Y_\Dc \repb{8_F}^{\dot{\rm VII} \,,\dot{2}} \rep{56_F} \repb{28_H}_{\,,\dot{\rm VII} \,,\dot{2} } + {\rm H.c.}$  leads to the vectorial masses of  $(\DG^{\prime \prime \prime} \,, \eG^{\prime\prime\prime } \,, \nG^{\prime\prime\prime })$, and  the term of $Y_\Dc \repb{8_F}^{\dot{\rm IX}} \rep{56_F} \repb{28_H}_{ \,,\dot{\rm IX} } + {\rm H.c.}$ leads to the vectorial masses of  $(\DG^{\prime \prime \prime\prime} \,,\eG^{\prime\prime\prime\prime\prime } \,,\nG^{\prime\prime\prime\prime\prime }   )$, respectively. 
The remaining massless fermions of the $\Gc_{\rm SM}$ are listed as follows:
\begin{eqnarray}\label{eq:321B_fermions}
	&& \Big[ ( \repb{3}\,, \rep{1}\,, +\frac{1}{3} )_{ \mathbf{F}}^\Omega \oplus ( \rep{1}\,, \rep{1}\,, 0 )_{ \mathbf{F}}^\Omega \Big]  \oplus \Big[ ( \rep{1}\,, \repb{2}\,, -\frac{1}{2} )_{ \mathbf{F}}^\Omega \oplus ( \rep{1}\,, \rep{1}\,, 0 )_{ \mathbf{F}}^{\Omega^{\prime}} \oplus ( \rep{1}\,, \rep{1}\,, 0 )_{ \mathbf{F}}^{\Omega^{\prime\prime}}   \Big]  \subset \repb{8_F}^\Omega \,, \quad \Omega = ( \dot 1\,, \dot 2\,, 3 ) \,, \non
	&& ( \rep{1}\,, \rep{1}\,, 0)_{ \mathbf{F}}^{{\rm V} } \oplus ... \oplus ( \rep{1}\,, \rep{1}\,, 0)_{ \mathbf{F}}^{\dot {\rm IX} } \subset \repb{8_F}^{\Omega } \,,  \non
	&& ( \rep{1}\,, \rep{1}\,, 0)_{ \mathbf{F}}^{{\rm VI}^\prime } \oplus  ( \rep{1}\,, \rep{1}\,, 0)_{ \mathbf{F}}^{\dot {\rm VIII}^\prime } \oplus ( \rep{1}\,, \rep{1}\,, 0)_{ \mathbf{F}}^{\dot {\rm IX}^\prime } \subset \repb{8_F}^{\Omega^\prime } \,,  \non
	&& ( \rep{1}\,, \rep{1}\,, 0)_{ \mathbf{F}}^{{\rm IV}^{ \prime \prime} } \oplus ... \oplus ( \rep{1}\,, \rep{1}\,, 0)_{ \mathbf{F}}^{\dot {\rm IX}^{ \prime \prime } } \subset \repb{8_F}^{\Omega^{ \prime\prime} } \,,  \non
	&& \Big[ \bcancel{ ( \rep{3}\,, \rep{1}\,, -\frac{1}{3} )_{ \mathbf{F}} } \oplus ( \repb{3}\,, \rep{1}\,, -\frac{2}{3} )_{ \mathbf{F}} \Big]  \oplus \Big[ \cancel{ ( \rep{1}\,, \rep{2}\,, +\frac{1}{2} )_{ \mathbf{F}} \oplus ( \rep{1}\,, \rep{1}\,, 0 )_{ \mathbf{F}} } \oplus \xcancel{ ( \rep{1}\,, \repb{2}\,, +\frac{1}{2} )_{ \mathbf{F}}^\prime }  \oplus ( \rep{1}\,, \rep{1}\,, +1 )_{ \mathbf{F}} \Big] \non
	&\oplus& \Big[ ( \rep{3}\,, \rep{2}\,, +\frac{1}{6} )_{ \mathbf{F}} \oplus \xcancel{ ( \rep{3}\,, \rep{1}\,, -\frac{1}{3} )_{ \mathbf{F}}^\prime } \oplus \bcancel{ ( \rep{1}\,, \rep{2}\,, +\frac{1}{2} )_{ \mathbf{F}}^{ \prime \prime} \oplus ( \rep{1}\,, \rep{1}\,, 0 )_{ \mathbf{F}}^{ \prime } } \oplus \cancel{ ( \rep{3}\,, \rep{1}\,, -\frac{1}{3} )_{ \mathbf{F}}^{\prime \prime }  \oplus  ( \rep{1}\,, \rep{1}\,, 0 )_{ \mathbf{F}}^{ \prime \prime }} \Big]  \subset \rep{28_F}\,, \non
	&&\Big[ \xcancel{ ( \rep{1}\,, \repb{2}\,, +\frac{1}{2} )_{ \mathbf{F}}^{\prime\prime\prime }} \oplus  ( \rep{1}\,, \rep{1}\,, +1 )_{ \mathbf{F}}^\prime \oplus  \bcancel{ ( \rep{1}\,, \rep{1}\,, +1 )_{ \mathbf{F}}^{\prime\prime} } \Big] \oplus  \Big[ ( \repb{3}\,, \rep{1}\,, -\frac{2}{3} )_{ \mathbf{F}}^\prime \oplus \bcancel{ ( \rep{1}\,, \rep{1}\,, -1 )_{ \mathbf{F}}  } \Big] \non
	&\oplus&  \Big[ ( \rep{3}\,, \rep{2}\,, +\frac{1}{6} )_{ \mathbf{F}}^\prime \oplus \xcancel{ ( \rep{3}\,, \rep{1}\,, -\frac{1}{3} )_{ \mathbf{F}}^{\prime\prime \prime} } \oplus \bcancel{ ( \rep{1}\,, \rep{2}\,, +\frac{1}{2} )_{ \mathbf{F}}^{\prime\prime\prime\prime} \oplus  ( \rep{1}\,, \rep{1}\,, 0 )_{ \mathbf{F}}^{\prime\prime \prime}  }  \non
	&\oplus& \bcancel{  ( \rep{3}\,, \repb{2}\,, +\frac{1}{6} )_{ \mathbf{F}}^{\prime\prime} \oplus ( \rep{3}\,, \rep{1}\,, +\frac{2}{3} )_{ \mathbf{F}} } \oplus \xcancel{ ( \rep{1}\,, \repb{2}\,, +\frac{1}{2})_{ \mathbf{F}}^{\prime\prime\prime\prime \prime} } \oplus ( \rep{1}\,, \rep{1}\,, +1)_{ \mathbf{F}}^{\prime\prime \prime} \Big] \non
	&\oplus& \Big[ ( \rep{3}\,, \rep{2}\,, +\frac{1}{6} )_{ \mathbf{F}}^{\prime\prime \prime} \oplus \xcancel{ ( \rep{3}\,, \rep{1}\,, -\frac{1}{3})_{ \mathbf{F}}^{\prime\prime\prime \prime} }  \oplus \bcancel{ ( \repb{3}\,, \rep{2}\,, -\frac{1}{6})_{ \mathbf{F}}   \oplus ( \repb{3}\,, \rep{1}\,, -\frac{2}{3})_{ \mathbf{F}}^{\prime\prime} } \non
	&\oplus& \bcancel{ ( \rep{3}\,, \rep{1}\,, -\frac{1}{3})_{ \mathbf{F}}^{\prime\prime\prime \prime \prime }  } \oplus ( \repb{3}\,, \rep{1}\,, -\frac{2}{3})_{ \mathbf{F}}^{\prime\prime\prime} \Big]   \subset \rep{56_F} \,.
\end{eqnarray}
The fermions that become massive at this stage are further crossed out.
After this stage of symmetry breaking, there are three-generational massless SM fermions together with $23$ left-handed massless sterile neutrinos.~\footnote{The number of residual left-handed massless sterile neutrinos have been precisely obtained through the `t Hooft anomaly matching in Ref.~\cite{Chen:2023qxi}.}
The third-generational SM fermions are from the rank-two chiral IRAFFS of $\[ \repb{8_F}^\omega \oplus \rep{28_F} \]$, while the first- and second-generational SM fermions are from the rank-three chiral IRAFFS of $\[ \repb{8_F}^{\dot \omega }\oplus \rep{56_F}\]$.

\subsection{The $d=5$ bilinear fermion operators}

\para
We proceed to analyze the $d=5$ bilinear fermion operators in Eqs.~\eqref{eqs:d5_ffHH} along the WSW symmetry breaking pattern.
To be brief, we only include the biliner mass terms involving the SM quarks and leptons. 
The operator of $\Oc_\Fc^{ ( 3\,, 2) }$ is decomposed as
	\beqn\label{eq:OFB_32b}
	&&\frac{c_3}{M_{\rm{pl}}}\repb{8_F}^{\dot \omega}\rep{56_F} \cdot \repb{28_H}_{,\dot \kappa }^{\dag}  \cdot \rep{70_H}^\dag+ H.c. \non
	&\supset&\frac{c_3}{M_{\rm{pl}}}\Big[ (\repb{4}\,,\rep{1}\,,+\frac{1}{4})_{\mathbf{F}}^{\dot \omega}\otimes(\rep{4}\,,\rep{6}\,,+\frac{1}{4})_{\mathbf{F}} \oplus (\rep{1}\,,\repb{4}\,,-\frac{1}{4})_{\mathbf{F}}^{\dot \omega} \otimes(\rep{1}\,,\repb{4}\,,+\frac{3}{4})_{\mathbf{F}} \Big]\non
	&\otimes&(\repb{4}\,,\repb{4}\,,0)_{\mathbf{H},\dot \kappa}^\dag \otimes (\rep{4}\,,\repb{4}\,,+\frac{1}{2})_{\mathbf{H}}^{\dag}+H.c.\non
	&\supset&\frac{c_3}{M_{\rm{pl}}}\Big[ (\repb{4}\,,\rep{1}\,,+\frac{1}{4})_{\mathbf{F}}^{\dot \omega}\otimes(\rep{4}\,,\rep{3}\,,+\frac{1}{12})_{\mathbf{F}} \oplus (\rep{1}\,,\repb{3}\,,-\frac{1}{3})_{\mathbf{F}}^{\dot \omega} \otimes(\rep{1}\,,\repb{3}\,,+\frac{2}{3})_{\mathbf{F}}^\prime \Big]\non
	&\otimes&(\repb{4}\,,\repb{3}\,,-\frac{1}{12})_{\mathbf{H},\dot \kappa}^\dag \otimes (\rep{4}\,,\repb{3}\,,+\frac{5}{12})_{\mathbf{H}}^{\dag}+H.c.\non
	&\supset&\frac{c_3}{M_{\rm{pl}}}\Big[ (\repb{3}\,,\rep{1}\,,+\frac{1}{3})_{\mathbf{F}}^{\dot \omega}\otimes(\rep{3}\,,\rep{3}\,,0)_{\mathbf{F}}^\prime \oplus (\rep{1}\,,\rep{1}\,,0)_{\mathbf{F}}^{\dot \omega} \otimes(\rep{1}\,,\rep{3}\,,+\frac{1}{3})_{\mathbf{F}}^\prime \oplus(\rep{1}\,,\repb{3}\,,-\frac{1}{3})_{\mathbf{F}}^{\dot \omega} \otimes(\rep{1}\,,\repb{3}\,,+\frac{2}{3})_{\mathbf{F}}^\prime\Big]\non
	&\otimes& \langle (\rep{1}\,,\repb{3}\,,-\frac{1}{3})_{\mathbf{H},\dot \kappa}^{\dag} \rangle \otimes \langle (\rep{1}\,,\repb{3}\,,+\frac{2}{3})_{\mathbf{H}}^{\prime\prime\prime\,\dag}\rangle+H.c.\non
	%
	%
	&\Rightarrow&\frac{c_3}{2}\dot{\zeta_3}  ( \underline{ d_L {  \Dc_R^{\dot\omega } }^c  } +\check{\Nc}_L^{\dot\omega} {\nG_R^{\prime\prime\prime\prime}}^c - \underline{\Ec_L^{\dot\omega} { e_R}^c } +  \check{\Nc}_L^{\dot{\omega}^\prime} { \nG_R^{\prime\prime\prime } }^c  ) v_{\rm EW}+H.c. \,,  
	\eeqn
	and
	\beqn\label{eq:OFB_32c}
	&&\frac{c_3}{M_{\rm{pl}}}\repb{8_F}^{\dot \omega}\rep{56_F} \cdot \repb{28_H}_{,\dot \kappa }^{\dag}  \cdot \rep{70_H}^\dag+ H.c. \non
	&\supset&\frac{c_3}{M_{\rm{pl}}}\Big[ (\repb{4}\,,\rep{1}\,,+\frac{1}{4})_{\mathbf{F}}^{\dot \omega}\otimes(\rep{6}\,,\rep{4}\,,-\frac{1}{4})_{\mathbf{F}} \oplus    (\rep{1}\,,\repb{4}\,,-\frac{1}{4})_{\mathbf{F}}^{\dot \omega}\otimes(\rep{4}\,,\rep{6}\,,+\frac{1}{4})_{\mathbf{F}}     \Big] \non
	&\otimes&(\rep{1}\,, \rep{6}\,, -\frac{1}{2})_{\mathbf{H} ,\dot \kappa}^\dag \otimes (\rep{4}\,,\repb{4}\,,+\frac{1}{2})_{\mathbf{H}}^{\dag}+H.c.\non
	&\supset&\frac{c_3}{M_{\rm{pl}}}\Big[ (\repb{4}\,,\rep{1}\,,+\frac{1}{4})_{\mathbf{F}}^{\dot \omega}\otimes(\rep{6}\,,\rep{3}\,,-\frac{1}{6})_{\mathbf{F}} \oplus    (\rep{1}\,,\repb{3}\,,-\frac{1}{3})_{\mathbf{F}}^{\dot \omega}\otimes(\rep{4}\,,\repb{3}\,,+\frac{5}{12})_{\mathbf{F}}   \oplus    (\rep{1}\,,\rep{1}\,,0)_{\mathbf{F}}^{\dot \omega^{\prime\prime}}\otimes(\rep{4}\,,\rep{3}\,,+\frac{1}{12})_{\mathbf{F}}\Big] \non
	&\otimes&(\rep{1}\,,\repb{3}\,,-\frac{1}{3})_{\mathbf{H} ,\dot \kappa}^{\prime\,\dagger} \otimes (\rep{4}\,,\repb{3}\,,+\frac{5}{12})_{\mathbf{H}}^{\dag}+H.c.\non
	&\supset&\frac{c_3}{M_{\rm{pl}}}\Big[ (\repb{3}\,,\rep{1}\,,+\frac{1}{3})_{\mathbf{F}}^{\dot \omega}\otimes(\rep{3}\,,\rep{3}\,,0)_{\mathbf{F}}^{\prime\prime} \oplus    (\rep{1}\,,\repb{3}\,,-\frac{1}{3})_{\mathbf{F}}^{\dot \omega}\otimes(\rep{1}\,,\repb{3}\,,+\frac{2}{3})_{\mathbf{F}}^{\prime\prime}   \oplus    (\rep{1}\,,\rep{1}\,,0)_{\mathbf{F}}^{\dot \omega^{\prime\prime}}\otimes(\rep{1}\,,\rep{3}\,,+\frac{1}{3})_{\mathbf{F}}^\prime\Big] \non
	&\otimes&\langle (\rep{1}\,,\repb{3}\,,-\frac{1}{3})_{\mathbf{H} ,\dot \kappa}^{\prime\,\dagger}  \rangle \otimes \langle(\rep{1}\,,\repb{3}\,,+\frac{2}{3})_{\mathbf{H}}^{\prime\prime\prime\,\dag}\rangle+H.c.\non
	%
	%
	&\Rightarrow&\frac{c_3}{2}  \dot \zeta_3^\prime (  \underline{ s_L { \Dc_R^{\dot\omega} }^c  } -\underline{\Ec_L^{\dot  \omega}  { \mu_R}^c } +  \check{\Nc}_L^{\dot\omega^\prime}{ \nG_R^{\prime\prime\prime\prime\prime } }^c + \check{\Nc}_L^{\dot\omega^{\prime\prime}} {\nG_R^{\prime\prime\prime\prime}}^c  )  v_{\rm EW}+H.c.\,.
\end{eqnarray}
By taking the possible flavor indices of $\dot \omega = \dot 1\,, \dot 2$ in Eqs.~\eqref{eq:OFB_32b} and \eqref{eq:OFB_32c}, one finds the following set of mass matrices of the $(d\,, s)$ and $(e\,, \mu)$:
\beqs
\beqn
&&\Big( \Mc_d \Big)_{2 \times 2}^{\rm direct} = \frac{ c_3 }{2 }  \left( \ba{cc}  
\dot \zeta_3  & \dot \zeta_3  \\
\dot \zeta_3^\prime &  \dot \zeta_3^\prime   \\ \ea  \right)   v_{\rm EW} \,, \label{eq:B_ds_direct}\\[1mm]
&& \Big( \Mc_e \Big)_{2 \times 2}^{\rm direct} = - \frac{ c_3 }{2 } \left( \ba{cc}  
\dot \zeta_3 &  \dot \zeta_3^\prime   \\
\dot \zeta_3  & \dot \zeta_3^\prime  \\  \ea  \right)  v_{\rm EW}  \,,\label{eq:B_emu_direct}
\eeqn
\eeqs
which leave the down quark and electron massless.

\para
For the operator of $\Oc_\Fc^{ (4\,,1) }$, it is decomposed as
\beqs  \label{eqs:OFB_41}
\begin{eqnarray}
	&&\frac{c_4}{M_{\rm{pl}}}\rep{56_F}\rep{56_F} \cdot \repb{28_H}_{,\dot \omega } \cdot  \rep{70_H} + H.c. \non
	&\supset&\frac{c_4}{M_{\rm{pl}}}\Big[ (\repb{4}\,,\rep{1}\,,-\frac{3}{4})_{\mathbf{F}}\otimes(\rep{4}\,,\rep{6}\,,+\frac{1}{4})_{\mathbf{F}} \oplus \cancel{ (\rep{6}\,,\rep{4}\,,-\frac{1}{4})_{\mathbf{F}} \otimes (\rep{6}\,,\rep{4}\,,-\frac{1}{4})_{\mathbf{F}} } \Big] \otimes (\repb{4}\,,\repb{4}\,,0)_{\mathbf{H},\dot \omega} \otimes (\rep{4}\,,\repb{4}\,,+\frac{1}{2})_{\mathbf{H}}+H.c. \non
	&\supset&\frac{c_4}{M_{\rm{pl}}}   (\repb{4}\,,\rep{1}\,,-\frac{3}{4})_{\mathbf{F}} \otimes (\rep{4}\,,\rep{3}\,,+\frac{1}{12})_{\mathbf{F}}  \otimes \langle (\repb{4}\,,\rep{1}\,,+\frac{1}{4})_{\mathbf{H},\dot \omega}  \rangle \otimes (\rep{4}\,,\repb{3}\,,+\frac{5}{12})_{\mathbf{H}}+H.c.\non
	&\supset& c_4 \frac{w_{\repb{4},\dot{\rm VII}}}{\sqrt{2}M_{\rm{pl}}} \Big[ (\repb{3}\,,\rep{1}\,,-\frac{2}{3})_{\mathbf{F}}^{\prime} \otimes (\rep{3}\,,\rep{3}\,,0)_{\mathbf{F}}^{\prime} \oplus (\rep{1}\,,\rep{1}\,,-1)_{\mathbf{F}} \otimes (\rep{1}\,,\rep{3}\,,+\frac{1}{3})_{\mathbf{F}}^{\prime} \Big]\otimes \langle(\rep{1}\,,\repb{3}\,,+\frac{2}{3})_{\mathbf{H}}^{\prime \prime \prime}\rangle + H.c.\non
	%
	%
	&\Rightarrow& \frac{c_4}{2}  \dot{\zeta_2}  ( \underline{ u_L {c_R}^c }+ \EG_L {\eG_R^{\prime\prime\prime\prime } }^c ) v_{\rm EW} + H.c. \,. \label{eq:OFB_41a} 
\end{eqnarray}
\eeqs

\para
The bilinear fermion product of $(\rep{6}\,,\rep{4}\,,-\frac{1}{4})_{\mathbf{F}} \otimes (\rep{6}\,,\rep{4}\,,-\frac{1}{4})_{\mathbf{F}} \otimes (\repb{4}\,,\repb{4}\,,0)_{\mathbf{H},\dot \omega} \otimes (\rep{4}\,,\repb{4}\,,+\frac{1}{2})_{\mathbf{H}}+{\rm H.c.}$ was previously found to vanish due to their antisymmetric properties~\cite{Chen:2024cht}.

\para
For the operator of $\Oc_\Fc^{ (5\,,1)}$, it is decomposed as
\beqs\label{eqs:OFB_51}
\begin{eqnarray}
	&&\frac{c_5}{M_{\rm{pl}}}\rep{28_F}\rep{56_F} \cdot \repb{8_H}_{ \,, \omega } \cdot \rep{70_H}  + H.c. \non
	&\supset&\frac{c_5}{M_{\rm{pl}}}\Big[(\rep{6}\,,\rep{1}\,,-\frac{1}{2})_\mathbf{F}\otimes(\rep{4}\,,\rep{6}\,,+\frac{1}{4})_\mathbf{F} \oplus (\rep{1}\,,\rep{6}\,,+\frac{1}{2})_\mathbf{F}\otimes(\rep{1}\,,\repb{4}\,,-\frac{3}{4})_\mathbf{F}\oplus(\rep{4}\,,\rep{4}\,,0)_\mathbf{F}\otimes(\rep{6}\,,\rep{4}\,,-\frac{1}{4})_\mathbf{F}\Big]\non
	&\otimes&  \langle (\rep{1}\,,\repb{4}\,,-\frac{1}{4})_{\mathbf{H}, \omega }  \rangle \otimes (\rep{4}\,,\repb{4}\,,+\frac{1}{2})_\mathbf{H} + H.c. \non
	&\supset& c_5\frac{W_{\repb{4},\rm{IV}}}{\sqrt{2}M_{\rm{pl}}}\Big[(\rep{6}\,,\rep{1}\,,-\frac{1}{2})_\mathbf{F}\otimes(\rep{4}\,,\rep{3}\,,+\frac{1}{12})_\mathbf{F} \oplus (\rep{1}\,,\rep{3}\,,+\frac{1}{3})_\mathbf{F}\otimes(\repb{4}\,,\rep{1}\,,-\frac{3}{4})_\mathbf{F}\oplus(\rep{4}\,,\rep{3}\,,+\frac{1}{12})_\mathbf{F}\otimes(\rep{6}\,,\rep{1}\,,-\frac{1}{2})_\mathbf{F}^{\prime\prime} \non
	&\oplus&  (\rep{4}\,,\rep{1}\,,-\frac{1}{4})_\mathbf{F} \otimes (\rep{6}\,,\rep{3}\,,-\frac{1}{6})_\mathbf{F} \Big] \otimes(\rep{4}\,,\repb{3}\,,+\frac{5}{12})_\mathbf{H} + H.c. \non
	&\supset& \frac{ c_5}{ \sqrt{2} } \zeta_1 \Big[(\repb{3}\,,\rep{1}\,,-\frac{2}{3})_\mathbf{F}\otimes(\rep{3}\,,\rep{3}\,,0)_\mathbf{F}^\prime \oplus (\rep{1}\,,\rep{3}\,,+\frac{1}{3})_\mathbf{F}\otimes(\rep{1}\,,\rep{1}\,,-1)_\mathbf{F}\oplus(\rep{3}\,,\rep{3}\,,0)_\mathbf{F}\otimes(\repb{3}\,,\rep{1}\,,-\frac{2}{3})_\mathbf{F}^{\prime\prime\prime} \non
	&\oplus&  (\rep{3}\,,\rep{1}\,,-\frac{1}{3})_\mathbf{F}^{\prime\prime} \otimes (\repb{3}\,,\rep{3}\,,-\frac{1}{3})_\mathbf{F} \Big] \otimes\langle(\rep{1}\,,\repb{3}\,,+\frac{2}{3})_\mathbf{H}^{\prime \prime \prime}\rangle + H.c. \non
	&\Rightarrow&\frac{ c_5 }{2} \zeta_1 ( \underline{ u_L {t_R}^c} + \EG_L {\eG_R }^c +\underline{t_L {u_R}^c } + \DG_L^{\prime \prime } {\dG_R}^c ) v_{\rm EW} + H.c.  \,, \label{eq:OFB_51a} \\[1mm]
	&&\frac{c_5}{M_{\rm{pl}}}\rep{28_F}\rep{56_F} \cdot \repb{8_H}_{ \,, \omega } \cdot \rep{70_H}+H.c.\non
	&\supset&\frac{c_5}{M_{\rm{pl}}}\Big[  (\rep{6}\,,\rep{1}\,,-\frac{1}{2})_\mathbf{F}\otimes (\rep{6}\,,\rep{4}\,,-\frac{1}{4})_\mathbf{F} \oplus (\rep{4}\,,\rep{4}\,,0)_\mathbf{F} \otimes (\repb{4}\,,\rep{1}\,,-\frac{3}{4})_\mathbf{F}   \Big] \otimes (\repb{4}\,,\rep{1}\,,+\frac{1}{4})_{\mathbf{H} \,, \omega } \otimes (\rep{4}\,,\repb{4}\,,+\frac{1}{2})_{\mathbf{H}} + H.c. \non
	&\supset&\frac{c_5}{M_{\rm{pl}}}\Big[  (\rep{6}\,,\rep{1}\,,-\frac{1}{2})_\mathbf{F}\otimes (\rep{6}\,,\rep{3}\,,-\frac{1}{6})_\mathbf{F} \oplus (\rep{4}\,,\rep{3}\,,+\frac{1}{12})_\mathbf{F} \otimes (\repb{4}\,,\rep{1}\,,-\frac{3}{4})_\mathbf{F}   \Big] \non
	&\otimes& \langle (\repb{4}\,,\rep{1}\,,+\frac{1}{4})_{\mathbf{H} \,,  \omega } \rangle \otimes  (\rep{4}\,,\repb{3}\,,+\frac{5}{12})_{\mathbf{H}} + H.c. \non
	&\supset&c_5\frac{w_{\repb{4},\rm{V}}}{\sqrt{2}M_{\rm{pl}}}\Big[  (\rep{3}\,,\rep{1}\,,-\frac{1}{3})_\mathbf{F}\otimes (\repb{3}\,,\rep{3}\,,-\frac{1}{3})_\mathbf{F} \oplus (\repb{3}\,,\rep{1}\,,-\frac{2}{3})_\mathbf{F} \otimes (\rep{3}\,,\rep{3}\,,0)_\mathbf{F}^{\prime\prime}   \oplus  (\rep{3}\,,\rep{3}\,,0)_\mathbf{F} \otimes (\repb{3}\,,\rep{1}\,,-\frac{2}{3})_\mathbf{F}^\prime\non
	&\oplus&  (\rep{1}\,, \rep{3}\,,+\frac{1}{3})_\mathbf{F}^{\prime\prime} \otimes (\rep{1}\,,\rep{1}\,,-1)_\mathbf{F}  \Big]\otimes\langle(\rep{1}\,,\repb{3}\,,+\frac{2}{3})_{\mathbf{H}}^{\prime \prime \prime} \rangle+ H.c. \non
	&\Rightarrow&\frac{c_5}{2} \zeta_2 ( \DG_L {\dG_R}^c +\underline{ c_L  {t_R}^c}+\underline{t_L  {c_R}^c} + \EG_L{\eG_R}^c ) v_{\rm EW}+H.c. \,. \label{eq:OFB_51b} 
\end{eqnarray}
\eeqs

\subsection{The $d=5$ irreducible Higgs mixing operators}

\para
We further decompose the $d=5$ irreducible Higgs mixing operators along the WSW symmetry breaking pattern.
For the Yukawa coupling of $ \repb{8_F}^{\omega_1}  \rep{28_F} \repb{8_H}_{\,,\omega_1}$, we find the mass terms of
\begin{eqnarray}\label{eq:WSW_btau_indirect}
	&&Y_\Bc \repb{8_F}^{\omega_1}  \rep{28_F} \repb{8_H}_{\,,\omega_1} \times   \frac{d_{\mathscr{A}}}{M_{\rm{pl}}} \epsilon_{\omega_1 \omega_2\omega_3\omega_4 }  \repb{8_H}_{\,,\omega_1}^\dag  \repb{8_H}_{\,,\omega_2}^\dag  \repb{8_H}_{\,,\omega_3}^\dag   \repb{8_H}_{\,,\omega_4}^\dag   \rep{70_H}^{\dag}+H.c. \non
	&\supset& Y_\Bc\left[ (\repb{4 }\,,\rep{1}\,,+\frac{1}{4})_\mathbf{F}^{\omega_1} \otimes (\rep{ 4}\,,\rep{4}\,,0)_\mathbf{F} \oplus  (\rep{1}\,,\repb{4}\,,-\frac{1}{4})_\mathbf{F}^{\omega_1}\otimes  (\rep{1}\,,\rep{6}\,,+\frac{1}{2})_\mathbf{F} \right] \otimes  (\rep{1}\,,\repb{4}\,,-\frac{1}{4})_{\mathbf{H}\,,\omega_1} \non
	&\times& \frac{d_{\mathscr{A}}}{M_{\rm{pl}}}  (\rep{1}\,,\repb{4}\,,-\frac{1}{4})_{\mathbf{H}\,,\omega_1}^\dag \otimes(\rep{1}\,,\repb{4}\,,-\frac{1}{4})_{\mathbf{H}\,,\omega_2}^\dag \otimes(\rep{1}\,,\repb{4}\,,-\frac{1}{4})_{\mathbf{H}\,,\omega_3}^\dag\otimes (\repb{4}\,,\rep{1}\,,+\frac{1}{4})_{\mathbf{H}\,,\omega_4}^\dag \otimes (\rep{4}\,,\repb{4}\,,+\frac{1}{2})_\mathbf{H}^\dag+H.c. \non
	&\supset &Y_\Bc\left[ (\repb{4 }\,,\rep{1}\,,+\frac{1}{4})_\mathbf{F}^{\omega_1} \otimes (\rep{ 4}\,,\rep{3}\,,+\frac{1}{12})_\mathbf{F} \oplus  (\rep{1}\,,\repb{3}\,,-\frac{1}{3})_\mathbf{F}^{\omega_1}\otimes  (\rep{1}\,,\rep{3}\,,+\frac{2}{3})_\mathbf{F} \oplus(\rep{1}\,,\rep{1}\,,0)_\mathbf{F}^{\omega_1^{\prime\prime}}\otimes  (\rep{1}\,,\rep{3}\,,+\frac{1}{3})_\mathbf{F}  \right] \non
	&\otimes&  (\rep{1}\,,\repb{3}\,,-\frac{1}{3})_{\mathbf{H}\,,\omega_1} \times \frac{d_{\mathscr{A}}W_{\repb{4}\,\rm{IV}}}{\sqrt{2}M_{\rm{pl}}}  (\rep{1}\,,\repb{3}\,,-\frac{1}{3})_{\mathbf{H}\,,\omega_1}^\dag  \otimes(\rep{1}\,,\repb{3}\,,-\frac{1}{3})_{\mathbf{H}\,,\omega_3}^\dag\otimes (\repb{4}\,,\rep{1}\,,+\frac{1}{4})_{\mathbf{H}\,,\omega_4}^\dag \otimes (\rep{4}\,,\repb{3}\,,+\frac{5}{12})_\mathbf{H}^\dag\non
	&+&H.c. \non
	&\supset &Y_\Bc\Big[ (\repb{3}\,,\rep{1}\,,+\frac{1}{3})_\mathbf{F}^{\omega_1} \otimes (\rep{3}\,,\rep{3}\,,0)_\mathbf{F} \oplus(\rep{1}\,,\rep{1}\,,0)_\mathbf{F}^{\omega_1} \otimes (\rep{1}\,,\rep{3}\,,+\frac{1}{3})_\mathbf{F}^{\prime\prime} \oplus  (\rep{1}\,,\repb{3}\,,-\frac{1}{3})_\mathbf{F}^{\omega_1}\otimes  (\rep{1}\,,\rep{3}\,,+\frac{2}{3})_\mathbf{F} \non
	&\oplus& (\rep{1}\,,\rep{1}\,,0)_\mathbf{F}^{\omega_1^{\prime\prime}}\otimes  (\rep{1}\,,\rep{3}\,,+\frac{1}{3})_\mathbf{F}  \Big]  (\rep{1}\,,\repb{3}\,,-\frac{1}{3})_{\mathbf{H}\,,\omega_1}  \times \frac{d_{\mathscr{A}}W_{\repb{4}\,\rm{IV}} \omega_{\repb{4},\rm{V}}}{2M_{\rm{pl}}}  (\rep{1}\,,\repb{3}\,,-\frac{1}{3})_{\mathbf{H}\,,\omega_1}^\dag  \otimes\langle(\rep{1}\,,\repb{3}\,,-\frac{1}{3})_{\mathbf{H}\,,\omega_3}^\dag \rangle  \non
	&\otimes& \langle(\rep{1}\,,\repb{3}\,,+\frac{2}{3})_\mathbf{H}^{\prime \prime \prime \, \dag}\rangle  +H.c. \non
	&\Rightarrow& \frac{ Y_\Bc   d_{ \mathscr A}}{ 4 } \frac{ W_{ \repb{4}\,,{\rm IV} }  w_{ \repb{4}\,,{\rm V} } V_{ \repb{3}\,,{\rm VI} } }{ M_{\rm pl}  m_{( \rep{1}\,, \repb{4 }\,, - \frac{1}{4 })_{ \rep{H}\,,3} }^2}  ( \underline{ b_L {b_R}^c  } +  \underline{ \tau_L {\tau_R}^c } + \check \Nc_L^3 {\nG_R^{\prime\prime}}^c   - \check \Nc_L^{ 3^\prime} {\nG_R^{\prime}}^c +\check \Nc_L^{ 3^{\prime \prime}} {\nG_R}^{c}  ) v_{\rm EW}  + H.c. \,.
\end{eqnarray}
If we consider all three possibilities of the propagator masses, each of them leads to the $(b\,,\tau)$ masses as follows
\beqs \label{eq:SU8_B_H3mass}
\beqn
m_{( \rep{1}\,, \repb{4 }\,, - \frac{1}{4 })_{ \rep{H}\,,3} } \sim \Oc(v_{441}) ~&:&~  m_b = m_\tau = \frac{ Y_\Bc  d_{ \mathscr A} }{ 4 }  \zeta_2 \zeta_{13} v_{\rm EW}  \,,\\[1mm]
m_{( \rep{1}\,, \repb{4 }\,, - \frac{1}{4 })_{ \rep{H}\,,3} } \sim \Oc(v_{431}) ~&:&~ m_b = m_\tau = \frac{ Y_\Bc  d_{ \mathscr A} }{ 4}  \zeta_1 \zeta_{23} v_{\rm EW} \,, \\[1mm]
m_{( \rep{1}\,, \repb{4 }\,, - \frac{1}{4 })_{ \rep{H}\,,3} } \sim \Oc(v_{331}) ~&:&~  m_b = m_\tau = \frac{ Y_\Bc  d_{ \mathscr A}  }{ 4 }  \frac{ \zeta_1 }{\zeta_{23} } v_{\rm EW}  \,.
\eeqn
\eeqs
The last choice coincides with our previous result in Ref.~\cite{Chen:2024cht}.

\para
The indirect Yukawa couplings from the operator in Eq.~\eqref{eq:d5_Hmixing_B} are expected to generate the first- and second-generational down-type quark and charged lepton masses.
The gauge-invariant subset of $\Large( \repb{28_H}_{\,,\dot 1 }^\dag \repb{28_H}_{\,,\dot {\rm VII} }  \Large)$ can develop the VEV of $\langle  \repb{28_H}_{\,,\dot 1 }^\dag \repb{28_H}_{\,,\dot {\rm VII} } \rangle = \frac{1}{2} w_{\repb{4}\,,  \dot 1} w_{\repb{4}\,,  \dot {\rm VII} } \sim \Oc ( v_{431}^2 )$ according to the VEV assignments in Eq.~\eqref{eq:SU8_WSW_Higgs_VEVs_mini02}.
Similar to the indirect Yukawa couplings in Eq.~\eqref{eq:WSW_btau_indirect}, we should look for the EWSB components from the $\repb{28_{H}}_{\,, \dot 1 \,, \dot 2 }$ here.
For the Yukawa coupling of $\repb{8_F}^{\dot \omega_1} \rep{56_F}  \repb{28_{H}}_{\,, \dot \omega_1 }$, we find the mass terms of
\begin{eqnarray}\label{eq:WSW_de_indirect}
	&& Y_\Dc   \repb{8_F}^{\dot \omega_1} \rep{56_F}  \repb{28_{H}}_{\,, \dot \omega_1 }  \times \frac{ d_{ \mathscr B} }{ M_{\rm pl} }   \repb{28_{H}}_{\,, \dot \omega_1 }^\dag   \repb{28_{H}}_{\,, \dot \omega_2 }^\dag  \rep{70_{H}}^\dag  \Large( \repb{28_H}_{\,,\dot 1 }^\dag \repb{28_H}_{\,,\dot {\rm VII} } \Large) + H.c. \non
	&\supset& Y_\Dc \Big[ ( \repb{4 } \,, \rep{1 } \,,  + \frac{1 }{4 } )_{ \rep{F}}^{\dot \omega_1} \otimes ( \rep{4 } \,, \rep{6 } \,,  + \frac{1 }{4 } )_{ \rep{F}} \oplus ( \rep{1 } \,, \repb{4 } \,,  - \frac{1 }{4 } )_{ \rep{F}}^{\dot \omega_1 } \otimes ( \rep{1 } \,, \repb{4 } \,,  + \frac{3 }{4 } )_{ \rep{F}} \Big]  \otimes ( \rep{1 } \,, \rep{6 } \,,  - \frac{1 }{2 } )_{ \rep{H}\,, \dot \omega_1} \non
	&\times&  \frac{  d_{ \mathscr B} }{ M_{\rm pl} } ( \rep{1 } \,, \rep{6 } \,,  - \frac{1 }{2 } )_{ \rep{H}\,, \dot \omega_1}^\dag \otimes ( \repb{4 } \,, \repb{4 } \,,  0 )_{ \rep{H}\,, \dot \omega_2 }^\dag \otimes ( \rep{4 } \,, \repb{4 } \,,  + \frac{1 }{2 } )_{ \rep{H}}^\dag \otimes \langle  \repb{28_H}_{\,,\dot 1 }^\dag \repb{28_H}_{\,,\dot {\rm VII} }  \rangle  + H.c. \non
	&\supset& Y_\Dc \Big[ ( \repb{4 } \,, \rep{1 } \,,  + \frac{1 }{4 } )_{ \rep{F}}^{\dot \omega_1} \otimes ( \rep{4 } \,, \rep{3 } \,,  + \frac{1 }{12 } )_{ \rep{F}} \oplus ( \rep{1 } \,, \repb{3 } \,,  - \frac{1 }{3 } )_{ \rep{F}}^{\dot \omega_1 } \otimes ( \rep{1 } \,, \repb{3 } \,,  + \frac{2 }{3 } )_{ \rep{F}}^{\prime} \Big]  \otimes ( \rep{1 } \,, \repb{3 } \,,  - \frac{1 }{3 } )_{ \rep{H}\,, \dot \omega_1}^{\prime} \non
	&\times&  \frac{  d_{ \mathscr B} }{ M_{\rm pl} } ( \rep{1 } \,, \repb{3 } \,,  - \frac{1 }{3 } )_{ \rep{H}\,, \dot \omega_1}^{\prime\,\dag} \otimes ( \repb{4 } \,, \repb{3 } \,,  -\frac{1}{12} )_{ \rep{H}\,, \dot \omega_2 }^\dag \otimes ( \rep{4 } \,, \repb{3 } \,,  + \frac{5 }{12 } )_{ \rep{H}}^\dag \otimes \langle  \repb{28_H}_{\,,\dot 1 }^\dag \repb{28_H}_{\,,\dot {\rm VII} }  \rangle  + H.c. \non
	&\supset& Y_\Dc d_{ \mathscr B}  \frac{ w_{\repb{4}\,,  \dot 1} w_{\repb{4}\,,  \dot {\rm VII} }  }{ 2 M_{\rm pl} m_{ ( \rep{1 } \,, \rep{6 } \,,  - \frac{1 }{ 2 } )_{ \rep{H}\,, \dot  \omega_1 } }^{2} }    \Big[  ( \repb{3 } \,, \rep{1 } \,,  + \frac{1 }{3 } )_{ \rep{F}}^{\dot \omega_1} \otimes ( \rep{3 } \,, \rep{3 } \,,  0 )_{ \rep{F}}^\prime \oplus ( \rep{1 } \,, \rep{1 } \,, 0 )_{ \rep{F}}^{\dot \omega_1} \otimes ( \rep{1 } \,, \rep{3 } \,,  + \frac{1  }{3 } )_{ \rep{F}}^{ \prime }  \non
	&\oplus& ( \rep{1 } \,, \repb{3 } \,,  - \frac{1 }{3 } )_{ \rep{F}}^{\dot \omega_1 } \otimes ( \rep{1 } \,, \repb{3 } \,,  +\frac{2 }{3 } )_{ \rep{F}}^{\prime } \Big] \otimes   \langle ( \rep{1 } \,, \repb{3 } \,,  - \frac{1 }{3 } )_{ \rep{H}\,, \dot \omega_2 }^{  \dag} \rangle  \otimes  \langle ( \rep{1 } \,, \repb{3 } \,,  + \frac{2 }{3 } )_{ \rep{H}}^{\prime \prime \prime \,  \dag} \rangle + H.c. \non
	&\Rightarrow&  \frac{ Y_\Dc d_{ \mathscr B} }{4 }  \dot \zeta_3   \Big[ \frac{   w_{\repb{4}\,,  \dot 1} w_{\repb{4}\,,  \dot {\rm VII} }}{  m_{ ( \rep{1 } \,, \rep{6 } \,,  - \frac{1 }{ 2 })_{ \rep{H}\,, \dot  1 }  }^2 }   ( \underline{ d_L {d_R}^c } +  \underline{ e_L {e_R}^c  }  )  + \frac{  w_{\repb{4}\,,  \dot 1} w_{\repb{4}\,,  \dot {\rm VII} }  }{  m_{ ( \rep{1 } \,, \rep{6 } \,,  - \frac{1 }{ 2 })_{ \rep{H}\,, \dot  2 }  }^2 }   ( \underline{ d_L {s_R}^c } +  \underline{ \mu_L {e_R}^c  }  )  \Big] v_{\rm EW}  + H.c. \,,
\end{eqnarray}
where the SM quark/lepton components from the $\repb{8_F}^{\dot \omega_1= \dot 1}$/$\repb{8_F}^{\dot \omega_1= \dot 2}$ correspond to the $({d_R}^c \,, e_L)$ and $({s_R}^c \,, \mu_L)$, respectively.
The other mass terms read
\begin{eqnarray}\label{eq:WSW_smu_indirect}
	&& Y_\Dc  \repb{8_F}^{\dot \omega_1} \rep{56_F}  \repb{28_{H}}_{\,, \dot \omega_1 } \times \frac{ d_{ \mathscr B} }{ M_{\rm pl} }   \repb{28_{H}}_{\,, \dot \omega_1 }^\dag   \repb{28_{H}}_{\,, \dot \omega_2 }^\dag  \rep{70_{H}}^\dag \Large( \repb{28_H}_{\,,\dot 1 }^\dag \repb{28_H}_{\,,\dot {\rm VII} } \Large)  + H.c. \non
	&\supset& Y_\Dc  \Big[  ( \repb{4 } \,, \rep{1 } \,,  + \frac{1 }{4 } )_{ \rep{F}}^{\dot \omega_1} \otimes ( \rep{6 } \,, \rep{4 } \,,  - \frac{1 }{4 } )_{ \rep{F}} \oplus ( \rep{1 } \,, \repb{4 } \,,  - \frac{1 }{4 } )_{ \rep{F}}^{\dot \omega_1 } \otimes ( \rep{4 } \,, \rep{6 } \,,  + \frac{1 }{4 } )_{ \rep{F}} \Big]  \otimes ( \repb{4 } \,, \repb{4 } \,,  0 )_{ \rep{H}\,, \dot \omega_1} \non
	&\times& \frac{ d_{ \mathscr B} }{ M_{\rm pl} }  ( \repb{4 } \,, \repb{4 } \,,  0 )_{ \rep{H}\,, \dot \omega_1}^\dag \otimes ( \rep{1 } \,, \rep{6 } \,,  - \frac{1 }{2 } )_{ \rep{H}\,, \dot \omega_2 }^\dag  \otimes ( \rep{4 } \,, \repb{4 } \,,  + \frac{1 }{2 } )_{ \rep{H}}^\dag \otimes \langle  \repb{28_H}_{\,,\dot 1 }^\dag \repb{28_H}_{\,,\dot {\rm VII} } \rangle + H.c. \non
	&\supset& Y_\Dc  \Big[  ( \repb{4 } \,, \rep{1 } \,,  + \frac{1 }{4 } )_{ \rep{F}}^{\dot \omega_1} \otimes ( \rep{6 } \,, \rep{3 } \,,  - \frac{1 }{6 } )_{ \rep{F}} \oplus ( \rep{1 } \,, \repb{3 } \,,  - \frac{1 }{3 } )_{ \rep{F}}^{\dot \omega_1 } \otimes ( \rep{4 } \,, \repb{3 } \,,  + \frac{5 }{12 } )_{ \rep{F}}  \oplus ( \rep{1 } \,, \rep{1} \,,  0 )_{ \rep{F}}^{\dot \omega_1^{\prime\prime} } \otimes ( \rep{4 } \,, \rep{3 } \,,  + \frac{1 }{12 } )_{ \rep{F}} \Big]  \non
	&\otimes& ( \repb{4 } \,, \repb{3 } \,,  -\frac{1}{12} )_{ \rep{H}\,, \dot \omega_1} \times\frac{ d_{ \mathscr B}  }{  M_{\rm pl} }  ( \repb{4 } \,, \repb{3 } \,,  -\frac{1}{12} )_{ \rep{H}\,, \dot \omega_1}^{\dag} \otimes ( \rep{1 } \,, \repb{3 } \,,  -\frac{1}{3} )_{ \rep{H}\,, \dot \omega_2}^{\prime\,\dag}\otimes( \rep{4 } \,, \repb{3 } \,,  +\frac{5}{12} )_{ \rep{H}\,, \dot \omega_1}^{\dag}  \non
	&\otimes& \langle  \repb{28_H}_{\,,\dot 1 }^\dag \repb{28_H}_{\,,\dot {\rm VII} } \rangle+H.c.\non
	&\supset& Y_\Dc  \Big[  ( \repb{3 } \,, \rep{1 } \,,  + \frac{1 }{3 } )_{ \rep{F}}^{\dot \omega_1} \otimes ( \rep{3 } \,, \rep{3 } \,,  0 )_{ \rep{F}}^{ \prime \prime} \oplus  ( \rep{1 } \,, \repb{3 } \,,  - \frac{1 }{3 } )_{ \rep{F}}^{\dot \omega_1} \otimes   ( \rep{1 } \,, \repb{3 } \,,  + \frac{2 }{3 } )_{ \rep{F}}^{\prime \prime } \oplus ( \rep{1 } \,, \rep{1 } \,,  0 )_{ \rep{F}}^{\dot \omega_1^{ \prime \prime } } \otimes   ( \rep{1 } \,, \rep{3 } \,,  + \frac{1 }{3 } )_{ \rep{F}}^{\prime }  \Big]\non
	&\otimes& ( \rep{1 } \,, \repb{3 } \,,  - \frac{1 }{ 3} )_{ \rep{H}\,, \dot \omega_1}^{  }   \times d_{ \mathscr B} \frac{w_{\repb{4}\,,  \dot 1} w_{\repb{4}\,,  \dot {\rm VII} }  }{ 2 M_{\rm pl} } ( \rep{1 } \,, \repb{3 } \,,  - \frac{1 }{ 3} )_{ \rep{H}\,, \dot \omega_1}^{\dag}  \langle ( \rep{1 } \,, \repb{3 } \,,  - \frac{1 }{ 3} )_{ \rep{H}\,, \dot \omega_2}^{\,\dag } \rangle \otimes  ( \rep{1 } \,, \repb{3 } \,,  + \frac{2 }{3 } )_{ \rep{H}}^{\prime \prime \prime \,\dag }  + H.c.  \non
	&\supset& Y_\Dc d_{ \mathscr B}  \frac{ w_{\repb{4}\,,  \dot 1} w_{\repb{4}\,,  \dot {\rm VII} } V_{ \repb{3}\,, \dot {\rm VIII} }^{ \prime} }{ 2 \sqrt{2} M_{\rm pl} m_{  (\repb{4 } \,, \repb{4 } \,,  0  )_{ \rep{H}\,, \dot  \omega_1}   }^2 } \Big[  ( \repb{3 } \,, \rep{1 } \,,  + \frac{1 }{3 } )_{ \rep{F}}^{\dot \omega_1} \otimes ( \rep{3 } \,, \rep{2 } \,,  +\frac{1 }{6 } )_{ \rep{F}}^{ \prime \prime \prime } \oplus  ( \rep{1 } \,, \repb{2 } \,,  - \frac{1 }{2 } )_{ \rep{F}}^{\dot \omega_1} \otimes   ( \rep{1 } \,, \rep{1 } \,,  + 1 )_{ \rep{F}}^{\prime \prime \prime } \non
	&\oplus&  ( \rep{1 } \,, \rep{1 } \,,  0 )_{ \rep{F}}^{\dot \omega_1^{ \prime } } \otimes   ( \rep{1 } \,, \repb{2 } \,,  + \frac{1 }{2 } )_{ \rep{F}}^{\prime \prime \prime \prime \prime }   \oplus  ( \rep{1 } \,, \rep{1 } \,,  0 )_{ \rep{F}}^{\dot \omega_1^{ \prime \prime } } \otimes   ( \rep{1 } \,, \rep{2 } \,,  + \frac{1 }{2 } )_{ \rep{F}}^{\prime\prime \prime\prime }   \Big] \otimes  \langle ( \rep{1 } \,, \repb{2 } \,,  + \frac{1 }{2 } )_{ \rep{H}}^{\prime \prime \prime \,\dag } \rangle  + H.c. \non
	&\Rightarrow& \frac{ Y_\Dc d_{ \mathscr B} }{4 } \dot \zeta_3^\prime   \Big[ \frac{w_{\repb{4}\,,  \dot 1} w_{\repb{4}\,,  \dot {\rm VII} }   }{  m_{ ( \repb{4} \,, \repb{4 } \,,  0 )_{ \rep{H}\,, \dot  1}   }^2 }   ( \underline{ s_L {d_R}^c } + \underline{ e_L {\mu_R}^c } )  + \frac{ w_{\repb{4}\,,  \dot 1} w_{\repb{4}\,,  \dot {\rm VII} }  }{  m_{ ( \repb{4 } \,, \repb{4 } \,,   0)_{ \rep{H}\,, \dot  2 }   }^2 }   ( \underline{ s_L {s_R}^c } + \underline{ \mu_L {\mu_R}^c } )   \Big] v_{\rm EW} + H.c.  \,.
\end{eqnarray}
With the Higgs VEV assignments in \eqref{eq:SU8_WSW_Higgs_VEVs_mini03}, one expects the natural propagator masses of
\begin{eqnarray}  \label{eq:SU8_B_H1H2mass}
	&& m_{ ( \repb{4 } \,, \repb{4 } \,,  0 )_{ \rep{H}\,, \dot  1} } \sim m_{ ( \rep{1 } \,, \rep{6 } \,,  - \frac{1 }{2 } )_{ \rep{H}\,, \dot  1} } \sim \Oc( v_{431 } ) \,, \non
	&&  m_{ ( \rep{1 } \,, \rep{6 } \,,  - \frac{1 }{2 } )_{ \rep{H}\,, \dot  2}  } \sim m_{ ( \repb{4 } \,, \repb{4 } \,,  0 )_{ \rep{H}\,, \dot  2} } \sim  \Oc( v_{331 } ) \,.
\end{eqnarray}
For convenience, we parametrize the following ratios of
\begin{eqnarray}  \label{eq:SU8_B_Delta}
	&& \Delta_{ \dot \omega } \equiv \frac{w_{\repb{4}\,,  \dot 1} w_{\repb{4}\,,  \dot {\rm VII} }   }{  m_{ ( \repb{4} \,, \repb{4 } \,, 0 )_{ \rep{H}\,, \dot {\omega}}   }^2 }        \,, \quad \Delta_{ \dot \omega}^\prime \equiv \frac{ w_{\repb{4}\,,  \dot 1} w_{\repb{4}\,,  \dot {\rm VII} }  }{  m_{ ( \rep{1 } \,, \rep{6 } \,,   -\frac{ 1}{ 2} )_{ \rep{H}\,, \dot  {\omega} }   }^2 }\,.
\end{eqnarray}
%
%

\subsection{The SM quark/lepton masses and the CKM mixing}

\para
For all up-type quarks with $Q_e=+\frac{2}{3}$, we write down the following tree-level masses from both the renormalizable Yukawa couplings and the gravity-induced terms in the basis of $\Uc \equiv (u\,,c\,,t)$:
\beqs\label{eqs:WSW_Uquark_masses}
\beqn
\Mc_u   &=&   \frac{1}{\sqrt{2} }  \left( \ba{ccc}  
0 &   c_4  \dot \zeta_2 /\sqrt{2}   & c_5 \zeta_1 /\sqrt{2} \\
0 & 0  & c_5 \zeta_2/ \sqrt{2}   \\
c_5 \zeta_1 /\sqrt{2}  &  c_5 \zeta_2/\sqrt{2}  &   Y_\Tc \\  \ea  \right) v_{\rm EW}  \approx \Mc_u^{ (0)} + \Mc_u^{ (1 ) } + \Mc_u^{( 2)}   \,, \\[1mm]
\Mc_u^{ (0)}    &=& \frac{1}{\sqrt{2} }  \left( \ba{ccc}  
0 & 0   & 0   \\
0   &  0  & 0   \\
0  &  0  &   Y_\Tc   \\  \ea  \right)  v_{\rm EW} \,, \label{eq:WSW_Uquark_mass00} \\[1mm]
\Mc_u^{ (1)}    &=& \frac{1}{\sqrt{2} }  \left( \ba{ccc}  
0&  0  &  c_5 \zeta_1 /\sqrt{2}   \\
0   &   0 & 0    \\
c_5 \zeta_1 /\sqrt{2} &  0 &   0   \\   \ea  \right) v_{\rm EW}   \,,\label{eq:WSW_Uquark_mass01}  \\[1mm]
\Mc_u^{ (2)}    &=&  \frac{1}{\sqrt{2} }  \left( \ba{ccc}  
0 &  c_4 \dot \zeta_2  /\sqrt{2}  &  0  \\
0 & 0 & c_5 \zeta_2 /\sqrt{2}  \\
0  &  c_5 \zeta_2 /\sqrt{2} &   0  \\  \ea  \right) v_{\rm EW}  \,, \label{eq:WSW_Uquark_mass02}
\eeqn
\eeqs
where we have neglected the $\sim \Oc(\zeta_3\,v_{\rm EW})$ terms in the above expansions.
One obvious feature is that the gauge eigenstates of up quark and the charm quark do not obtain tree-level masses through the $d=5$ operators with the SM Higgs doublet.
Instead, there are only off-diagonal mass mixing terms in Eqs.~\eqref{eq:WSW_Uquark_mass01} and \eqref{eq:WSW_Uquark_mass02}.
Accordingly, we find that
\begin{eqnarray}  \label{eq:SU8_B_mt2}
	{\rm det}^\prime \[ \Mc_u^{ (0)} \Mc_u^{ (0)\, \dag}   \] &=& \frac{1}{2}  Y_\Tc^2 \,  v_{\rm EW}^2  \Rightarrow m_t^2 \approx \frac{ 1 }{  2 } Y_\Tc^2 \, v_{\rm EW}^2 \,.
\end{eqnarray}
Here and below, we use the ${\rm det}^\prime$ to denote the matrix determinant that is equal to the products of all nonzero eigenvalues.
Next, we find the charm quark mass squared of
\begin{eqnarray} \label{eq:SU8_B_mc2}
	m_c^2 &=& {\rm det}^\prime \[ \Big( \Mc_u^{ (0)} + \Mc_u^{ (1)}  \Big) \cdot \Big( \Mc_u^{ (0)\, \dag} + \Mc_u^{ (1)\, \dag} \Big)  \]   \Big/ {\rm det}^\prime \[ \Mc_u^{ (0)} \Mc_u^{ (0)\, \dag}   \]  \approx   c_5^4  \frac{  \zeta_1^4 }{8 Y_\Tc^2 } \, v_{\rm EW}^2 \,.
\end{eqnarray}
The up quark mass squared can be similarly obtained by
\begin{eqnarray}\label{eq:SU8_B_mu2}
	m_u^2 &=& {\rm det} \[  \Big( \Mc_u^{ (0)} + \Mc_u^{ (1)} + \Mc_u^{ (2)}  \Big) \cdot \Big( \Mc_u^{ (0)\, \dag} + \Mc_u^{ (1)\, \dag} + \Mc_u^{ (2)\, \dag} \Big)  \]  \non
	&&  \Big/ {\rm det}^\prime \[ \Big( \Mc_u^{ (0)} + \Mc_u^{ (1)}  \Big) \cdot \Big( \Mc_u^{ (0)\, \dag} + \Mc_u^{ (1)\, \dag} \Big)  \]   \approx c_4^2 \frac{ \zeta_2^2 \dot \zeta_2^2  }{4 \zeta_1^2 }\, v_{\rm EW}^2 \,.
\end{eqnarray}
To summarize, all SM up-type quark masses are expressed as follows
\begin{eqnarray}\label{eq:SU8_WSW_SMumasses}
	&&  m_u \approx c_4 \frac{ \zeta_2 \dot \zeta_2 }{ 2 \zeta_1 } v_{\rm EW} \,,\quad m_c \approx c_5^2  \frac{ \zeta_1^2 }{  2 \sqrt{2} Y_\Tc } v_{\rm EW}  \,, \quad  m_t \approx \frac{ Y_\Tc }{ \sqrt{2} } v_{\rm EW} \,.
\end{eqnarray}

\para
For all down-type quarks with $Q_e=-\frac{1}{3}$, we find the following tree-level SM mass matrix 
\begin{eqnarray}\label{eq:WSW_Dquark_massdd}
	&&  \Big( \Mc_d \Big)_{3\times 3}   \approx  \frac{1}{4} \left( \ba{ccc}
	( 2 c_3  +  Y_\Dc d_{\mathscr B}  )  \dot \zeta_3  &  (  2 c_3  +  Y_\Dc d_{\mathscr B}   \zeta_{23 }^{-2 } ) \dot \zeta_3   &  0   \\
	( 2 c_3 + Y_\Dc d_{\mathscr B}  )\dot \zeta_3^\prime  &  ( 2 c_3  +  Y_\Dc d_{\mathscr B} \zeta_{23 }^{-2 }) \dot \zeta_3^\prime  & 0   \\
	0 & 0  &  Y_\Bc d_{\mathscr A}   \zeta_{23}^{-1} \zeta_1   \\  \ea  \right) v_{\rm EW}    \,.
\end{eqnarray}
For convenience, we parametrize all $\Gc_{331}$ breaking VEVs as follows:
\begin{eqnarray}\label{eq:WSW_331VEVratio}
	&& \zeta_3 = \dot \zeta_3^\prime =  \frac{ \dot \zeta_3 }{ \tan \lambda } \, .
\end{eqnarray}
It is straightforward to find the following SM down-type quark masses of
\beqs\label{eqs:WSW_SMdmasses}
\beqn
m_b &\approx&   \frac{ 1 }{ 4  }  Y_\Bc d_{\mathscr A} \zeta_{ 23 }^{-1} \zeta_1 \, v_{\rm EW} \,,  \\[1mm]
m_s  &\approx& \frac{1 }{4 } ( 2c_3 + Y_\Dc d_{\mathscr B} \zeta_{ 23 }^{-2 } ) \dot \zeta_3 \, v_{\rm EW} \,,  \\[1mm]
m_d &\approx& \hf c_3 \dot \zeta_3     \, v_{\rm EW} \,,
\eeqn
\eeqs
from Eq.~\eqref{eq:WSW_Dquark_massdd}.

\para
For all charged leptons with $Q_e=-1$, their tree-level mass matrix is correlated with the down-type quark mass matrix as
\begin{eqnarray} \label{eq:WSW_Lepton_massll}
	\Big( \Mc_\ell \Big)_{ 3\times 3} &=&   \Big( \Mc_d^T \Big)_{3\times 3} \non
	&=& \frac{1}{4} \left( \ba{ccc}
	( 2 c_3  +  Y_\Dc d_{\mathscr B} \Delta_{ \dot 1}^\prime ) \dot \zeta_3  & ( 2 c_3 + Y_\Dc d_{\mathscr B}  )\dot \zeta_3^\prime  &  0   \\
	(  2 c_3  +  Y_\Dc d_{\mathscr B}   \zeta_{23 }^{-2 } ) \dot \zeta_3   & ( 2 c_3  +  Y_\Dc d_{\mathscr B}   \Delta_{ \dot 2}   )  \dot \zeta_3^\prime  & 0   \\
	0 & 0  &  Y_\Bc d_{\mathscr A}   \zeta_{23}^{-1} \zeta_1   \\  \ea  \right) v_{\rm EW}   \,.
\end{eqnarray}
Thus, it is straightforward to find the tree-level mass relations of
\begin{eqnarray} \label{eq:WSW_SMleptonmasses}
	&& m_\tau = m_b \,,\quad m_\mu = m_s \,,\quad m_e = m_d  \,.
\end{eqnarray}

\section{The intermediate stages along the WWS symmetry breaking pattern}\label{section:WWS_process}

\subsection{The first stage}

\para
The first symmetry breaking stage of $\Gc_{441} \to \Gc_{431}$ along the WWS symmetry breaking pattern is completely identical to what we have described in Appendix~\ref{section:WSW_stage1}.
Thus, the remaining massless fermions can be found in Eq.~\eqref{eq:431B_fermions}.

\subsection{The second stage}

\begin{table}[htp]
	\begin{center}
		\begin{tabular}{c|cccc}
			\hline\hline
			$\repb{8_F}^\Omega $ & $( \repb{4}\,, \rep{1}\,, +\frac{1}{4} )_{\rep{F}}^\Omega $ & $( \rep{1}\,, \repb{3}\,,  -\frac{1}{3} )_{\rep{F}}^\Omega $  &  $( \rep{1}\,, \rep{1}\,, 0 )_{\rep{F}}^{\Omega^{ \prime\prime}}$   &   \\[1mm]
			\hline
			$\Tc^{\prime\prime}$  &  $ -2t$ & $ -4t$  &  $ - 4 t$   &    \\[1mm]
			\hline
			$\rep{28_F}$  &  $( \rep{6}\,, \rep{1}\,,  -\frac{1}{2 } )_{\rep{F}}$ &  $( \rep{1}\,, \repb{3}\,,  + \frac{2}{3} )_{\rep{F}}$ & $( \rep{ 4 }\,, \rep{3}\,, + \frac{1 }{12 } )_{\rep{F}}$  &    \\[1mm]
			\hline
			$\Tc^{\prime\prime}$   & $0$  &  $ +4 t$  &  $+2t$  &    \\[1mm]
			\hline
			$\rep{56_F}$ & $( \rep{1}\,, \repb{3}\,,  + \frac{2 }{3} )_{\rep{F}}^\prime$ & $( \rep{1}\,,  \rep{1}\,,  +1 )_{\rep{F}}^{\prime\prime}$ &  $( \repb{4}\,,  \rep{1}\,,  - \frac{3}{4} )_{\rep{F}}$  &     \\[1mm]
			\hline
			$\Tc^{\prime\prime}$  &  $+4 t$  &  $+4t$  & $-2t$  &     \\[1mm]
			\hline
			& $( \rep{4}\,,  \rep{3}\,,  +\frac{1}{12} )_{\rep{F}}^{\prime}$ & $( \rep{4}\,,  \repb{3}\,,  +\frac{5}{12} )_{\rep{F}}$ & $( \rep{6}\,,  \rep{3}\,,  -\frac{1}{ 6} )_{\rep{F}}$  &   $( \rep{6}\,,  \rep{1}\,,  -\frac{1}{2} )_{\rep{F}}^{\prime\prime}$   \\[1mm]
			\hline
			$\Tc^{\prime\prime}$ & $ +2t$ & $+ 2t$    & $0$ &   $ 0$ \\[1mm]
			\hline\hline
			$\repb{8_H}_{\,, \omega}$  &  $( \rep{1}\,, \repb{3}\,,  -\frac{1}{3} )_{\rep{H}\,,\omega}$  & $( \repb{4}\,, \rep{1}\,,  +\frac{1}{4} )_{\rep{H}\,,\omega}$  &  &    \\[1mm]
			\hline
			$\Tc^{\prime\prime}$  &  $0$  & $+ 2t$  &   &     \\[1mm]
			\hline
			$\repb{28_H}_{\,, \dot \omega }$  & $( \repb{4}\,, \repb{3}\,,  -\frac{1}{12} )_{\rep{H}\,,\dot \omega }$  &  $( \repb{4}\,, \rep{1}\,,  +\frac{1}{4} )_{\rep{H}\,,\dot \omega }^\prime$  & $( \rep{1}\,, \repb{3}\,,  -\frac{1}{3} )_{\rep{H}\,,\dot \omega }^\prime $  & $( \rep{1}\,, \rep{3}\,,  -\frac{2}{3} )_{\rep{H}\,,\dot \omega }$    \\[1mm]
			\hline
			$\Tc^{\prime\prime}$  &  $ +2t$  &  $+2t $  & $ 0$ &  $ 0$   \\[1mm]
			\hline
			$\rep{70_H}$ &  $( \rep{4 }\,, \repb{3}\,,  +\frac{5}{12} )_{\rep{H}}$  & $( \repb{4 }\,, \rep{3}\,,  -\frac{5}{12} )_{\rep{H}}$   &  &    \\[1mm]
			\hline
			$\Tc^{\prime\prime}$  & $-2t$  &  $ -6t$  &   &     \\[1mm]
			\hline\hline
		\end{tabular}
	\end{center}
	\caption{The $\widetilde{ {\rm U}}(1)_{T^{\prime \prime} }$ charges for massless fermions and possible symmetry breaking Higgs components in the $\Gc_{431}$ theory according to Eq.~\eqref{eq:U1T_defC}.
	}
	\label{tab:CG431_Tcharges}
\end{table}

\para
The second symmetry breaking stage of $\Gc_{431} \to \Gc_{421}$ along the WWS symmetry breaking pattern can be achieved by Higgs fields of $( \rep{1} \,, \repb{3} \,, -\frac{1}{3} )_{\mathbf{H}\,, \rm V } \subset \repb{8_H}_{\,,\rm V}$ and $( \rep{1} \,, \repb{3} \,, -\frac{1}{3} )_{\mathbf{H}\,, \dot{\rm VII}\,,\dot{1} }^\prime \subset ( \rep{1} \,, \rep{6} \,, -\frac{1}{2} )_{\mathbf{H}\,, \dot{\rm VII}\,,\dot{1} } \subset \repb{28_H}_{\,,\dot{\rm VII}\,,\dot{1}}$, according to the decompositions in Eqs.~\eqref{eq:SU8C_Higgs_Br01} and \eqref{eq:SU8C_Higgs_Br02}, as well as their $\widetilde{ {\rm U}}(1)_{T^{ \prime \prime}}$ charges in Table~\ref{tab:CG431_Tcharges}.
Accordingly, the term of $ Y_\Bc \repb{8_F}^{\rm V} \rep{28_F} \repb{8_H}_{\,,{\rm V }} + {\rm H.c.}$ leads to the vectorial masses of $(\DG^{\prime}\,, \check \nG^{\prime} \,, \eG^{\prime}\,, \nG^{\prime}  )$,  and the term of $  Y_\Dc \repb{8_F}^{\dot{\rm VII}} \rep{56_F} \repb{28_H}_{\,,\dot{\rm VII} } + {\rm H.c.}$ leads to the vectorial masses of $(\DG^{\prime\prime \prime }  \,, \check \nG^{\prime\prime\prime }  \,, \eG^{ \prime  \prime\prime }  \,,  \nG^{\prime \prime  \prime }    )$, respectively.
After integrating out the massive fermions, the remaining massless fermions expressed in terms of the $\Gc_{421}$ IRs are the following:
\begin{eqnarray} \label{eq:421C_fermions}
	&& ( \repb{4}\,, \rep{1}\,, +\frac{1}{4} )_{ \mathbf{F}}^\Omega \oplus \Big[  ( \rep{1}\,, \repb{2}\,, -\frac{1}{2} )_{ \mathbf{F}}^\Omega \oplus ( \rep{1}\,, \rep{1}\,, 0 )_{ \mathbf{F}}^{\Omega ^\prime} \Big]  \oplus  ( \rep{1}\,, \rep{1}\,, 0 )_{ \mathbf{F}}^{\Omega^{\prime \prime }}  \subset \repb{8_F}^\Omega \,, \non
	&& \Omega = ( \omega\,, \dot \omega ) \,, \quad \omega = (3\,, {\rm VI})\,,  \quad \dot \omega = (\dot 1\,, \dot 2\,, \dot {\rm VIII}\,, \dot {\rm IX} )\,,\non
	&& ( \rep{1}\,, \rep{1}\,, 0)_{ \mathbf{F}}^{{\rm IV}^{\prime\prime} } \subset \repb{8_F}^{{\rm IV} } \,, \quad ( \rep{1}\,, \rep{1}\,, 0)_{ \mathbf{F}}^{{\rm V}^{ \prime}  } \oplus ( \rep{1}\,, \rep{1}\,, 0)_{ \mathbf{F}}^{ {\rm V}^{ \prime \prime} } \subset \repb{8_F}^{ \rm V } \,,  \non
	&&  ( \rep{1}\,, \rep{1}\,, 0)_{ \mathbf{F}}^{\dot {\rm VII}^{ \prime}  } \oplus ( \rep{1}\,, \rep{1}\,, 0)_{ \mathbf{F}}^{\dot {\rm VII}^{ \prime \prime} } \subset \repb{8_F}^{\dot {\rm VII} } \,,  \non
	&& ( \rep{6}\,, \rep{1}\,, -\frac{1}{2} )_{ \mathbf{F}}  \oplus \Big[   \cancel{ ( \rep{1}\,, \rep{2}\,, +\frac{1}{2} )_{ \mathbf{F}} \oplus ( \rep{1}\,, \rep{1}\,, 0 )_{ \mathbf{F}} } \oplus    \bcancel{( \rep{1}\,, \repb{2}\,, +\frac{1}{2} )_{ \mathbf{F}}} \oplus ( \rep{1}\,, \rep{1}\,, +1 )_{ \mathbf{F}} \Big] \non
	&\oplus& \Big[ ( \rep{4}\,, \rep{2}\,, +\frac{1}{4} )_{ \mathbf{F}} \oplus \bcancel{ ( \rep{4}\,, \rep{1}\,, -\frac{1}{4} )_{ \mathbf{F}}^{} } \Big]  \oplus \cancel{ ( \rep{4 }\,, \rep{1}\,, -\frac{1}{4} )_{ \mathbf{F}}^{\prime\prime} }  \subset \rep{28_F}\,, \non
	&& \Big[ \bcancel{( \rep{1}\,, \repb{2}\,, +\frac{1}{2} )_{ \mathbf{F}}^{\prime \prime \prime } }\oplus  ( \rep{1}\,, \rep{1}\,, +1 )_{ \mathbf{F}}^{\prime}   \oplus  ( \rep{1}\,, \rep{1}\,, +1 )_{ \mathbf{F}}^{\prime\prime}  \Big] \oplus ( \repb{4}\,, \rep{1}\,, -\frac{3}{4} )_{ \mathbf{F}}  \non
	&\oplus& \Big[  ( \rep{4}\,, \rep{2}\,, +\frac{1}{4} )_{ \mathbf{F}}^\prime \oplus  \bcancel{ ( \rep{4}\,, \rep{1}\,, -\frac{1}{4} )_{ \mathbf{F}}^{\prime \prime  } }  \oplus  ( \rep{4}\,, \repb{2}\,, +\frac{1}{4} )_{ \mathbf{F}} \oplus   ( \rep{4}\,, \rep{1}\,, +\frac{3}{4} )_{ \mathbf{F}}^{}  \Big]  \non
	&\oplus& \Big[  ( \rep{6}\,, \rep{2}\,, 0 )_{ \mathbf{F}} \oplus   ( \rep{6}\,, \rep{1}\,, -\frac{1}{2} )_{ \mathbf{F}}^{\prime }  \oplus ( \rep{6}\,, \rep{1}\,, -\frac{1}{2})_{ \mathbf{F}}^{\prime\prime} \Big]  \subset \rep{56_F}\,.
\end{eqnarray}
We use the slashes and the back slashes to cross out massive fermions at the first and the second stages, respectively.
From the anomaly-free conditions of $[ {\rm SU}(4)_c]^2 \cdot {\rm U}(1)_{X_2}=0$, $\[ {\rm SU}(2)_W \]^2 \cdot {\rm U}(1)_{X_2}=0$, and $\[ {\rm U}(1)_{X_2} \]^3=0$, we find that one  $\repb{8_F}^{\rm V}$ and one $\repb{8_F}^{\dot {\rm VII}}$ are integrated out.

\subsection{The third stage}

\begin{table}[htp]
	\begin{center}
		\begin{tabular}{c|ccccc}
			\hline\hline
			$\repb{8_F}^\Omega$ & $( \repb{4}\,, \rep{1}\,, +\frac{1}{4} )_{\rep{F}}^\Omega$ & $( \rep{1}\,, \repb{2}\,,  -\frac{1}{2} )_{\rep{F}}^\Omega$  &  $( \rep{1}\,, \rep{1}\,,  0 )_{\rep{F}}^{\Omega^{ \prime} }$  &  $( \rep{1}\,, \rep{1}\,,  0 )_{\rep{F}}^{\Omega^{ \prime\prime} }$  &  \\[1mm]
			\hline
			$\Tc^{ \prime \prime \prime }$  &  $-4t$ & $0$  &  $- 4 t$  & $ -4 t$  &   \\[1mm]
			\hline
			$\rep{28_F}$ & $( \rep{6}\,, \rep{1}\,, - \frac{1}{2} )_{\rep{F}}$  &  $( \rep{1}\,, \rep{1}\,,  +1  )_{\rep{F}}$  &  $( \rep{4}\,, \rep{2 }\,,  +\frac{1}{4} )_{\rep{F}}$   &   & \\[1mm]
			\hline
			$\Tc^{ \prime \prime \prime }$  & $+4t$  &  $-4t$  &  $0$ &  &  \\[1mm]
			\hline
			$\rep{56_F}$  & $( \rep{1}\,,  \rep{1}\,, +1 )_{\rep{F}}^\prime$ &  $( \rep{1}\,,  \rep{1}\,, +1 )_{\rep{F}}^{\prime\prime}$  &  $( \repb{4}\,, \rep{1 }\,,  -\frac{3}{4} )_{\rep{F}}$  &   $( \rep{4}\,, \rep{2 }\,,  +\frac{1}{4} )_{\rep{F}}^\prime  $  &  \\[1mm]
			\hline
			$\Tc^{ \prime \prime \prime }$  &  $ -4 t$  &  $  -4t$ &  $+4t$ & $0$  &  \\[1mm]
			\hline
			& $( \rep{4}\,, \repb{2 }\,,  +\frac{1}{4} )_{\rep{F}}$ & $( \rep{4}\,, \rep{1}\,,  +\frac{3}{4} )_{\rep{F}}$ &  $( \rep{6}\,,  \rep{2}\,, 0 )_{\rep{F}} $  & $( \rep{6 }\,,   \rep{1}\,,  -\frac{1}{2 } )_{\rep{F}}^{\prime}$   &  $( \rep{6}\,, \rep{1}\,,  -\frac{1}{2} )_{\rep{F}}^{\prime\prime}$  \\[1mm]
			\hline
			$\Tc^{ \prime \prime \prime }$  &  $0$  &  $  -4 t$  &  $0$  &  $+4t$  &  $+4t$ \\[1mm]
			\hline\hline
			$\repb{8_H}_{ \,, \omega }$   &  $( \rep{1}\,, \repb{2}\,,  - \frac{1}{2} )_{\rep{H}\,,\omega }$  & $( \repb{4}\,, \rep{1}\,,  + \frac{1}{4} )_{\rep{H}\,,\omega }$  &   &   &   \\[1mm]
			\hline
			$\Tc^{ \prime \prime \prime }$  &  $+4t$  & $0$  &  &  &   \\[1mm]
			\hline
			$\repb{28_H}_{ \,, \dot \omega }$  &  $( \repb{4}\,,  \rep{1}\,, +\frac{1}{ 4} )_{\rep{H}\,, \dot \omega }$  & $( \repb{4}\,,  \repb{2}\,, -\frac{1 }{ 4} )_{\rep{H}\,, \dot \omega }$   &  $( \repb{4}\,,  \rep{1}\,, +\frac{1 }{ 4} )_{\rep{H}\,, \dot \omega }^\prime$  &  $( \rep{1}\,,  \repb{2}\,, -\frac{1}{ 2} )_{\rep{H}\,, \dot \omega }$ & $( \rep{1}\,,  \rep{2}\,, -\frac{1 }{ 2} )_{\rep{H}\,, \dot \omega }$   \\[1mm]
			\hline
			$\Tc^{ \prime \prime \prime }$  & $0$  & $+4t$  & $0$ & $+4t$  & $+4t$  \\[1mm]
			\hline
			$\rep{70_H}$  &  $( \rep{4}\,, \repb{2}\,,  + \frac{1}{4} )_{\rep{H}}^{}$  &  $( \repb{4}\,, \rep{2}\,,  - \frac{1}{4} )_{\rep{H}}^{} $  &  &  &  \\[1mm]
			\hline
			$\Tc^{ \prime \prime \prime }$  &   $-4t$ &   $-4t$ &  &  &   \\[1mm]
			\hline\hline
		\end{tabular}
	\end{center}
	\caption{The $\widetilde{ {\rm U}}(1)_{T^{\prime \prime \prime } }$ charges for massless fermions and possible symmetry breaking Higgs components in the $\Gc_{421}$ theory according to Eq.~\eqref{eq:U1T_defC}.
	}
	\label{tab:CG421_Tcharges}
\end{table}

\para
The third symmetry breaking stage of $\Gc_{ 421} \to \Gc_{\rm SM}$ can be achieved by Higgs fields of $( \repb{4}\,,   \rep{1}\,,  + \frac{1}{4 } )_{\rep{H}\,,\rm{VI}\,,3  }\subset \repb{8_H}_{ \,,\rm{VI}\,,3 }$ and $\[ (   \repb{4}\,, \rep{1}\,, +\frac{1}{ 4 } )_{\rep{H}\,, \dot{\rm{VIII}} \,,\dot{2} }  \oplus (  \repb{4}\,,   \rep{1}\,,  + \frac{1}{ 4} )_{\rep{H}\,, \dot{\rm{IX} } }^\prime  \] \subset (   \repb{4}\,, \repb{4}\,, 0 )_{\rep{H}\,, \dot{\rm{VIII}} \,,\dot{2}\,,\dot{\rm{IX}} } \subset \repb{28_H}_{ \,, \dot{\rm{VIII}} \,,\dot{2}\,,\dot{\rm{IX}}  }$, according to the decompositions in Eqs.~\eqref{eq:SU8C_Higgs_Br01} and \eqref{eq:SU8C_Higgs_Br02}, as well as their $\widetilde{ {\rm U}}(1)_{T^{\prime \prime \prime } }$ charges in Table~\ref{tab:CG421_Tcharges}.
Accordingly, the term of $  Y_\Bc \repb{8_F}^{\rm VI \,,3 } \rep{28_F} \repb{8_H}_{\,,\rm VI\,,3 } + {\rm H.c.}$ leads to the vectorial masses of  $(\DG \,, \eG^{\prime \prime}\,,  \nG^{\prime \prime} )$, the term of $ Y_\Dc \repb{8_F}^{\dot{\rm{VIII}} \,,\dot{2}} \rep{56_F} \repb{28_H}_{\,,\dot{\rm{VIII}} \,,\dot{2} } + {\rm H.c.}$ leads to the vectorial masses of $( \DG^{\prime \prime \prime \prime } \,, \eG^{\prime\prime\prime \prime \prime} \,, \nG^{\prime\prime\prime \prime\prime }    )$, the term of  $ Y_\Dc \repb{8_F}^{\dot{\rm{IX}}} \rep{56_F} \repb{28_H}_{\,,\dot{\rm{IX}} } + {\rm H.c.}$ leads to the vectorial masses of $(  \DG^{\prime \prime \prime\prime\prime} \,, \eG^{\prime\prime\prime\prime }   \,, \nG^{\prime\prime\prime\prime } )$, and the term of $\frac{c_4}{M_{\rm{pl}}}\rep{56_F}\rep{56_F} \langle  \rep{63_H} \rangle \repb{28_H}_{\,,\dot{\rm{IX}}}^{\dag} + {\rm H.c.}$ leads to the vectorial masses of  $( \EG\,,\dG\,, \uG\,,\UG   )$, respectively.
The remaining massless fermions of the $\Gc_{\rm SM}$ are listed as follows:
\begin{eqnarray} \label{eq:321C_fermions}
	&& \Big[ ( \repb{3}\,, \rep{1}\,, +\frac{1}{3} )_{ \mathbf{F}}^\Omega \oplus ( \rep{1}\,, \rep{1}\,, 0 )_{ \mathbf{F}}^\Omega \Big]  \oplus \Big[ ( \rep{1}\,, \repb{2}\,, -\frac{1}{2} )_{ \mathbf{F}}^\Omega \oplus ( \rep{1}\,, \rep{1}\,, 0 )_{ \mathbf{F}}^{\Omega^{\prime}} \oplus ( \rep{1}\,, \rep{1}\,, 0 )_{ \mathbf{F}}^{\Omega^{\prime\prime}}   \Big]  \subset \repb{8_F}^\Omega \,, \quad \Omega = ( \dot 1\,, \dot 2\,, 3 ) \,, \non
	&& ( \rep{1}\,, \rep{1}\,, 0)_{ \mathbf{F}}^{{\Omega} }  \subset \repb{8_F}^\Omega \,, \quad \Omega = ( \rm VI\,, \dot{\rm VIII} \,, \dot{\rm IX} ) \,, \non
	&& ( \rep{1}\,, \rep{1}\,, 0)_{ \mathbf{F}}^{{\Omega}^\prime } \subset \repb{8_F}^\Omega \,, \quad \Omega = ( \rm V\,,\rm VI\,, \dot{\rm VII} \,,\dot{\rm VIII} \,, \dot{\rm IX}) \,, \non
	&& ( \rep{1}\,, \rep{1}\,, 0)_{ \mathbf{F}}^{{\Omega}^{ \prime \prime} } \subset \repb{8_F}^\Omega \,, \quad \Omega = (\rm IV \,,  \rm V\,,\rm VI\,, \dot{\rm VII} \,,\dot{\rm VIII} \,, \dot{\rm IX} ) \,, \non
	&& \Big[ \xcancel{( \rep{3}\,, \rep{1}\,, -\frac{1}{3} )_{ \mathbf{F}}} \oplus ( \repb{3}\,, \rep{1}\,, -\frac{2}{3} )_{ \mathbf{F}}  \Big] \oplus \Big[    \cancel{ ( \rep{1}\,, \rep{2}\,, +\frac{1}{2} )_{ \mathbf{F}} \oplus ( \rep{1}\,, \rep{1}\,, 0 )_{ \mathbf{F}} }  \oplus   \bcancel{( \rep{1}\,, \repb{2}\,, +\frac{1}{2} )_{ \mathbf{F}}^{\prime}} \oplus ( \rep{1}\,, \rep{1}\,, +1 )_{ \mathbf{F}} \Big]\non
	&&\oplus \Big[  ( \rep{3}\,, \rep{2}\,, +\frac{1}{6} )_{ \mathbf{F}} \oplus \xcancel{( \rep{1}\,, \rep{2}\,, +\frac{1}{2} )_{ \mathbf{F}}^{\prime \prime}}  \oplus \bcancel{ ( \rep{3}\,, \rep{1}\,, -\frac{1}{3} )_{ \mathbf{F}}^{\prime} \oplus  ( \rep{1}\,, \rep{1}\,,0 )_{ \mathbf{F}}^{\prime} } \Big]  \oplus \Big[ \cancel{ ( \rep{3 }\,, \rep{1}\,, -\frac{1}{3} )_{ \mathbf{F}}^{\prime\prime} \oplus  ( \rep{1}\,, \rep{1}\,, 0 )_{ \mathbf{F}}^{\prime\prime} } \Big] \subset \rep{28_F}\,, \non
	&&\Big[  \bcancel{( \rep{1}\,, \repb{2}\,, +\frac{1}{2} )_{ \mathbf{F}}^{\prime\prime\prime} }\oplus  ( \rep{1}\,, \rep{1}\,, +1 )_{ \mathbf{F}}^{\prime}  \oplus \xcancel{ ( \rep{1}\,, \rep{1}\,, +1 )_{ \mathbf{F}}^{\prime\prime} } \Big] \oplus \Big[ ( \repb{3}\,, \rep{1}\,, -\frac{2}{3} )_{ \mathbf{F}}^{\prime} \oplus \xcancel{ ( \rep{1}\,, \rep{1}\,, -1 )_{ \mathbf{F}} } \Big] \non
	&&\oplus \Big[ ( \rep{3}\,, \rep{2}\,, +\frac{1}{6} )_{ \mathbf{F}}^\prime \oplus \xcancel{ ( \rep{1}\,, \rep{2}\,, +\frac{1}{2} )_{ \mathbf{F}}^{\prime \prime \prime \prime} } \oplus  \bcancel{ ( \rep{3}\,, \rep{1}\,, -\frac{1}{3} )_{ \mathbf{F}}^{\prime \prime \prime} \oplus ( \rep{1}\,, \rep{1}\,, 0 )_{ \mathbf{F}}^{\prime \prime \prime} }  \non
	&&\oplus \xcancel{( \rep{3}\,, \repb{2}\,, +\frac{1}{6} )_{ \mathbf{F}}^{\prime \prime}} \oplus \xcancel{( \rep{1}\,, \repb{2}\,, +\frac{1}{2} )_{ \mathbf{F}}^{\prime\prime\prime\prime\prime}} \oplus  \xcancel{ ( \rep{3}\,, \rep{1}\,, +\frac{2}{3} )_{ \mathbf{F}}^{}} \oplus ( \rep{1}\,, \rep{1}\,, +1 )_{ \mathbf{F}}^{\prime\prime\prime}  \Big] \non
	&&\oplus \Big[  ( \rep{3}\,, \rep{2}\,, +\frac{1}{6} )_{ \mathbf{F}}^{\prime \prime \prime} \oplus \xcancel{( \repb{3}\,, \rep{2}\,, -\frac{1}{6} )_{ \mathbf{F}} } \oplus  \xcancel{ ( \rep{3}\,, \rep{1}\,, -\frac{1}{3} )_{ \mathbf{F}}^{\prime\prime\prime\prime} } \oplus \xcancel{ ( \repb{3}\,, \rep{1}\,, -\frac{2}{3} )_{ \mathbf{F}}^{\prime\prime} }  \non
	&&\oplus  \xcancel{ ( \rep{3}\,, \rep{1}\,, -\frac{1}{3} )_{ \mathbf{F}}^{\prime\prime\prime\prime\prime} } \oplus  ( \repb{3}\,, \rep{1}\,, -\frac{2}{3} )_{ \mathbf{F}}^{\prime\prime\prime}  \Big]  \subset \rep{56_F}\,.
\end{eqnarray}
The fermions that become massive at this stage are further crossed out.
After this stage of symmetry breaking, there are three-generational massless SM fermions together with 23 left-handed massless sterile neutrinos.
The third-generational SM fermions are from the rank-two chiral IRAFFS of $\[ \repb{8_F}^\omega \oplus \rep{28_F} \]$, while the first- and second-generational SM fermions are from the rank-three chiral IRAFFS of $\[ \repb{8_F}^{\dot \omega }\oplus \rep{56_F}\]$.

\subsection{The $d=5$ bilinear fermion operators}

\para
We proceed to analyze the $d=5$ bilinear fermion operators in Eqs.~\eqref{eqs:d5_ffHH} along the WWS symmetry breaking pattern.
To be brief, we only include the biliner mass terms involving the SM quarks and leptons. 
The operator of $\Oc_\Fc^{ ( 3\,, 2) }$ is decomposed as
\begin{eqnarray}\label{eq:WWS_OF_32b}
	&&\frac{c_3}{M_{\rm{pl}}}\repb{8_F}^{\dot \omega}\rep{56_F}\cdot \repb{28_H}_{,\dot \kappa }^{\dag} \cdot \rep{70_H}^\dag+ H.c. \non
	&\supset&\frac{c_3}{M_{\rm{pl}}}\Big[ (\repb{4}\,,\rep{1}\,,+\frac{1}{4})_{\mathbf{F}}^{\dot \omega}\otimes(\rep{4}\,,\rep{6}\,,+\frac{1}{4})_{\mathbf{F}} \oplus (\rep{1}\,,\repb{4}\,,-\frac{1}{4})_{\mathbf{F}}^{\dot \omega} \otimes(\rep{1}\,,\repb{4}\,,+\frac{3}{4})_{\mathbf{F}} \Big]\non
	&\otimes&(\repb{4}\,,\repb{4}\,,0)_{\mathbf{H},\dot \kappa}^\dag \otimes (\rep{4}\,,\repb{4}\,,+\frac{1}{2})_{\mathbf{H}}^{\dag}+H.c.\non
	&\supset&\frac{c_3}{M_{\rm{pl}}}\Big[ (\repb{4}\,,\rep{1}\,,+\frac{1}{4})_{\mathbf{F}}^{\dot \omega}\otimes(\rep{4}\,,\rep{3}\,,+\frac{1}{12})_{\mathbf{F}}^{\prime} \oplus (\rep{1}\,,\repb{3}\,,-\frac{1}{3})_{\mathbf{F}}^{\dot \omega} \otimes(\rep{1}\,,\repb{3}\,,+\frac{2}{3})_{\mathbf{F}}^\prime \Big]\non
	&\otimes&(\repb{4}\,,\repb{3}\,,-\frac{1}{12})_{\mathbf{H} ,\dot \kappa}^\dag \otimes (\rep{4}\,,\repb{3}\,,+\frac{5}{12})_{\mathbf{H}}^{\dag}+H.c.\non
	&\supset&\frac{c_3}{M_{\rm{pl}}}\Big[ (\repb{4}\,,\rep{1}\,,+\frac{1}{4})_{\mathbf{F}}^{\dot \omega}\otimes(\rep{4}\,,\rep{2}\,,+\frac{1}{4})_{\mathbf{F}}^{\prime} \oplus(\rep{1}\,,\repb{2}\,,-\frac{1}{2})_{\mathbf{F}}^{\dot \omega} \otimes(\rep{1}\,,\rep{1}\,,+1)_{\mathbf{F}}^\prime  \oplus(\rep{1}\,,\rep{1}\,,0)_{\mathbf{F}}^{\dot \omega^\prime} \otimes(\rep{1}\,,\repb{2}\,,+\frac{1}{2})_{\mathbf{F}}^{\prime\prime\prime} \Big]\non
	&\otimes&\langle  (\repb{4}\,,\rep{1}\,,-\frac{1}{4})_{\mathbf{H} ,\dot \kappa}^\dag \rangle  \otimes \langle(\rep{4}\,,\repb{2}\,,+\frac{1}{4})_{\mathbf{H}}^{\dag}\rangle+H.c.\non
	%
	%
	%
	&\Rightarrow&\frac{c_3}{2}\dot{\zeta_3}(   \underline{d_L  {\Dc_R^{\dot\omega}  }^c } +\check{\Nc}_L^{\dot\omega}   {   \nG_R^{\prime\prime\prime\prime}  }^c-\underline{\Ec_L^{\dot\omega} {e_R}^c} +\check{\Nc}_L^{\dot{\omega}^\prime}{\nG_R^{\prime\prime\prime }  }^c  )  v_{\rm EW}+H.c.\,,
\end{eqnarray}
and
\begin{eqnarray}\label{eq:WWS_OF_32c}
	&&\frac{c_3}{M_{\rm{pl}}}\repb{8_F}^{\dot \omega}\rep{56_F}\cdot  \repb{28_H}_{,\dot \kappa }^{\dag}  \cdot  \rep{70_H}^\dag+ H.c. \non
	&\supset&\frac{c_3}{M_{\rm{pl}}}\Big[ (\repb{4}\,,\rep{1}\,,+\frac{1}{4})_{\mathbf{F}}^{\dot \omega}\otimes(\rep{6}\,,\rep{4}\,,-\frac{1}{4})_{\mathbf{F}} \oplus    (\rep{1}\,,\repb{4}\,,-\frac{1}{4})_{\mathbf{F}}^{\dot \omega}\otimes(\rep{4}\,,\rep{6}\,,+\frac{1}{4})_{\mathbf{F}}     \Big] \non
	&\otimes&(\rep{1}\,,\rep{6}\,,-\frac{1}{2})_{\mathbf{H} ,\dot \kappa}^\dag \otimes (\rep{4}\,,\repb{4}\,,+\frac{1}{2})_{\mathbf{H}}^{\dag}+H.c.\non
	&\supset&\frac{c_3}{M_{\rm{pl}}}\Big[ (\repb{4}\,,\rep{1}\,,+\frac{1}{4})_{\mathbf{F}}^{\dot \omega}\otimes(\rep{6}\,,\rep{3}\,,-\frac{1}{6})_{\mathbf{F}} \oplus    (\rep{1}\,,\repb{3}\,,-\frac{1}{3})_{\mathbf{F}}^{\dot \omega}\otimes(\rep{4}\,,\repb{3}\,,+\frac{5}{12})_{\mathbf{F}}   \oplus    (\rep{1}\,,\rep{1}\,,0)_{\mathbf{F}}^{\dot \omega^{\prime\prime}}\otimes(\rep{4}\,,\rep{3}\,,+\frac{1}{12})_{\mathbf{F}}\Big] \non
	&\otimes&\langle  (\rep{1}\,,\repb{3}\,,-\frac{1}{3})_{\mathbf{H},\dot \kappa}^\dag \rangle  \otimes (\rep{4}\,,\repb{3}\,,+\frac{5}{12})_{\mathbf{H}}^{\dag}+H.c.\non
	&\supset&c_3\frac{w_{\repb{3},\dot{\rm VII}}}{\sqrt{2}M_{\rm{pl}}}\Big[ (\repb{4}\,,\rep{1}\,,+\frac{1}{4})_{\mathbf{F}}^{\dot \omega}\otimes(\rep{6}\,,\rep{2}\,,0 )_{\mathbf{F}} \oplus    (\rep{1}\,,\repb{2}\,,-\frac{1}{2})_{\mathbf{F}}^{\dot \omega}\otimes(\rep{1}\,,\rep{1}\,,+1)_{\mathbf{F}}^{\prime\prime\prime}   \oplus    (\rep{1}\,,\rep{1}\,,0)_{\mathbf{F}}^{\dot \omega^{\prime}}\otimes(\rep{4}\,,\repb{2}\,,+\frac{1}{4})_{\mathbf{F}}^{\prime\prime\prime\prime\prime}  \non
	&\oplus&  (\rep{1}\,,\rep{1}\,,0)_{\mathbf{F}}^{\dot \omega^{\prime\prime}}\otimes(\rep{1}\,,\rep{2}\,,+\frac{1}{2})_{\mathbf{F}}^{\prime\prime\prime\prime}\Big] \otimes  \langle(\rep{4}\,,\repb{2}\,,+\frac{1}{4})_{\mathbf{H}}^{\dag}\rangle +H.c.\non
	%
	%
	%
	&\Rightarrow&\frac{c_3}{2}\dot{\zeta_2} (\underline{   s_L  {\Dc_R^{\dot\omega}}^2   } -\underline{\Ec_L^{\dot\omega} {\mu_R}^c}+\check{\Nc}_L^{\dot\omega^\prime}  { \nG_R^{\prime\prime\prime\prime\prime}   }^c+\check{\Nc}_L^{\dot\omega^{\prime\prime}}  {\nG_R^{\prime\prime\prime\prime}}^c  )v_{\rm EW}  +H.c. \,.
\end{eqnarray}
By taking the possible flavor indices of $\dot \omega = \dot 1\,, \dot 2$ in mass terms from Eqs.~\eqref{eq:WWS_OF_32b} and \eqref{eq:WWS_OF_32c}, one finds the following set of mass matrices of the $(d\,, s)$ and $(e\,, \mu)$:
\beqs
\beqn
&&\Big( \Mc_d \Big)_{2 \times 2}^{\rm direct} = \frac{ c_3 }{2 }  \left( \ba{cc}  
\dot \zeta_3 &  \dot \zeta_3   \\
\dot \zeta_2  & \dot \zeta_2  \\  \ea  \right)   v_{\rm EW} \,, \label{eq:C_ds_direct}\\[1mm]
&& \Big( \Mc_e \Big)_{2 \times 2}^{\rm direct} = - \frac{ c_3 }{2 } \left( \ba{cc}  
\dot \zeta_3 &  \dot \zeta_2   \\
\dot \zeta_3  & \dot \zeta_2  \\  \ea  \right)  v_{\rm EW}  \,,\label{eq:C_emu_direct}
\eeqn
\eeqs
which leave the down quark and electron massless.

\para
For the operator of $\Oc_\Fc^{ (4\,,1)}$, it is decomposed as
\begin{eqnarray}\label{eq:WWS_OF_41a}
	&&\frac{c_4}{M_{\rm{pl}}}\rep{56_F}\rep{56_F}  \cdot \repb{28_H}_{,\dot \omega }   \cdot \rep{70_H} + H.c. \non
	&\supset& \frac{c_4}{M_{\rm{pl}}}\Big[ (\repb{4}\,,\rep{1}\,,-\frac{3}{4})_{\mathbf{F}}\otimes(\rep{4}\,,\rep{6}\,,+\frac{1}{4})_{\mathbf{F}}  \oplus \cancel{ (\rep{6}\,,\rep{4}\,,-\frac{1}{4})_{\mathbf{F}} \otimes (\rep{6}\,,\rep{4}\,,-\frac{1}{4})_{\mathbf{F}} }    \Big] \otimes (\repb{4}\,,\repb{4}\,,0)_{\mathbf{H} ,\dot \omega} \otimes (\rep{4}\,,\repb{4}\,,+\frac{1}{2})_{\mathbf{H}}+H.c. \non
	&\supset&\frac{c_4}{M_{\rm{pl}}}\Big[ (\repb{4}\,,\rep{1}\,,-\frac{3}{4})_{\mathbf{F}} \otimes (\rep{4}\,,\rep{3}\,,+\frac{1}{12})_{\mathbf{F}}^{\prime} \Big] \otimes  (\repb{4}\,,\rep{1}\,,+\frac{1}{4})_{\mathbf{H},\dot \omega}^{\prime}  \otimes (\rep{4}\,,\repb{3}\,,+\frac{5}{12})_{\mathbf{H}}+H.c.\non
	&\supset&  \frac{c_4}{M_{\rm{pl}}} \Big[ (\repb{4}\,,\rep{1}\,,-\frac{3}{4})_{\mathbf{F}} \otimes (\rep{4}\,,\rep{2}\,,+\frac{1}{4})_{\mathbf{F}}^{\prime} \Big] \otimes \langle  (\repb{4}\,,\rep{1}\,,+\frac{1}{4})_{\mathbf{H},\dot \omega}^{\prime} \rangle \otimes \langle(\rep{4}\,,\repb{2}\,,+\frac{1}{4})_{\mathbf{H}}\rangle+ H.c. \non
	&\Rightarrow& \frac{c_4}{2}  \dot{\zeta_3}^\prime ( \underline{u_L {c_R}^c} + \EG_L {\eG_R^{\prime\prime\prime\prime } }^c )  v_{\rm EW} + H.c. \, .
\end{eqnarray}

\para
For the operator of $\Oc_\Fc^{ (5\,,1)}$, it is decomposed as
\begin{eqnarray}\label{eq:WWS_OF_51a}
	&&\frac{c_5}{M_{\rm{pl}}}\rep{28_F}\rep{56_F}\cdot \repb{8_H}_{,\dot \omega } \cdot \rep{70_H}+H.c. \,\non
	&\supset&\frac{c_5}{M_{\rm{pl}}}\Big[(\rep{6}\,,\rep{1}\,,-\frac{1}{2})_\mathbf{F}\otimes(\rep{4}\,,\rep{6}\,,+\frac{1}{4})_\mathbf{F} \oplus (\rep{1}\,,\rep{6}\,,+\frac{1}{2})_\mathbf{F}\otimes(\rep{1}\,,\repb{4}\,,-\frac{3}{4})_\mathbf{F}\oplus(\rep{4}\,,\rep{4}\,,0)_\mathbf{F}\otimes(\rep{6}\,,\rep{4}\,,-\frac{1}{4})_\mathbf{F}\Big]\non
	&\otimes&(\rep{1}\,,\repb{4}\,,-\frac{1}{4})_{\mathbf{H},\dot{\omega}} \otimes(\rep{4}\,,\repb{4}\,,+\frac{1}{2})_\mathbf{H}\non
	&\supset&c_5\frac{W_{\repb{4},\rm{IV}}}{\sqrt{2}M_{\rm{pl}}}\Big[(\rep{6}\,,\rep{1}\,,-\frac{1}{2})_\mathbf{F}\otimes(\rep{4}\,,\rep{3}\,,+\frac{1}{12})_\mathbf{F}^{\prime} \oplus (\rep{1}\,,\rep{3}\,,+\frac{1}{3})_\mathbf{F}\otimes(\repb{4}\,,\rep{1}\,,-\frac{3}{4})_\mathbf{F}\oplus(\rep{4}\,,\rep{3}\,,+\frac{1}{12})_\mathbf{F}\otimes(\rep{6}\,,\rep{1}\,,-\frac{1}{2})_\mathbf{F}^{\prime\prime} \non
	&\oplus&  (\rep{4}\,,\rep{1}\,,-\frac{1}{4})_\mathbf{F} \otimes (\rep{6}\,,\rep{3}\,,-\frac{1}{6})_\mathbf{F} \Big] \otimes(\rep{4}\,,\repb{3}\,,+\frac{5}{12})_\mathbf{H}\non
	&\supset&c_5\frac{W_{\repb{4},\rm{IV}}}{\sqrt{2}M_{\rm{pl}}}\Big[(\rep{6}\,,\rep{1}\,,-\frac{1}{2})_\mathbf{F}\otimes(\rep{4}\,,\rep{2}\,,+\frac{1}{4})_\mathbf{F}^{\prime} \oplus (\rep{1}\,,\rep{2}\,,+\frac{1}{2})_\mathbf{F}\otimes(\repb{4}\,,\rep{1}\,,-\frac{3}{4})_\mathbf{F}\oplus(\rep{4}\,,\rep{2}\,,+\frac{1}{4})_\mathbf{F}\otimes(\rep{6}\,,\rep{1}\,,-\frac{1}{2})_\mathbf{F}^{\prime\prime} \non
	&\oplus&  (\rep{4}\,,\rep{1}\,,-\frac{1}{4})_\mathbf{F} \otimes (\rep{6}\,,\rep{2}\,,0)_\mathbf{F} \Big] \otimes\langle(\rep{4}\,,\repb{2}\,,+\frac{1}{4})_\mathbf{H}\rangle\non
	%
	%
	%
	&\Rightarrow&\frac{ c_5 }{2} \zeta_1    (   \underline{ u_L {t_R}^c} + \EG_L {\eG_R }^c +\underline{t_L  {u_R}^c   } + \DG_L^{\prime \prime } {\dG_R}^c )  v_{\rm EW} + H.c. \, ,
\end{eqnarray}
and
\begin{eqnarray}\label{eq:WWS_OF_51b}
	&&\frac{c_5}{M_{\rm{pl}}}\rep{28_F}\rep{56_F}\cdot\repb{8_H}_{,\dot \omega }\cdot \rep{70_H}+H.c.\non
	&\supset&\frac{c_5}{M_{\rm{pl}}}\Big[  (\rep{6}\,,\rep{1}\,,-\frac{1}{2})_\mathbf{F}\otimes (\rep{6}\,,\rep{4}\,,-\frac{1}{4})_\mathbf{F} \oplus (\rep{4}\,,\rep{4}\,,0)_\mathbf{F} \otimes (\repb{4}\,,\rep{1}\,,-\frac{3}{4})_\mathbf{F}   \Big]\otimes (\repb{4}\,,\rep{1}\,,+\frac{1}{4})_{\mathbf{H},\dot{\omega}} \otimes  (\rep{4}\,,\repb{4}\,,+\frac{1}{2})_{\mathbf{H}} \non
	&\supset&\frac{c_5}{M_{\rm{pl}}}\Big[  (\rep{6}\,,\rep{1}\,,-\frac{1}{2})_\mathbf{F}\otimes (\rep{6}\,,\rep{3}\,,-\frac{1}{6})_\mathbf{F} \oplus (\rep{4}\,,\rep{3}\,,+\frac{1}{12})_\mathbf{F} \otimes (\repb{4}\,,\rep{1}\,,-\frac{3}{4})_\mathbf{F}   \Big]\otimes (\repb{4}\,,\rep{1}\,,+\frac{1}{4})_{\mathbf{H},\dot{\omega}}  \otimes  (\rep{4}\,,\repb{3}\,,+\frac{5}{12})_{\mathbf{H}} \non
	&\supset&\frac{c_5}{M_{\rm{pl}}}\Big[  (\rep{6}\,,\rep{1}\,,-\frac{1}{2})_\mathbf{F}\otimes (\rep{6}\,,\rep{2}\,,0)_\mathbf{F} \oplus (\rep{4}\,,\rep{2}\,,+\frac{1}{4})_\mathbf{F} \otimes (\repb{4}\,,\rep{1}\,,-\frac{3}{4})_\mathbf{F}   \Big]\otimes \langle(\repb{4}\,,\rep{1}\,,+\frac{1}{4})_{\mathbf{H},\dot{\omega}} \rangle \otimes  \langle(\rep{4}\,,\repb{2}\,,+\frac{1}{4})_{\mathbf{H}}\rangle \non
	%
	%
	%
	&\Rightarrow&\frac{c_5}{2} \zeta_3 (  \DG_L {\dG_R}^c +\underline{c_L {t_R}^c}+\underline{t_L {c_R}^c} + \EG_L  {\eG_R}^c   ) v_{\rm EW}+H.c. \, .
\end{eqnarray}
%
%

\subsection{The $d=5$ irreducible Higgs mixing operators}

\para
For the Yukawa coupling of $\repb{8_F}^{\omega_1}  \rep{28_F} \repb{8_H}_{\,,\omega_1} $, we find the mass terms of
\begin{eqnarray} \label{eq:WWS_btau_indirect}
	&&Y_\Bc \repb{8_F}^{\omega_1}  \rep{28_F} \repb{8_H}_{\,,\omega_1} \times   \frac{d_{\mathscr{A}}}{M_{\rm{pl}}} \epsilon_{\omega_1 \omega_2\omega_3\omega_4 }  \repb{8_H}_{\,,\omega_1}^\dag  \repb{8_H}_{\,,\omega_2}^\dag  \repb{8_H}_{\,,\omega_3}^\dag   \repb{8_H}_{\,,\omega_4}^\dag   \rep{70_H}^{\dag}+H.c. \non
	&\supset& Y_\Bc\left[ (\repb{4 }\,,\rep{1}\,,+\frac{1}{4})_\mathbf{F}^{\omega_1} \otimes (\rep{ 4}\,,\rep{4}\,,0)_\mathbf{F} \oplus  (\rep{1}\,,\repb{4}\,,-\frac{1}{4})_\mathbf{F}^{\omega_1}\otimes  (\rep{1}\,,\rep{6}\,,+\frac{1}{2})_\mathbf{F} \right] \otimes  (\rep{1}\,,\repb{4}\,,-\frac{1}{4})_{\mathbf{H}\,,\omega_1} \non
	&\times& \frac{d_{\mathscr{A}}}{M_{\rm{pl}}}  (\rep{1}\,,\repb{4}\,,-\frac{1}{4})_{\mathbf{H}\,,\omega_1}^\dag \otimes(\rep{1}\,,\repb{4}\,,-\frac{1}{4})_{\mathbf{H}\,,\omega_2}^\dag \otimes(\rep{1}\,,\repb{4}\,,-\frac{1}{4})_{\mathbf{H}\,,\omega_3}^\dag\otimes (\repb{4}\,,\rep{1}\,,+\frac{1}{4})_{\mathbf{H}\,,\omega_4}^\dag \otimes (\rep{4}\,,\repb{4}\,,+\frac{1}{2})_\mathbf{H}^\dag+H.c. \non
	&\supset &Y_\Bc\left[ (\repb{4 }\,,\rep{1}\,,+\frac{1}{4})_\mathbf{F}^{\omega_1} \otimes (\rep{ 4}\,,\rep{3}\,,+\frac{1}{12})_\mathbf{F} \oplus  (\rep{1}\,,\repb{3}\,,-\frac{1}{3})_\mathbf{F}^{\omega_1}\otimes  (\rep{1}\,,\rep{3}\,,+\frac{2}{3})_\mathbf{F} \oplus(\rep{1}\,,\rep{1}\,,0)_\mathbf{F}^{\omega_1^{\prime\prime}}\otimes  (\rep{1}\,,\rep{3}\,,+\frac{1}{3})_\mathbf{F}  \right] \non
	&\otimes&  (\rep{1}\,,\repb{3}\,,-\frac{1}{3})_{\mathbf{H}\,,\omega_1} \times \frac{d_{\mathscr{A}}W_{\repb{4}\,\rm{IV}}}{\sqrt{2}M_{\rm{pl}}}  (\rep{1}\,,\repb{3}\,,-\frac{1}{3})_{\mathbf{H}\,,\omega_1}^\dag  \otimes(\rep{1}\,,\repb{3}\,,-\frac{1}{3})_{\mathbf{H}\,,\omega_3}^\dag\otimes (\repb{4}\,,\rep{1}\,,+\frac{1}{4})_{\mathbf{H}\,,\omega_4}^\dag  \otimes (\rep{4}\,,\repb{3}\,,+\frac{5}{12})_\mathbf{H}^\dag + H.c. \non
	&\supset &Y_\Bc\Big[ (\repb{4}\,,\rep{1}\,,+\frac{1}{4})_\mathbf{F}^{\omega_1} \otimes (\rep{4}\,,\rep{2}\,,+\frac{1}{4})_\mathbf{F} \oplus(\rep{1}\,,\repb{2}\,,-\frac{1}{2})_\mathbf{F}^{\omega_1} \otimes (\rep{1}\,,\rep{1}\,,+1)_\mathbf{F}^{} \oplus  (\rep{1}\,,\rep{1}\,,0)_\mathbf{F}^{\omega_1^{\prime}} \otimes  (\rep{1}\,,\repb{2}\,,+\frac{1}{2})_\mathbf{F}^{\prime} \non
	& \oplus&  (\rep{1}\,,\rep{1}\,,0)_\mathbf{F}^{\omega_1^{\prime\prime}}\otimes  (\rep{1}\,,\rep{2}\,,+\frac{1}{2})_\mathbf{F}  \Big] \otimes (\rep{1}\,,\repb{2}\,,-\frac{1}{2})_{\mathbf{H}\,,\omega_1}  \non
	&\times& \frac{d_{\mathscr{A}}W_{\repb{4}\,\rm{IV}} w_{\repb{3},\rm{V}}}{2M_{\rm{pl}}}  (\rep{1}\,,\repb{2}\,,-\frac{1}{2})_{\mathbf{H}\,,\omega_1}^\dag  \otimes(\repb{4}\,,\rep{1}\,,+\frac{1}{4})_{\mathbf{H}\,,\omega_4}^\dag  \otimes (\rep{4}\,,\repb{2}\,,+\frac{1}{4})_\mathbf{H}^\dag+H.c. \non
	&\supset &Y_\Bc\Big[ (\repb{3}\,,\rep{1}\,,+\frac{1}{3})_\mathbf{F}^{\omega_1} \otimes (\rep{3}\,,\rep{2}\,,+\frac{1}{6})_\mathbf{F} \oplus (\rep{1}\,,\rep{1}\,,0)_\mathbf{F}^{\omega_1}  \otimes  (\rep{1}\,,\rep{2}\,,+\frac{1}{2})_\mathbf{F} \oplus(\rep{1}\,,\repb{2}\,,-\frac{1}{2})_\mathbf{F}^{\omega_1} \otimes (\rep{1}\,,\rep{1}\,,+1)_\mathbf{F}^{}  \non
	&& \oplus  (\rep{1}\,,\rep{1}\,,0)_\mathbf{F}^{\omega_1^{\prime}} \otimes  (\rep{1}\,,\repb{2}\,,+\frac{1}{2})_\mathbf{F}^{\prime} \oplus (\rep{1}\,,\rep{1}\,,0)_\mathbf{F}^{\omega_1^{\prime\prime}}\otimes  (\rep{1}\,,\rep{2}\,,+\frac{1}{2})_\mathbf{F}  \Big] \otimes (\rep{1}\,,\repb{2}\,,-\frac{1}{2})_{\mathbf{H}\,,\omega_1}  \non && \times \frac{d_{\mathscr{A}}W_{ \repb{4}\,,{\rm IV} }  w_{ \repb{3}\,,{\rm V} } V_{ \repb{4}\,,{\rm VI} }  }{2\sqrt{2}M_{\rm{pl}}}  (\rep{1}\,,\repb{2}\,,-\frac{1}{2})_{\mathbf{H}\,,\omega_1}^\dag \otimes\langle (\rep{1}\,,\repb{2}\,,+\frac{1}{2})_\mathbf{H}^{\prime\prime \prime \dag}\rangle+H.c. \non
	&\Rightarrow& \frac{ Y_\Bc   d_{ \mathscr A}}{ 4 } \frac{ W_{ \repb{4}\,,{\rm IV} }  w_{ \repb{3}\,,{\rm V} } V_{ \repb{4}\,,{\rm VI} } }{ M_{\rm pl}  m_{( \rep{1}\,, \repb{4 }\,, - \frac{1}{4 })_{ \rep{H}\,,3} }^2}  ( \underline{ b_L {b_R}^c  } +  \underline{ \tau_L {\tau_R}^c } + \check \Nc_L^3 {\nG_R^{\prime\prime}}^c   - \check \Nc_L^{ 3^\prime} {\nG_R^{\prime}}^c +\check \Nc_L^{ 3^{\prime \prime}} {\nG_R}^{c}  ) v_{\rm EW}  + H.c. \,.
\end{eqnarray}
If we consider all three possibilities of the propagator masses, each of them leads to the $(b\,,\tau)$ masses as follows
\beqs   \label{eq:SU8_C_H3mass}
\beqn
m_{( \rep{1}\,, \repb{4 }\,, - \frac{1}{2 })_{ \rep{H}\,,3} } \sim \Oc(v_{441}) ~&:&~  m_b = m_\tau = \frac{ Y_\Bc  d_{ \mathscr A} }{ 4 }  \zeta_2 \zeta_{13} v_{\rm EW}  \,,\\[1mm]
m_{( \rep{1}\,, \repb{4}\,, - \frac{1}{2})_{ \rep{H}\,,3} } \sim \Oc(v_{431}) ~&:&~ m_b = m_\tau = \frac{ Y_\Bc  d_{ \mathscr A} }{ 4}  \zeta_1 \zeta_{23} v_{\rm EW} \,, \\[1mm]
m_{( \rep{1}\,, \repb{4 }\,, - \frac{1}{2 })_{ \rep{H}\,,3} } \sim \Oc(v_{421}) ~&:&~  m_b = m_\tau = \frac{ Y_\Bc  d_{ \mathscr A}  }{ 4 }  \frac{ \zeta_1 }{\zeta_{23} } v_{\rm EW}  \,.
\eeqn
\eeqs

\para
The indirect Yukawa couplings from the operator in Eq.~\eqref{eq:d5_Hmixing_B} are expected to generate the first- and second-generational down-type quark and charged lepton masses.
The gauge-invariant subset of $\Large( \repb{28_H}_{\,,\dot 1 }^\dag \repb{28_H}_{\,,\dot {\rm VII} }  \Large)$ can develop the VEV of $\langle  \repb{28_H}_{\,,\dot 1 }^\dag \repb{28_H}_{\,,\dot {\rm VII} } \rangle = \frac{1}{2} w_{\repb{3}\,,  \dot 1} w_{\repb{3}\,,  \dot {\rm VII} } \sim \Oc ( v_{431}^2 )$ according to the VEV assignments in Eq.~\eqref{eq:SU8_WWS_Higgs_VEVs_mini02}.
Similar to the indirect Yukawa couplings in Eq.~\eqref{eq:WSW_btau_indirect}, we should look for the EWSB components from the $\repb{28_{H}}_{\,, \dot 1 \,, \dot 2 }$ here.
For the Yukawa coupling of $\repb{8_F}^{\dot \omega_1} \rep{56_F}  \repb{28_{H}}_{\,, \dot \omega_1 }$, we find the mass terms of
\begin{eqnarray}\label{eq:C-smu_indirect}
	&& Y_\Dc   \repb{8_F}^{\dot \omega_1} \rep{56_F}  \repb{28_{H}}_{\,, \dot \omega_1 }  \times \frac{ d_{ \mathscr B} }{ M_{\rm pl} }   \repb{28_{H}}_{\,, \dot \omega_1 }^\dag   \repb{28_{H}}_{\,, \dot \omega_2 }^\dag  \rep{70_{H}}^\dag  \Large( \repb{28_H}_{\,,\dot 1 }^\dag \repb{28_H}_{\,,\dot {\rm VII} } \Large) + H.c. \non
	&\supset& Y_\Dc \Big[ ( \repb{4 } \,, \rep{1 } \,,  + \frac{1 }{4 } )_{ \rep{F}}^{\dot \omega_1} \otimes ( \rep{4 } \,, \rep{6 } \,,  + \frac{1 }{4 } )_{ \rep{F}} \oplus ( \rep{1 } \,, \repb{4 } \,,  - \frac{1 }{4 } )_{ \rep{F}}^{\dot \omega_1 } \otimes ( \rep{1 } \,, \repb{4 } \,,  + \frac{3 }{4 } )_{ \rep{F}} \Big]  \otimes ( \rep{1 } \,, \rep{6 } \,,  - \frac{1 }{2 } )_{ \rep{H}\,, \dot \omega_1} \non
	&\times&  \frac{  d_{ \mathscr B} }{ M_{\rm pl} } ( \rep{1 } \,, \rep{6 } \,,  - \frac{1 }{2 } )_{ \rep{H}\,, \dot \omega_1}^\dag \otimes ( \repb{4 } \,, \repb{4 } \,,  0 )_{ \rep{H}\,, \dot \omega_2 }^\dag \otimes ( \rep{4 } \,, \repb{4 } \,,  + \frac{1 }{2 } )_{ \rep{H}}^\dag \otimes \langle  \repb{28_H}_{\,,\dot 1 }^\dag \repb{28_H}_{\,,\dot {\rm VII} }  \rangle  + H.c. \non
	&\supset& Y_\Dc \Big[ ( \repb{4 } \,, \rep{1 } \,,  + \frac{1 }{4 } )_{ \rep{F}}^{\dot \omega_1} \otimes ( \rep{4 } \,, \rep{3 } \,,  + \frac{1 }{12 } )_{ \rep{F}} \oplus ( \rep{1 } \,, \repb{3 } \,,  - \frac{1 }{3 } )_{ \rep{F}}^{\dot \omega_1 } \otimes ( \rep{1 } \,, \repb{3 } \,,  + \frac{2 }{3 } )_{ \rep{F}}^{\prime} \Big]  \otimes ( \rep{1 } \,, \repb{3 } \,,  - \frac{1 }{3 } )_{ \rep{H}\,, \dot \omega_1} \non
	&\times&  \frac{  d_{ \mathscr B} }{ M_{\rm pl} } ( \rep{1 } \,, \repb{3 } \,,  - \frac{1 }{3 } )_{ \rep{H}\,, \dot \omega_1}^{\dag} \otimes ( \repb{4 } \,, \repb{3 } \,,  -\frac{1}{12} )_{ \rep{H}\,, \dot \omega_2 }^\dag \otimes ( \rep{4 } \,, \repb{3 } \,,  + \frac{5 }{12 } )_{ \rep{H}}^\dag \otimes \langle  \repb{28_H}_{\,,\dot 1 }^\dag \repb{28_H}_{\,,\dot {\rm VII} }  \rangle  + H.c. \non
	&\supset& Y_\Dc d_{ \mathscr B}  \frac{ w_{\repb{3}\,,  \dot 1} w_{\repb{3}\,,  \dot {\rm VII} }  }{ 2 M_{\rm pl} m_{ ( \rep{1 } \,, \rep{6 } \,,  - \frac{1 }{ 2 } )_{ \rep{H}\,, \dot  \omega_1 }  }^2 }    \Big[  ( \repb{4 } \,, \rep{1 } \,,  + \frac{1 }{4 } )_{ \rep{F}}^{\dot \omega_1} \otimes ( \rep{4 } \,, \rep{2 } \,,  +\frac{1}{4} )_{ \rep{F}}^\prime \oplus ( \rep{1 } \,, \repb{2 } \,, -\frac{1}{2} )_{ \rep{F}}^{\dot \omega_1} \otimes ( \rep{1 } \,, \rep{1 } \,,  + 1 )_{ \rep{F}}^{ \prime }  \non
	&\oplus& ( \rep{1 } \,, \rep{1 } \,,  0 )_{ \rep{F}}^{\dot \omega_1 ^\prime } \otimes ( \rep{1 } \,, \repb{2 } \,,  +\frac{1 }{2 } )_{ \rep{F}}^{\prime\prime\prime } \Big] \otimes   \langle ( \repb{4 } \,, \rep{1 } \,,  + \frac{1 }{4 } )_{ \rep{H}\,, \dot \omega_2 }^\dag \rangle  \otimes  \langle ( \rep{4 } \,, \repb{2 } \,,  + \frac{1 }{4 } )_{ \rep{H}}^{ \dag} \rangle + H.c. \non
	&\Rightarrow&  \frac{ Y_\Dc d_{ \mathscr B} }{4 }  \dot {\zeta}_3   \Big[ \frac{   w_{\repb{3}\,,  \dot 1}  w_{\repb{3}\,,  \dot {\rm VII} } }{  m_{ ( \rep{1 } \,, \rep{6 } \,,  - \frac{1 }{ 2 })_{ \rep{H}\,, \dot  1 }   }^2 }   ( \underline{ d_L {d_R}^c } +  \underline{ e_L {e_R}^c  }  )  + \frac{  w_{\repb{3}\,,  \dot 1} w_{\repb{3}\,,  \dot {\rm VII} } }{  m_{ ( \rep{1 } \,, \rep{6 } \,,  - \frac{1 }{ 2 })_{ \rep{H}\,, \dot  2 }  }^2 }   ( \underline{ d_L {s_R}^c } +  \underline{ \mu_L {e_R}^c  }  )  \Big] v_{\rm EW}  + H.c. \,,
\end{eqnarray}
where the SM quark/lepton components from the $\repb{8_F}^{\dot \omega_1= \dot 1}$/$\repb{8_F}^{\dot \omega_1= \dot 2}$ correspond to the $({d_R}^c \,, e_L)$ and $({s_R}^c \,, \mu_L)$, respectively.
The other mass terms read
\begin{eqnarray}\label{eq:C-de_indirect}
	&& Y_\Dc  \repb{8_F}^{\dot \omega_1} \rep{56_F}  \repb{28_{H}}_{\,, \dot \omega_1 } \times \frac{ d_{ \mathscr B} }{ M_{\rm pl} }   \repb{28_{H}}_{\,, \dot \omega_1 }^\dag   \repb{28_{H}}_{\,, \dot \omega_2 }^\dag  \rep{70_{H}}^\dag \Large( \repb{28_H}_{\,,\dot 2 }^\dag \repb{28_H}_{\,,\dot {\rm VII} } \Large)  + H.c. \non
	&\supset& Y_\Dc  \Big[  ( \repb{4 } \,, \rep{1 } \,,  + \frac{1 }{4 } )_{ \rep{F}}^{\dot \omega_1} \otimes ( \rep{6 } \,, \rep{4 } \,,  - \frac{1 }{4 } )_{ \rep{F}} \oplus ( \rep{1 } \,, \repb{4 } \,,  - \frac{1 }{4 } )_{ \rep{F}}^{\dot \omega_1 } \otimes ( \rep{4 } \,, \rep{6 } \,,  + \frac{1 }{4 } )_{ \rep{F}} \Big]  \otimes ( \repb{4 } \,, \repb{4 } \,,  0 )_{ \rep{H}\,, \dot \omega_1} \non
	&& \times \frac{ d_{ \mathscr B} }{ M_{\rm pl} }  ( \repb{4 } \,, \repb{4 } \,,  0 )_{ \rep{H}\,, \dot \omega_1}^\dag \otimes ( \rep{1 } \,, \rep{6 } \,,  - \frac{1 }{2 } )_{ \rep{H}\,, \dot \omega_2 }^\dag  \otimes ( \rep{4 } \,, \repb{4 } \,,  + \frac{1 }{2 } )_{ \rep{H}}^\dag \otimes \langle  \repb{28_H}_{\,,\dot 2 }^\dag \repb{28_H}_{\,,\dot {\rm VII} } \rangle + H.c. \non
	&\supset& Y_\Dc  \Big[  ( \repb{4 } \,, \rep{1 } \,,  + \frac{1 }{4 } )_{ \rep{F}}^{\dot \omega_1} \otimes ( \rep{6 } \,, \rep{3 } \,,  - \frac{1 }{6 } )_{ \rep{F}} \oplus ( \rep{1 } \,, \repb{3 } \,,  - \frac{1 }{3 } )_{ \rep{F}}^{\dot \omega_1 } \otimes ( \rep{4 } \,, \repb{3 } \,,  + \frac{5 }{12 } )_{ \rep{F}}  \oplus ( \rep{1 } \,, \rep{1} \,,  0 )_{ \rep{F}}^{\dot \omega_1^{\prime\prime} } \otimes ( \rep{4 } \,, \rep{3 } \,,  + \frac{1 }{12 } )_{ \rep{F}} \Big]  \non
	&& \otimes ( \repb{4 } \,, \repb{3 } \,,  -\frac{1}{12} )_{ \rep{H}\,, \dot \omega_1} \times\frac{ d_{ \mathscr B}  }{  M_{\rm pl} }  ( \repb{4 } \,, \repb{3 } \,,  -\frac{1}{12} )_{ \rep{H}\,, \dot \omega_1}^{\dag} \otimes ( \rep{1 } \,, \repb{3 } \,,  -\frac{1}{3} )_{ \rep{H}\,, \dot \omega_2}^{\dag}\otimes( \rep{4 } \,, \repb{3 } \,,  +\frac{5}{12} )_{ \rep{H}\,, \dot \omega_1}^{\dag}  \non
	&& \otimes \langle  \repb{28_H}_{\,,\dot 2 }^\dag \repb{28_H}_{\,,\dot {\rm VII} } \rangle+H.c.\non
	&\supset& Y_\Dc  \Big[  ( \repb{4 } \,, \rep{1 } \,,  + \frac{1 }{4 } )_{ \rep{F}}^{\dot \omega_1} \otimes ( \rep{6 } \,, \rep{2 } \,,  0 )_{ \rep{F}}^{} \oplus  ( \rep{1 } \,, \repb{2 } \,,  - \frac{1 }{2 } )_{ \rep{F}}^{\dot \omega_1} \otimes   ( \rep{4 } \,, \rep{1 } \,,  + \frac{3 }{4 } )_{ \rep{F}}^{} \oplus ( \rep{1 } \,, \rep{1 } \,,  0 )_{ \rep{F}}^{\dot \omega_1^{ \prime } } \otimes   ( \rep{4 } \,, \repb{2 } \,,  + \frac{1 }{4 } )_{ \rep{F}}^{ } \non
	&& \oplus ( \rep{1 } \,, \rep{1 } \,,  0 )_{ \rep{F}}^{\dot \omega_1^{ \prime \prime } } \otimes   ( \rep{4 } \,, \rep{2 } \,,  + \frac{1 }{4 } )_{ \rep{F}}^{\prime } \Big] \otimes ( \repb{4 } \,, \repb{2 } \,,  - \frac{1 }{ 4} )_{ \rep{H}\,, \dot \omega_1}^{  }    \times d_{ \mathscr B} \frac{ \omega_{\repb{3}\,,  \dot 2} \omega_{\repb{3}\,,  \dot {\rm VII} }^2  }{ 2\sqrt{2} M_{\rm pl} } ( \repb{4 } \,, \repb{2 } \,,  - \frac{1 }{ 4} )_{ \rep{H}\,, \dot \omega_1}^{\,\dag}  \langle ( \rep{4 } \,, \repb{2 } \,,  + \frac{1 }{ 4} )_{ \rep{H}}^{\,\dag } \rangle  + H.c.  \non
	&\supset& Y_\Dc d_{ \mathscr B}  \frac{ w_{\repb{3}\,,  \dot 1} w_{\repb{3}\,,  \dot {\rm VII} }^2  }{ 2 \sqrt{2} M_{\rm pl} m_{  (\repb{4 } \,, \repb{4} \,,   0  )_{ \rep{H}\,, \dot  \omega_1}  }^2 } \Big[  ( \repb{3 } \,, \rep{1 } \,,  + \frac{1 }{3 } )_{ \rep{F}}^{\dot \omega_1} \otimes ( \rep{3 } \,, \rep{2 } \,,  +\frac{1 }{6 } )_{ \rep{F}}^{ \prime \prime \prime } \oplus  ( \rep{1 } \,, \repb{2 } \,,  - \frac{1 }{2 } )_{ \rep{F}}^{\dot \omega_1} \otimes   ( \rep{1 } \,, \rep{1 } \,,  + 1 )_{ \rep{F}}^{\prime \prime \prime } \non
	&& \oplus  ( \rep{1 } \,, \rep{1 } \,,  0 )_{ \rep{F}}^{\dot \omega_1^{ \prime } } \otimes   ( \rep{1 } \,, \repb{2 } \,,  + \frac{1 }{2 } )_{ \rep{F}}^{\prime \prime \prime \prime \prime }   \oplus  ( \rep{1 } \,, \rep{1 } \,,  0 )_{ \rep{F}}^{\dot \omega_1^{ \prime \prime } } \otimes   ( \rep{1 } \,, \rep{2 } \,,  + \frac{1 }{2 } )_{ \rep{F}}^{\prime\prime \prime\prime }   \Big] \otimes  \langle ( \rep{1 } \,, \repb{2 } \,,  + \frac{1 }{2 } )_{ \rep{H}}^{\prime \prime \prime \,\dag } \rangle  + H.c. \non
	&\Rightarrow& \frac{ Y_\Dc d_{ \mathscr B} }{4 } \dot \zeta_2   \Big[ \frac{w_{\repb{3}\,,  \dot 1} w_{\repb{3}\,,  \dot {\rm VII} }   }{  m_{ ( \repb{4} \,, \repb{4 } \,,  0 )_{ \rep{H}\,, \dot  1}  }^2 }   ( \underline{ s_L {d_R}^c } + \underline{ e_L {\mu_R}^c } )  + \frac{ w_{\repb{3}\,,  \dot 1} w_{\repb{3}\,,  \dot {\rm VII} }   }{  m_{ ( \repb{4 } \,, \repb{4 } \,,  0 )_{ \rep{H}\,, \dot  2 }  }^2 }   ( \underline{ s_L {s_R}^c } + \underline{ \mu_L {\mu_R}^c } )   \Big] v_{\rm EW} + H.c.  \,.
\end{eqnarray}

\para
With the Higgs VEV assignments in \eqref{eq:SU8_WWS_Higgs_VEVs_mini03}, we expect the following natural propagator masses of
\begin{eqnarray} \label{eq:SU8_C_H1H2mass}
	&  m_{ ( \rep{1 } \,, \rep{6 } \,,  - \frac{1 }{2 } )_{ \rep{H}\,, \dot  1} } \sim \Oc( v_{431 } ) \,,
	&m_{ ( \rep{1 } \,, \rep{6 } \,,  - \frac{1 }{2 } )_{ \rep{H}\,, \dot  2} } \sim \Oc( v_{421 } ) \, ,\non
	& m_{ ( \repb{4 } \,, \repb{4 } \,,  0)_{ \rep{H}\,, \dot  1} } \sim \Oc( v_{431 } ) \,,
	& m_{ ( \repb{4 } \,, \repb{4 } \,,  0 )_{ \rep{H}\,, \dot  2} } \sim \Oc( v_{421 } ) \,.
\end{eqnarray}
For convenience, we parametrize the following ratios of
\begin{eqnarray} \label{eq:SU8_C_Delta}
	&& \Delta_{ \dot \omega } \equiv \frac{w_{\repb{4}\,,  \dot 1} w_{\repb{4}\,,  \dot {\rm VII} }   }{  m_{ ( \repb{4} \,, \repb{4 } \,,   0 )_{ \rep{H}\,, \dot {\omega}}   }^2 }        \,, \quad \Delta_{ \dot \omega}^\prime \equiv \frac{ w_{\repb{4}\,,  \dot 1} w_{\repb{4}\,,  \dot {\rm VII} }  }{  m_{ ( \rep{1 } \,, \rep{6 } \,,   -\frac{ 1}{ 2} )_{ \rep{H}\,, \dot  {\omega} }  }^2 }\,.
\end{eqnarray}
%
%
%

\subsection{The SM quark/lepton masses and the CKM mixing}

\para
For all up-type quarks with $Q_e=+\frac{2}{3}$, we write down the following tree-level masses from both the renormalizable Yukawa couplings and the gravity-induced terms in the basis of $\Uc\equiv (u\,,c\,,t)$:
\beqs\label{eqs:WWS_Uquark_masses}
\beqn
\Mc_u   &=&   \frac{1}{\sqrt{2} }  \left( \ba{ccc}  
0 & c_4  \dot \zeta_3^\prime/\sqrt{2} & c_5 \zeta_1 /\sqrt{2}    \\
0  &  0   & c_5 \zeta_3/\sqrt{2}  \\
c_5 \zeta_1/\sqrt{2}   &  c_5 \zeta_3/\sqrt{2}  &   Y_\Tc \\  \ea  \right) v_{\rm EW}  \approx \Mc_u^{ (0)} + \Mc_u^{ (1 ) } + \Mc_u^{( 2)}   \,, \\[1mm]
\Mc_u^{ (0)}    &=& \frac{1}{\sqrt{2} }  \left( \ba{ccc}  
0 & 0   & 0   \\
0   &  0  & 0   \\
0  &  0  &   Y_\Tc   \\  \ea  \right)  v_{\rm EW} \,, \label{eq:WWS_Uquark_mass00} \\[1mm]
\Mc_u^{ (1)}    &=&  \frac{1}{\sqrt{2} }  \left( \ba{ccc}  
0 &   0  & c_5 \zeta_1/\sqrt{2}  \\
0 & 0 & 0    \\
c_5 \zeta_1 /\sqrt{2}&  0  &0    \\  \ea  \right) v_{\rm EW}  \,, \label{eq:WWS_Uquark_mass01}  \\[1mm]
\Mc_u^{ (2)}    &=& \frac{1}{\sqrt{2} }  \left( \ba{ccc}  
0&  c_4 \dot \zeta_3^\prime/\sqrt{2}   &0     \\
0  &   0 & c_5 \zeta_3  /\sqrt{2}  \\
0 &  c_5 \zeta_3/\sqrt{2}  &   0   \\   \ea  \right) v_{\rm EW}   \,,\label{eq:WWS_Uquark_mass02}
\eeqn
\eeqs
where we have neglected the terms of $\sim \Oc(\zeta_3\,v_{\rm EW})$ in the expansion.
One obvious feature is that the gauge eigenstates of the up quark and the charm quark do not obtain tree-level masses through the $d=5$ operators with the SM Higgs doublet.
Instead, there are only off-diagonal mass mixing terms in Eqs.~\eqref{eq:WWS_Uquark_mass01} and \eqref{eq:WWS_Uquark_mass02}.
Accordingly, we find that
\begin{eqnarray} \label{eq:SU8_C_mt2}
	{\rm det}^\prime \[ \Mc_u^{ (0)} \Mc_u^{ (0)\, \dag}   \] &=& \frac{1}{2}  Y_\Tc^2 \,  v_{\rm EW}^2  \Rightarrow m_t^2 \approx \frac{ 1 }{  2 } Y_\Tc^2 \, v_{\rm EW}^2 \,.
\end{eqnarray}
Here and below, we use the ${\rm det}^\prime$ to denote the matrix determinant that is equal to the products of all nonzero eigenvalues.
Next, we find that
\begin{eqnarray} \label{eq:SU8_C_Mu_0_1}
	{\rm det}^\prime \[ \Big( \Mc_u^{ (0)} + \Mc_u^{ (1)}  \Big) \cdot  \Big( \Mc_u^{ (0)\, \dag} + \Mc_u^{ (1)\, \dag} \Big)  \] &\approx& \frac{1}{4 } c_5^4  \zeta_1^4  \,  v_{\rm EW}^4 \,.
\end{eqnarray}
Naturally, we expect the smaller eigenvalue above to be the charm quark mass squared of
\begin{eqnarray} \label{eq:SU8_C_mc2}
	m_c^2 &=& {\rm det}^\prime \[ \Big( \Mc_u^{ (0)} + \Mc_u^{ (1)}  \Big) \cdot \Big( \Mc_u^{ (0)\, \dag} + \Mc_u^{ (1)\, \dag} \Big)  \]   \Big/ {\rm det}^\prime \[ \Mc_u^{ (0)} \Mc_u^{ (0)\, \dag}   \] \approx    \frac{ c_5^4 \zeta_1^4 }{8 Y_\Tc^2 } \, v_{\rm EW}^2 \,.
\end{eqnarray}
The up quark mass squared can be similarly obtained by
\begin{eqnarray}\label{eq:SU8_C_mu2}
	m_u^2 &=& {\rm det} \[  \Big( \Mc_u^{ (0)} + \Mc_u^{ (1)} + \Mc_u^{ (2)}  \Big) \cdot \Big( \Mc_u^{ (0)\, \dag} + \Mc_u^{ (1)\, \dag} + \Mc_u^{ (2)\, \dag} \Big)  \]  \non
	&&  \Big/ {\rm det}^\prime \[ \Big( \Mc_u^{ (0)} + \Mc_u^{ (1)}  \Big) \cdot \Big( \Mc_u^{ (0)\, \dag} + \Mc_u^{ (1)\, \dag} \Big)  \]   \approx \frac{c_4^2 \zeta_3^2 \dot \zeta_3^{2} }{4 \zeta_1^2 }\, v_{\rm EW}^2 \,.
\end{eqnarray}
To summarize, all SM up-type quark masses are expressed as follows:
\begin{eqnarray}\label{eq:WWS_SMumasses}
	&&  m_u \approx c_4 \frac{ \zeta_3 \dot\zeta_3 }{ 2 \zeta_1 } v_{\rm EW} \,,\quad m_c \approx c_5^2  \frac{ \zeta_1^2 }{2 \sqrt{2} Y_\Tc } v_{\rm EW}  \,, \quad  m_t \approx \frac{ Y_\Tc }{ \sqrt{2} } v_{\rm EW} \,.
\end{eqnarray}

\para
For all down-type quarks with $Q_e=-\frac{1}{3}$, we find the following tree-level SM mass matrix:
\begin{eqnarray}\label{eq:WWS_Dquark_massMatrix}
	&&  \Big( \Mc_d \Big)_{3\times 3}   \approx  \frac{1}{4} \left( \ba{ccc}
	( 2 c_3  +  Y_\Dc d_{\mathscr B} )  \dot \zeta_3  &(  2 c_3  +  Y_\Dc d_{\mathscr B}   \zeta_{23 }^{-2 } ) \dot \zeta_3   &   0   \\
	( 2 c_3 + Y_\Dc d_{\mathscr B}  )\dot \zeta_2  &( 2 c_3  +  Y_\Dc d_{\mathscr B}  \zeta_{23 }^{-2 } ) \dot \zeta_2  &  0   \\
	0 & 0  &  Y_\Bc d_{\mathscr A}   \zeta_{23}^{-1} \zeta_1   \\
	\ea  \right) v_{\rm EW}    \,.
\end{eqnarray}
This mass matrix has the similar structure as in Eq.~\eqref{eq:WSW_Dquark_massdd} from the WSW pattern, while the terms in the second row receive the $\Gc_{431}$ breaking VEVs of $w_{\repb{3},\,\dot 1, \dot{\rm{VII}}  }$.
It is straightforward to find the following SM down-type quark masses of
\beqs\label{eqs:WWS_SMdmasses}
\beqn
m_b &\approx&   \frac{ 1 }{ 4  }  Y_\Bc d_{\mathscr A} \zeta_{ 23 }^{-1} \zeta_1 \, v_{\rm EW} \,,  \\[1mm]
m_s  &\approx& \frac{1 }{4 }\left( 2c_3+  Y_\Dc d_{\mathscr B} \zeta_{ 23 }^{-2 } \right) \dot \zeta_2 \, v_{\rm EW} \,,  \\[1mm]
m_d &\approx& \hf c_3 \dot \zeta_3   \, v_{\rm EW} \,,
\eeqn
\eeqs
from Eq.~\eqref{eq:WWS_Dquark_massMatrix}.

\para
For all charged leptons with $Q_e=-1$, their tree-level mass matrix is correlated with the down-type quark mass matrix as
\begin{eqnarray} \label{eq:WWS_Lepton_massll}
	\Big( \Mc_\ell \Big)_{ 3\times 3} &=&   \Big( \Mc_d^T \Big)_{3\times 3} \non
	&\approx& \frac{1}{4}  \left( \ba{ccc}
	( 2 c_3  +  Y_\Dc d_{\mathscr B} )  \dot \zeta_3  &( 2 c_3 + Y_\Dc d_{\mathscr B}  )\dot \zeta_2  &  0   \\
	(  2 c_3  +  Y_\Dc d_{\mathscr B}   \zeta_{23 }^{-2 } ) \dot \zeta_3    &( 2 c_3  +  Y_\Dc d_{\mathscr B}  \zeta_{23 }^{-2 } ) \dot \zeta_2 &   0   \\
	0 & 0  &  Y_\Bc d_{\mathscr A}   \zeta_{23}^{-1} \zeta_1   \\  \ea   \right) v_{\rm EW}  \,.
\end{eqnarray}
Thus, it is straightforward to find the tree-level mass relations of
\begin{eqnarray}  \label{eq:WWS_SMleptonmasses}
	&& m_\tau = m_b \,,\quad m_\mu = m_s \,,\quad m_e = m_d  \,.
\end{eqnarray}


\providecommand{\href}[2]{#2}\begingroup\raggedright\endgroup

\end{document}